\documentclass[12pt]{article}
\usepackage[toc,page]{appendix}

\usepackage[right=1in,left=1in,top=1in,bottom=1in]{geometry}
\usepackage{hyperref} 
\hypersetup{colorlinks, citecolor=blue, filecolor=blue, linkcolor=blue, urlcolor=blue}
\usepackage{graphicx}
\usepackage{xcolor}
\usepackage{url}
\usepackage{natbib}
 \bibpunct[, ]{(}{)}{,}{a}{}{,}%

\usepackage{amsmath,amsthm,amssymb} 
\usepackage{float}
\usepackage{bibunits}
\usepackage{bbm}
\usepackage{threeparttable}

\usepackage{booktabs}
\usepackage[labelsep=period]{caption} %
\definecolor{amethyst}{rgb}{0.6, 0.4, 0.8}
\definecolor{ao(english)}{rgb}{0.0, 0.5, 0.0}
\usepackage{subcaption}

\usepackage{multirow} %

\theoremstyle{plain}
\newtheorem{assumption}{Assumption}
\newtheorem{corollary}{Corollary}

\newtheorem{defn}{Definition}

\newtheorem{prop}{Proposition}

\usepackage{setspace}
\usepackage{sectsty}

\sectionfont{\large}
\subsectionfont{\normalsize}
\subsubsectionfont{\normalsize}

\newcommand{\squishlist}{
   \begin{list}{$\bullet$}
    { \setlength{\itemsep}{0pt} \setlength{\parsep}{1pt}
      \setlength{\topsep}{1pt} \setlength{\partopsep}{1pt}
      \setlength{\leftmargin}{1.5em} \setlength{\labelwidth}{1em}
      \setlength{\labelsep}{0.5em} } }

\newcommand{\squishlisttwo}{
   \begin{list}{$\bullet$}
    { \setlength{\itemsep}{0pt} \setlength{\parsep}{0pt}
      \setlength{\topsep}{0pt} \setlength{\partopsep}{0pt}
      \setlength{\leftmargin}{1em} \setlength{\labelwidth}{1.5em}
      \setlength{\labelsep}{0.5em} } }

\newcommand{\squishend}{
    \end{list}  }
\newcommand{\argmax}{\arg\!\max}

\usepackage{titlesec}

\titlespacing*{\section}{0pt}{6pt}{3pt}
\titlespacing*{\subsection}{0pt}{5pt}{2pt}
\titlespacing*{\subsubsection}{0pt}{4pt}{2pt}
\titlespacing*{\paragraph}{0pt}{4pt}{2pt}

\title{Auctions Meet Bandits: An Empirical Analysis}
\author{%
  \begin{minipage}[t]{0.3\textwidth}\centering
    Mohammad Rashid\thanks{All authors contributed equally and are listed in reverse alphabetical order. We would like to thank the participants of the 2025 Marketing Science Conference, 2026 UT Dallas BASS Forms conference, 2026 AI in Management (AIM) Conference, Imperial Business School, 2026 Workshop on Platform Analytics (WoPA). We also thank Ron Berman, Garrett Johnson, Pallavi Pal, Valeria Stourm, Eric Schwartz, Yuyan Wang, and Dennis Zhang for their detailed comments, which have significantly improved the paper. Please address all correspondence to: soheil.ghili@yale.edu, or83@cornell.edu, and rashid98@uw.edu.}\\
    \vspace{0.2em}
    \small University of Washington
  \end{minipage}
  \hfill
  \begin{minipage}[t]{0.3\textwidth}\centering
    Omid Rafieian\footnotemark[1]\\
    \vspace{0.2em}
    \small Cornell University
  \end{minipage}
  \hfill
  \begin{minipage}[t]{0.3\textwidth}\centering
    Soheil Ghili\footnotemark[1]\\
    \vspace{0.2em}
    \small Yale University
  \end{minipage}
}
\date{October 17, 2026}

\begin{document}
\maketitle

\begin{abstract}

Sponsored search positions are typically allocated through real-time auctions, where the outcomes depend on advertisers' quality-adjusted bids—the product of their bids and quality scores. Although quality scoring helps promote ads with higher conversion outcomes, setting these scores for new advertisers in any given market is challenging, leading to the cold-start problem. To address this, platforms incorporate multi-armed bandit algorithms in auctions to balance exploration and exploitation. However, little is known about the optimal exploration strategies in such auction environments. We utilize data from a leading Asian mobile app store that places sponsored ads for keywords. The platform employs a Thompson Sampling algorithm within a second-price auction to learn quality scores and allocate a single sponsored position for each keyword. We empirically quantify the gains from optimizing exploration under this combined auction-bandit model and show that this problem differs substantially from the canonical bandit problem. Drawing on these empirical insights, we propose a customized exploration strategy in which the platform adjusts the exploration levels for each keyword according to its characteristics. We derive the Pareto frontier for revenue and efficiency and provide actionable policies, demonstrating substantial gains for the platform on both metrics when using a tailored exploration approach.

\end{abstract}
\newpage

\section{Introduction}
\label{sec:intro}
The auction problem and the multi-armed bandit (MAB) problem are both fundamental problems in the fields of economics and computer science. In applied settings, firms often need to solve problems that combine elements from auctions and bandits. A prime example is sponsored search. In sponsored-search settings, for each search term, a platform auctions off advertisement slots to bidders whose ads have different probabilities of conversion (e.g., click, install). These conversion probabilities are unknown to the platform a priori, rendering the problem one of multi-armed bandit, in which the platform needs to balance exploration against exploitation to achieve near optimal performance. At the same time, advertisers have heterogeneous valuations for each conversion, known to them but unknown to the platform. Allocating the slot on the basis of these bids is an auction problem. The score for each bidder in such an auction is calculated by combining its bid amount with the platform's most current belief about the bidder's conversion probability. This renders the platform's problem one that combines auction and bandit features.

Although the auction and bandit literatures are more or less separate, the overlap between the two is a small but growing body of research \citep{devanur2009price, babaioff2015truthful, feng2023improved}. A common feature of all these studies is that they all investigate the theoretical properties of solutions to the auction-bandit problem. Our objective, however, is \textit{empirical}. In this paper, we empirically examine how introducing the bandit component and fine-tuning the extent of exploration in these algorithms influence market outcomes such as revenue and efficiency. Specifically, we quantify the benefits a platform can achieve by optimizing its exploration strategy under a combined auction-bandit model and highlight how this setting differs from a bandit-only problem. To our knowledge, this is the first attempt to document the empirical dynamics of the auction-bandit problem and connect them to market outcomes in a real-world environment.

We theoretically characterize a general auction-bandit model, where a seller auctions uncertain events with probabilities that can be learned sequentially. This framework encompasses classical auction design and bandit problems as special cases. We then characterize how the exploration challenge differs from the standard bandit problem, as rewards (prices) depend not only on the winning buyer, but also on other players in a competitive market environment. 

As an empirical application of the general problem, we analyze large-scale data from sponsored search auctions run by a leading app store in a major Asian country, where we observe complete information about all advertisers competing for an impression. The app store employs a Thompson Sampling algorithm to update advertisers' quality scores, integrated with a second-price auction.\footnote{Under Thompson Sampling, the platform starts each new ad with a prior belief about its conversion rate, updates this belief as it observes conversions, and ranks ads in each auction using a random draw from their current beliefs. The prior thus governs how much the platform explores: a more optimistic prior gives new ads more impressions before their performance is known.} The level of exploration in ad allocation is high enough to allow us to estimate advertisers' conversion rates, while still preserving room for strategic bidding behavior. This enables us to invert their equilibrium bids and structurally estimate advertisers' valuations per conversion, assuming they are utility-maximizing. Together, our setting allows us to identify advertisers' \textit{private valuation for conversion} and \textit{conversion rates} as the mechanism-invariant structural primitives, sufficient for evaluation of market outcomes under counterfactual mechanisms.

We focus on a class of Thompson Sampling Second-Price (TS-SP) auctions that differ from the mechanism used in the data only in their prior distribution of quality scores and evaluate market outcomes under each. More specifically, we consider the optimal choice of a prior belief about the conversion rate (``CVR'') of any new bidder, and we study how this optimal choice is shaped by the auction-bandit problem. To this end, we compare two metrics. First, we calculate the maximum revenue that the platform can achieve by optimally choosing the prior within a given range. This is the ``full'' problem in which both the auction and the bandit problems play roles in shaping the optimal choice. Second, we calculate the optimal prior that maximizes the efficiency (total surplus) and evaluate its revenue. We show that this second optimization treats the platform's problem as essentially a bandit-only problem. Comparing market outcomes under the revenue- and efficiency-maximizing priors therefore allows us to isolate the role of the auction component in the joint auction-bandit problem. 

We find that the interactions between the two sides of the platform's problem are indeed of empirical relevance. Specifically, we find that to maximize efficiency, the platform optimally chooses a prior mean of 0.002 for entrants' (new ads') conversion rate. This prior is above the median entrant's conversion rate in our data (0.0013), so even a platform that cares only about efficiency is somewhat optimistic about new ads in order to learn about them, as in a standard bandit problem. Beyond this level, however, further exploration misallocates the position and efficiency falls steadily. 

When we solve for the revenue-maximizing prior, however, the recommended prior is at the substantially larger level of 0.1. In addition, the ensuing expected revenue is about 32\% higher than the revenue arising from a prior of 0.002. We interpret this wide gap as affirmative of the relevance of the combined auction-bandit problem. We further investigate the mechanism behind this empirical result. In particular, we find that higher priors often help boost revenue by elevating the entrant's score to the second place and close to that of the incumbent winner, thereby allowing the auctioneer to extract additional surplus from the winner. This revenue channel does not exist in pure bandit problems. We further confirm the empirical relevance of this revenue channel by decomposing the platform revenue into revenue accrued from auctions where (i) the entrant wins, (ii) the entrant finishes second, and (iii) the entrant finishes third or lower. Our results show that the revenue maximizing prior of 0.1 improves upon the profitability of the efficiency maximizing prior of 0.002 by boosting the revenue in channel (ii) while revenue from channel (i) falls, though to a lesser degree.

We interpret our empirical finding as follows: a higher conversion rate prior in an auction-bandit problem helps make the market effectively ``thicker'' by pushing the second score closer to the first score. Consistent with this interpretation, we find that in auctions for keywords where the first and second scores are typically more distant (we term these keywords ``thinner'' markets), the revenue difference between the revenue-maximizing and efficiency-maximizing conversion-rate priors is the highest.  

Further, we investigate the role of customizing the prior policy for each separate keyword on which auctions are run. We find that customization in the auction-bandit problem  improves the performance not only by boosting revenue and efficiency individually (which happens by construction), but also by ``easing'' the trade-off between the two: targeting helps to boost revenue for auctions that benefit from higher priors without compromising on the efficiency in auctions where a high prior for entrant might adversely impact the allocation.

We extend our analysis by introducing Upper Confidence Bound Second-Price (UCB–SP) mechanisms as another common exploration method in auctions. Using the same setting and data, we find that a UCB-style exploration policy improves the revenue of the optimal TS–SP by about 25\%, a gain we trace to UCB’s optimistic quality-score assignments to non-winner advertisers. Importantly, the revenue–efficiency tradeoff is similar to that of TS–SP: more exploratory UCB policies yield higher revenue at the cost of efficiency. 

Lastly, we run a series of robustness checks in which we relax several assumptions from our main analysis. First, we relax our assumptions on advertisers' budgets and further constrain bidder spending by imposing a daily cap equal to each bidder's maximum observed daily spend. The results of this robustness check exhibit qualitative and quantitative patterns similar to those in our main analysis. Second, we allow advertisers' valuations per conversion to vary over time and show that the results remain essentially unchanged. Finally, we relax the assumption of static bidding and solve a dynamic structural model with forward-looking bidders. We find that the estimated discount factor is zero for the vast majority of bidders. We further document that our main insights remain qualitatively unchanged when we directly estimate advertisers' discount factors or artificially make bidders more forward-looking and perform counterfactual analyses. 

In summary, our paper makes contributions to the academic literature while offering practical implications for managers and policymakers. Our key substantive contribution lies in documenting the nuances of the combined auction-bandit problem and how it differs from traditional bandit or auction settings. We empirically show that framing the problem of prior optimization as a combined auction-bandit problem leads to substantially higher priors and significantly greater revenue compared to treating it solely as a bandit problem. Further, we show that customizing priors for keywords is an actionable approach that balances the revenue–efficiency trade-off and generates significant improvements in market outcomes. While the primary managerial implications concern advertising platforms, our framework applies broadly to any setting where an auctioneer sells an uncertain event and must learn bidder relevance scores. Finally, our findings carry important policy implications for pricing transparency, showing that platforms can exploit exploration as a revenue-generating tool, sometimes at the expense of overall market efficiency.

\section{Related Literature}
\label{sec:literature}
Our paper relates to the literature on online advertising platforms. Early theoretical studies in this area have examined the equilibrium properties of different auction formats in sponsored search \citep{edelman2007internet, varian2007position, lahaie2007sponsored, lahaie2007revenue}. Empirical research has explored various aspects of advertising auctions, including measuring market outcomes under different auctions \citep{yao2011dynamic, athey2010structural, kim2025nonparametric}, incorporating consumer search behavior \citep{chan2015consumer, choi2019monetizing}, audience externalities and ad avoidance \citep{wilbur2013correcting, stourm2017incorporating}, reserve price optimization \citep{ostrovsky2023reserve}, and dynamic allocation mechanisms \citep{rafieian2020revenue}. Our study extends this body of work by empirically evaluating different bandit-style exploration strategies that platforms use to learn quality scores over time. In particular, we show that the choice of exploration strategy meaningfully affects market outcomes, highlighting it as an important managerial lever for advertising platforms.

Our work is also related to the market thickness and revenue-efficiency trade-off literature. Early studies have highlighted the critical importance of market thickness in determining market outcomes \citep{levin2010online}. To increase market thickness, several methods have been proposed from a range of perspectives, such as utilizing randomized allocation \citep{celis2014buy}, adjusting the exponent on quality scores in standard GSP auctions—a modification motivated by \citep{lahaie2007revenue}, reducing the level of targeting \citep{rafieian2021targeting}, and giving extra exposure to new advertisers to reduce churn \citep{ye2022cold}. We contribute to this literature by examining how exploration increases revenue by improving market thickness and reducing dominant players’ informational rents. We demonstrate this by empirically benchmarking the performance of two prominent bandit algorithms: Thompson Sampling and UCB.

Finally, our work is closely related to the literature on the interplay between auctions and bandits. Recent research has made significant theoretical advances in developing algorithms that improve regret bounds in this context \citep{babaioff2009characterizing, devanur2009price, babaioff2015truthful, feng2023improved}. Our work adds to this literature in two ways. First, rather than focusing on asymptotic regret bounds, we provide finite-sample empirical estimates of market outcomes using large-scale data and quantify the benefits of considering the full auction-bandit problem compared to the bandit problem alone. Second, we extend the analysis of auction-bandit problems to Thompson Sampling in second-price auctions, providing valuable insights for practitioners looking to leverage Thompson Sampling algorithms due to their strong real-world performance \citep{chapelle2011empirical}.

\section{Model}
\label{sec:model}

\subsection{The General Auction-Bandit Problem}
\label{ssec:gen_aucband}

We consider a general problem in which a seller owns a sequence of items indexed by $t \in \{1, \ldots, T\}$ (e.g., advertising slots) and sells an uncertain event that may occur on each item (e.g., a click or conversion). There is a set of buyers $\mathcal{A} = \{1, \ldots, N\}$ interested in purchasing these events. Each buyer $a$ is characterized by two primitives:
\squishlist
\item A private valuation $v_{a,t}$ for the uncertain event on item $t$, known to the buyer but not to the seller.
\item An event probability $\mu_a$, the probability that the uncertain event occurs if item $t$ is allocated to buyer $a$. This probability is fixed across items but unknown to both the seller and the buyer ex ante.\footnote{The typical formulation of this problem treats $\mu_a$ as time-invariant, as we do throughout. Bandit learning can, however, accommodate structured forms of time-varying $\mu_{a,t}$ (e.g., smoothly drifting or periodic event probabilities).} The seller can learn $\mu_a$ over time by observing whether the event occurs when items are allocated to buyer $a$.
\squishend
In each period $t$, the seller runs a mechanism consisting of an allocation rule and a payment rule. Each buyer $a$ submits a bid $b_{a,t}$, which may or may not equal their true valuation $v_{a,t}$. Let $b_t = (b_{1,t}, \ldots, b_{N,t})$ denote the vector of bids in period $t$. The mechanism $M_t = (p_t, e_t)$ specifies:
\squishlist
\item An allocation rule $p_t(b_t, \hat{\mu}_t)$, where $p_{a,t}(b_t, \hat{\mu}_t) \in \{0, 1\}$ indicates whether buyer $a$ receives the item. Here $\hat{\mu}_t$ denotes the seller's current belief about the vector of event probabilities, updated as the seller learns over time.
\item A payment rule $e_t(b_t, \hat{\mu}_t)$, where $e_{a,t}(b_t, \hat{\mu}_t)$ is the per-event payment by buyer $a$ if allocated the item and if the event occurs.
\squishend
If buyer $a$ is allocated the item in period $t$, the event occurs with probability $\mu_a$. Conditional on the event occurring, buyer $a$ receives a net payoff of $v_{a,t} - e_{a,t}(b_t, \hat{\mu}_t)$ and the seller collects $e_{a,t}(b_t, \hat{\mu}_t)$. The seller's objective is to choose a sequence of mechanisms to maximize total expected revenue:
\begin{equation}\label{eq:general_objective}
    \max_{\{(p_t, e_t)\}_{t=1}^T} \mathbb{E} \left[ \sum_{t=1}^{T} \sum_{a \in \mathcal{A}} p_{a,t}(b_t, \hat{\mu}_t) \, \mu_a \, e_{a,t}(b_t, \hat{\mu}_t) \right]
\end{equation}
subject to incentive compatibility and individual rationality constraints.\footnote{Maximizing efficiency instead of revenue yields a weighted variant of the bandit problem: under a direct revelation mechanism in which bidders truthfully report their valuations, $e_{a,t}(b_t, \hat{\mu}_t)$ in the objective is replaced by the reported $b_{a,t} = v_{a,t}$.} Crucially, the seller's beliefs $\hat{\mu}_t$ evolve over time as the seller observes event outcomes for the buyers to whom items are allocated, creating a dynamic learning problem.
 
We term the above the \textit{auction-bandit problem} because the seller must simultaneously address two challenges: (i) eliciting buyers' private valuations and determining allocations and payments---the auction problem; and (ii) learning the unknown event probabilities $\mu_a$ through sequential allocation decisions---the bandit problem. To see how these two problems interact, it is instructive to examine two special cases in which the general problem reduces to each one individually:
\squishlist
\item \textit{The Auction Problem.} If the $\mu_a$'s are known to the seller, there is no learning problem: They enter the objective as known weights, and the problem reduces to standard optimal auction design \citep{myerson1981optimal}, with the seller's sole task being to design allocation and payment rules that extract surplus from buyers with heterogeneous private valuations.
\item \textit{The Bandit Problem.} If all buyers share a common valuation, $v_{a,t} = v > 0$ for all $a$ and $t$, the optimal per-event payment is trivially $v$, so the payment rule factors out, and the problem reduces to allocating each item to the buyer with the highest $\mu_a$. Since the $\mu_a$'s are unknown, this is the classical multi-armed bandit problem \citep{lattimore2020bandit}.\footnote{Marketing literature has looked at the bandit problem in different contexts, such as advertising \citep{schwartz2017customer}, pricing \citep{misra2019dynamic, jain2024effective}, movie recommendation \citep{rafieian2026rational}, and content design \citep{hauser2009website,ye2025lola}.}
\squishend
In the general auction-bandit problem, the seller faces both challenges simultaneously, and their interaction creates forces absent from either special case alone. In the canonical bandit problem, the optimal exploration strategy balances two opposing forces: under-exploration risks missing high-quality buyers, while over-exploration leads to inefficient allocation to low-quality buyers. The auction-bandit problem introduces a third force: because payments in an auction depend not only on the winning buyer but also on competition among buyers, exploration can increase market thickness, which translates into higher revenue through increased payments. This third force has no analogue in the pure bandit problem, where payments are fixed.
 
Characterizing the optimal solution to the general problem is difficult, as even the bandit special case lacks a closed-form optimal solution in general. Our approach is therefore empirical: we study a specific and practically important instance of the auction-bandit problem---sponsored search advertising---and quantify how the interaction between the auction and bandit components shapes optimal exploration and market outcomes.

\subsection{Our Empirical Application: Thompson Sampling Second-Price Auction}
 
Our data come from a leading Android app store in a large Asian country, which holds over 70\% of the market share in the Android app store category. The app store has a total of 100 million users, with more than 40 million active users each month. On the supply side, the app store hosts nearly one million mobile apps developed by 80,000 developers. Like most app stores, its primary function is to serve as a matchmaker between users and apps, and its main source of revenue comes from taking a share of transactions between users and apps facilitated through the platform. In this study, we focus on another key monetization strategy of the app store: sponsored search advertising.
 
The rest of this section develops our model of this setting. Section \ref{sssec:market_environment} describes the market environment and the mechanics of the platform's sponsored search auction. Section \ref{sssec:exploration} explains, both intuitively and formally, how the platform explores new advertisers using Thompson Sampling. Section \ref{sssec:tssp_mechanism} formalizes the resulting Thompson Sampling Second-Price (TS-SP) mechanism. Section \ref{sssec:bidding} then characterizes advertisers' equilibrium bidding behavior under this mechanism.
 
\subsubsection{The Auction Module: Single-Slot Sponsored Search Auction}
\label{sssec:market_environment}
 
\begin{figure}
    \centering
    \includegraphics[width=1\linewidth]{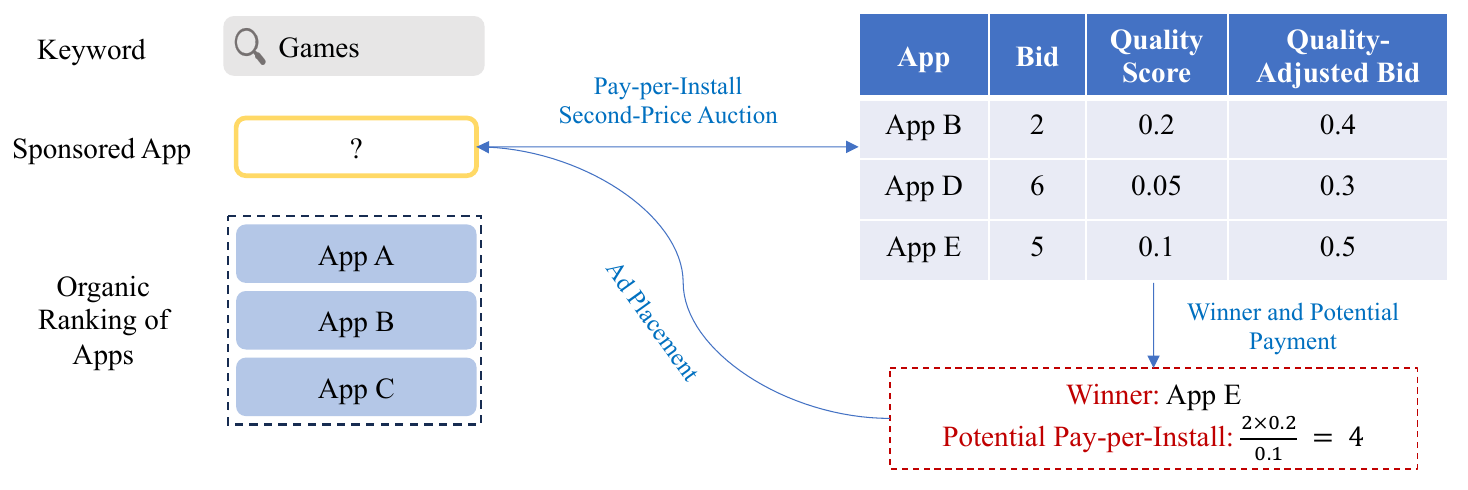}
    \caption{Sponsored position allocation through the auction used by the app store}
    \label{fig:auction}
\end{figure}

Users frequently search for apps by entering keywords, with nearly 5.5 million search sessions taking place on an average day. When a user submits a query, the app store's algorithm provides a rank-ordered list of organic search results relevant to the query, together with a single sponsored slot that allows apps to pay for greater visibility at the top of the results. The app store allocates this slot through an auction, making the app store the auctioneer and mobile apps the advertisers (bidders) in our setting; we use ``apps'' and ``ads'' interchangeably throughout. Concretely, the app store runs a pay-per-install second-price auction:
\begin{enumerate}
    \item When a user searches for a keyword, the app store's server receives a request to allocate the sponsored position through an auction.
    \item The app store solicits bids from participating apps. App $a$'s bid, $b_a$, represents the amount it is willing to pay per install.
    \item Each participating app $a$ is assigned a quality score $q_a$, reflecting its expected install probability (conversion rate). The app store computes a quality-adjusted bid, $b_a q_a$, for each app.
    \item The app with the highest quality-adjusted bid is selected. Without loss of generality, assume $b_1 q_1 > b_2 q_2 > \cdots > b_M q_M$. If $b_1 q_1$ exceeds the reserve price, app 1 wins the sponsored slot and is shown to the user.
    \item App 1 pays only if the user installs the app; clicks without a subsequent install trigger no payment. The per-install payment is the second price, the minimum bid needed to still win, so app 1 pays $b_2 q_2 / q_1$ per install.\footnote{If the reserve price exceeds the second-highest quality-adjusted bid, the payment is instead set so that the winning quality-adjusted bid just clears the reserve price.}
\end{enumerate}
Figure \ref{fig:auction} illustrates this process for a search for \textit{games}. Apps B, D, and E compete for the sponsored slot with bids of 2, 6, and 5; the app store assigns each a quality score, computes quality-adjusted bids, and awards the slot to App E, the highest quality-adjusted bidder. App E pays only if the user installs the app, and its per-install payment is the minimum amount needed to keep its quality-adjusted bid above App B's (the second-highest bidder) or above the reserve price, whichever binds.
 
We map this environment onto the general model as follows. The platform is the seller, and the items are impressions: opportunities to show a sponsored ad to a user who has searched for a keyword. There are multiple keywords, indexed by $k \in \mathcal{K}$, each generating its own sequence of impressions $t = 1, \ldots, T_k$. For impression $t$ of keyword $k$, a set of advertisers $\mathcal{A}_{k,t} \subseteq \mathcal{A}$ compete for the single sponsored slot, where $\mathcal{A}$ is the set of all advertisers. Each competing advertiser $a$ is characterized by a private valuation per install, $v_{a,k,t}$, and a true conversion rate $\mu_{a,k,t}$, the probability a user installs the app if shown ad $a$, which corresponds to the event probability $\mu_a$ in the general model. The quality score $q_a$ used operationally in the auction above is the platform's real-time estimate of this conversion rate. The platform does not know $\mu_{a,k,t}$ ex ante and must learn it through exploration, which is the source of the bandit problem in this context; we turn to how the platform forms and updates $q_a$ next.
 
\subsubsection{The Bandit Module: Learning via Thompson Sampling}
\label{sssec:exploration}
 
Because quality scores directly enter the allocation rule, they are essential to profitable allocation: in Figure \ref{fig:auction}, the highest bid does not win because its quality score is substantially lower than its competitors'. Advertising platforms typically fine-tune quality scores using large amounts of historical performance data. The challenge is that the app store has no such data for apps that are new to an auction, creating a cold-start problem: exploring new apps risks allocating impressions inefficiently, but never exploring them risks permanently missing high-quality advertisers.
 
The app store addresses this with a Thompson Sampling approach: it starts each new ad-keyword pair with a prior quality-score distribution and updates that distribution in a Bayesian fashion as impressions are allocated and outcomes are observed. Because conversion outcomes are binary, the platform models the quality score with a Beta distribution, which is conjugate to the Bernoulli conversion process. Let $\alpha_{a,k,t}$ and $\beta_{a,k,t}$ denote the Beta parameters for ad $a$ on keyword $k$ at impression $t$, so that the quality score used in the auction is drawn as
\[
q_{a,k,t} \sim \text{Beta}(\alpha_{a,k,t}, \beta_{a,k,t}).
\]
Let $A_{k,t}$ denote the ad allocated impression $t$ of keyword $k$, and let $Y_{k,t} \sim \text{Bern}(\mu_{a,k,t})$ denote the resulting conversion outcome when $A_{k,t}=a$. The parameters update only for the winning ad, and only based on the realized outcome:
\begin{equation}\label{eq:updating}
    (\alpha_{a,k,t+1}, \beta_{a,k,t+1}) = (\alpha_{a,k,t}, \beta_{a,k,t}) + \mathbbm{1}(A_{k,t} = a)(Y_{k,t}, 1 - Y_{k,t}),
\end{equation}
so that a win with a conversion adds one unit to $\alpha$, a win without a conversion adds one unit to $\beta$, and a loss leaves the parameters unchanged. Equivalently, at any point the parameters can be written in terms of an ad's cumulative track record,
\begin{equation}\label{eq:updating2}
    (\alpha_{a,k,t}, \beta_{a,k,t}) = (\alpha^0 + C_{a,k,t},\; \beta^0 + I_{a,k,t} - C_{a,k,t}),
\end{equation}
where $(\alpha^0,\beta^0)$ are the prior parameters and $C_{a,k,t}$, $I_{a,k,t}$ are the total conversions and impressions ad $a$ has accumulated on keyword $k$ before impression $t$.
 
\begin{figure}
    \centering
    \includegraphics[width=0.5\linewidth]{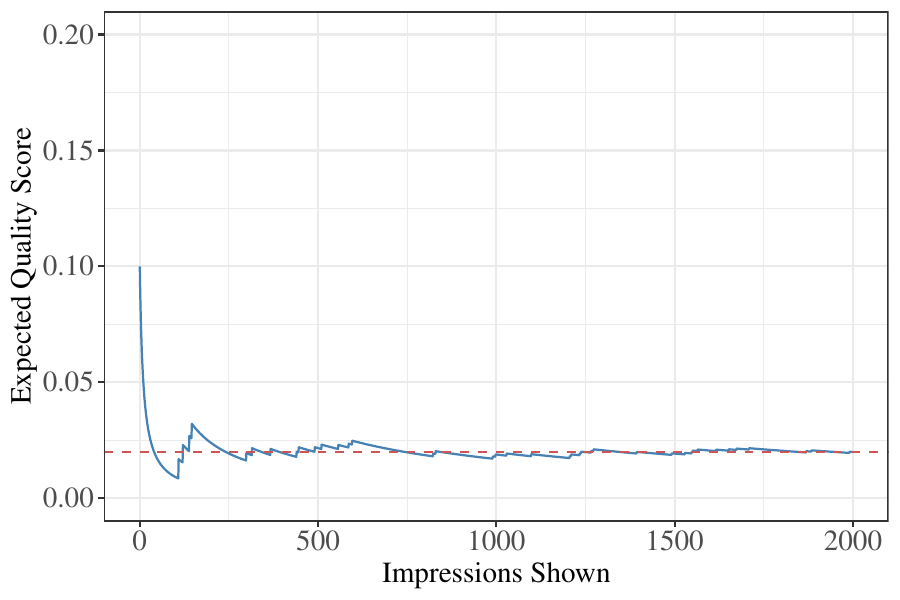}
    \caption{Posterior mean of quality score over time for one trajectory}
    \label{fig:tsupdate}
\end{figure}

The app store initializes every new ad-keyword pair from $\text{Beta}(1,9)$, a prior mean of $0.1$. To see how this converges, suppose an app accumulates 10,000 impressions and 200 conversions (a 0.02 conversion rate). Its posterior becomes $\text{Beta}(1+200,\, 9+10000-200) = \text{Beta}(201, 9809)$, with mean $201/10010 \approx 0.02$, indistinguishable from the true rate. Figure \ref{fig:tsupdate} plots the posterior mean along one such trajectory: it starts at $0.1$, swings noticeably in early periods when a single conversion still moves the estimate a lot, and settles down near the true conversion rate of $0.02$ as impressions accumulate.
 
The prior parameters $(\alpha^0, \beta^0)$ are what govern the extent of exploration, since they set how quickly, and from where, the quality score moves toward the truth. We formalize this as an exploration policy:
\begin{defn}\label{def:ts_policy}
The \textbf{Thompson Sampling exploration policy} is characterized by two functions $\alpha^0 : \mathcal{A} \times \mathcal{K} \to \mathbb{R}$ and $\beta^0 : \mathcal{A} \times \mathcal{K} \to \mathbb{R}$ that set the initial Beta parameters for an ad $a \in \mathcal{A}$ entering the market for keyword $k \in \mathcal{K}$: if ad $a$ enters the market for keyword $k$ at impression $t$, then $\alpha_{a,k,t} = \alpha^0(a,k)$ and $\beta_{a,k,t} = \beta^0(a,k)$.
\end{defn}
In our data, the platform applies a uniform policy, $\alpha^0(a,k) = 1$ and $\beta^0(a,k) = 9$ for all ad-keyword pairs. A key objective of our analysis is to evaluate market outcomes under alternative choices of $(\alpha^0, \beta^0)$.
 
\subsubsection{The TS-SP Auction Mechanism}
\label{sssec:tssp_mechanism}
 
The exploration policy can be embedded within any auction format. In our empirical context, it is embedded within the pay-per-install second-price auction described above, together with the reserve price that each auction must clear. We define the combined mechanism as follows:
\begin{defn}\label{def:tssp}
The \textbf{Thompson Sampling Second-Price (TS-SP)} auction mechanism $M(p, e; \alpha^0, \beta^0)$ is characterized by allocation rule $p$, payment rule $e$, and the Thompson Sampling exploration policy $(\alpha^0, \beta^0)$. For each impression, the mechanism proceeds in three steps:
\squishlist
\item \textit{Quality Score Sampling}: For each ad $a \in \mathcal{A}_{k,t}$, a quality score is drawn from its current posterior. Let $C_{a,k,t}^{M}$ and $I_{a,k,t}^{M}$ denote the total number of conversions and impressions ad $a$ receives in keyword $k$ before impression $t$ under mechanism $M$:
\[
q_{a,k,t}^M \sim \text{Beta}\left(\alpha^0(a,k) + C_{a,k,t}^M,\; \beta^0(a,k) + I_{a,k,t}^M - C_{a,k,t}^M\right)
\]
\item \textit{Allocation}: Let $b_{a,k,t}^{M}$ denote the bid submitted by ad $a$ in auction $t$ of keyword $k$ under mechanism $M$, and let $r_k$ denote the reserve price for keyword $k$, expressed in quality-adjusted units. The impression is allocated to the ad with the highest quality-adjusted bid, provided that bid clears the reserve; otherwise the slot goes unsold:
\[
p\left(\{b_{a,k,t}^M, q_{a,k,t}^M\}_{a \in \mathcal{A}_{k,t}};\, r_k\right) =
\begin{cases}
\displaystyle \argmax_{a \in \mathcal{A}_{k,t}} b_{a,k,t}^M \, q_{a,k,t}^M & \text{if } \displaystyle\max_{a \in \mathcal{A}_{k,t}} b_{a,k,t}^M \, q_{a,k,t}^M \geq r_k, \\[1.5ex]
\varnothing & \text{otherwise.}
\end{cases}
\]
We write $A_{k,t}^M = p\left(\{b_{a,k,t}^M, q_{a,k,t}^M\}_{a \in \mathcal{A}_{k,t}};\, r_k\right)$ for the winning ad whenever $p(\cdot) \neq \varnothing$.
\item \textit{Payment}: The winning ad pays only if a conversion occurs. The per-conversion payment is the minimum bid that would have guaranteed winning against both its rivals and the reserve:
\[
e\left(\{b_{a,k,t}^M, q_{a,k,t}^M\}_{a \in \mathcal{A}_{k,t}};\, r_k\right) = \frac{\max\left\{r_k,\; \max_{a \in \mathcal{A}_{k,t} \setminus\{A_{k,t}^M\}} b_{a,k,t}^M \, q_{a,k,t}^M\right\}}{q_{A_{k,t}^M, k, t}^M}
\]
The realized payment is $Y_{k,t}^{M}$ times this per-conversion price, where $Y_{k,t}^{M}$ is the conversion outcome for the winning ad under mechanism $M$. When the second-highest quality-adjusted bid exceeds the reserve, the winner pays that score divided by its own quality score; otherwise, the reserve binds and the winner pays $r_k/q_{A_{k,t}^M,k,t}^M$ instead. In either case, the payment is owed only if a conversion occurs.
\squishend
\end{defn}
\noindent Unlike the standard second-price auction with fixed quality scores, the allocation outcome of the TS-SP auction is probabilistic because quality scores are drawn from posterior distributions that evolve over time.
 
\subsubsection{Advertisers' Equilibrium Bidding}
\label{sssec:bidding}
 
We now characterize how a forward-looking advertiser should bid under the TS-SP mechanism. Fix advertiser $a$ and keyword $k$, and let $s_{k,t} = \{(\alpha_{a',k,t}, \beta_{a',k,t})\}_{a' \in \mathcal{A}}$ denote the belief states of all advertisers on keyword $k$ at impression $t$. If $a$ wins impression $t$ and a conversion occurs, it earns $v_{a,k,t} - e_{a,k,t}$; otherwise it earns nothing that period. Given the rivals' bidding policy profile $\sigma_{-a}$, and taking expectations over the quality-score draws and the conversion outcome, the per-period (stage) expected profit from bidding $b_{a,k,t}$ in state $s_{k,t}$ is
\begin{equation}\label{eq:stage_utility}
    \bar{\pi}^{\sigma_{-a}}_{a,k,t}(b_{a,k,t}, s_{k,t}) = \mu_{a,k,t} \left(v_{a,k,t} - \mathbb{E}\left[e_{a,k,t} \mid A_{k,t} = a, b_{a,k,t}, s_{k,t}, \sigma_{-a}\right]\right) \Pr\left(A_{k,t} = a \mid b_{a,k,t}, s_{k,t}, \sigma_{-a}\right),
\end{equation}
where $A_{k,t}$ is the winner under the allocation rule in Definition~\ref{def:tssp}. Because winning today changes the winner's Beta parameters through the updating rule in equation \eqref{eq:updating}, a forward-looking advertiser's bid also affects its continuation value. Let $V^{\sigma_{-a}}_{a,k,t}(s_{k,t})$ denote its optimal value given rivals' policies. Its problem can be written as a Bellman equation with discount factor $\delta$:
\begin{equation}\label{eq:bellman}
    V^{\sigma_{-a}}_{a,k,t}(s_{k,t}) = \max_{b_{a,k,t}} \mathbb{E} \Big[ \bar{\pi}^{\sigma_{-a}}_{a,k,t}(b_{a,k,t}, s_{k,t}) + \delta\, V^{\sigma_{-a}}_{a,k,t+1}(s_{k,t+1}) \,\Big|\, s_{k,t}, b_{a,k,t}, \sigma_{-a} \Big],
\end{equation}
where $s_{k,t+1}$ evolves according to equation \eqref{eq:updating}, which depends on the bid $b_{a,k,t}$ through the probability that $a$ wins in \eqref{eq:stage_utility}. This dynamic link is what makes bidding under exploration a genuinely dynamic problem: an advertiser may want to bid above or below its valuation today to influence its own future quality score, or to let a rival win so that the rival's score is revised.

That link disappears, however, if the advertiser is static and the current quality score is predetermined with respect to the current bid, as it is under the TS-SP mechanism, where quality scores are drawn in the Quality Score Sampling step before bids are used in the Allocation step. We formalize this next.
 
\begin{prop}\label{prop:truthful}
Suppose $\delta = 0$ and, in every period, the quality score $q_{a,k,t}$ used for allocation is determined independently of the contemporaneous bid $b_{a,k,t}$. Then truthful bidding, $b_{a,k,t} = v_{a,k,t}$ for every advertiser $a$, keyword $k$, and impression $t$, is a weakly dominant strategy, and truthful bidding by all advertisers is an equilibrium of the TS-SP mechanism.
\end{prop}
 
\begin{proof}
See Appendix \ref{appsec:proof}.
\end{proof}
 
Two remarks help interpret Proposition \ref{prop:truthful}. First, both hypotheses are needed: if $\delta > 0$, winning an impression changes future Beta parameters through \eqref{eq:updating}, so an advertiser's bid affects $V^{\sigma_{-a}}_{a,k,t+1}(s_{k,t+1})$, and the stage-game argument in the proof no longer characterizes the advertiser's full objective, because forward-looking advertisers may then deviate from truthful bidding to manage the informational consequences of winning, particularly early in a keyword's life when quality scores are still close to the prior. Second, although our setting is a sponsored search auction, this truthfulness result is distinct from the generalized second-price (GSP) auction with multiple slots, where truthful bidding is generally not an equilibrium because a bidder's payment and slot depend on the bids of others in a way that creates incentives to shade across positions. Here, the platform allocates a single sponsored slot via a second-price rule, so once the quality score is held fixed with respect to the bid, Vickrey's logic applies directly and the mechanism inherits the strategy-proofness of the classic single-item second-price auction with a reserve.

\section{Data and Preliminary Analysis}
\label{sec:setting_data}

\subsection{Data}
\label{ssec:data}
We have impression-level data from an app store over the span of six months. As discussed earlier, each impression is a sponsored position for a keyword searched by a user that is allocated to an app through the auction mechanism described above. A unique feature of our dataset is that, in addition to observing information about the winning bidder selected for the impression, we also observe details about the losing bidders. This is different from most high-frequency auction data sets where we only have information about the winning bidder, which allows us to conduct more robust counterfactual analysis. In $\S$\ref{sssec:variables}, we describe the variables we observe for each impression. Next, in $\S$\ref{sssec:sample}, we discuss our sampling strategy. 

\subsubsection{Variables}
\label{sssec:variables}
For each impression (auction) in our data, we observe the following sets of information:
\squishlist
\item \textit{Timestamp}: This presents the exact time of the impression.
\item \textit{Keyword-specific information}: This information set includes the information about the keyword searched by the user and whether it is a brand or non-brand keyword. Consistent with the definition used in the literature, a brand keyword is a keyword or phrase that directly references a specific app in the app store. 
\item \textit{Bidder-specific information}: For each bidder who participates in the auction, we observe their bid, the quality score drawn from their quality score distribution, and the $\alpha$ and $\beta$ parameters of their quality score distribution. 
\item \textit{Reserve price}: This is the price set by the app store as the minimum quality-adjusted bid needed for participating in the auction. This helps discard bidders whose quality-adjusted bid (product of bid and quality score) is lower than the reserve price.
\item \textit{Winning app}: This presents the bidder (app) who wins the auction. This is the bidder with the highest quality-adjusted bid (if any). There are auctions that are unallocated because no bidder's quality-adjusted bid surpassed the reserve price.
\item \textit{Conversion}: This indicates whether the impression of the winning app resulted in an install or conversion event. 
\item \textit{Potential revenue-per-conversion}: This is the amount that the winning app would be charged if a conversion event happens. In the second-price auction run by the app store, this is the lowest bid the winning app could have submitted to maintain the winning spot, as illustrated in Figure \ref{fig:auction}.
\item \textit{Revenue}: This information shows the amount of revenue earned by the app store for the impression. If the conversion event happens, this amount is equal to the potential revenue-per-conversion. Otherwise, it is equal to zero. 
\squishend
The detailed set of variables observed at the auction level allows us to evaluate counterfactual policies different from the one implemented in the data in terms of a variety of market outcomes. 

\subsubsection{Sample Construction}
\label{sssec:sample}
To reduce the computational burden of our analysis, we follow the convention in the literature and employ a sampling strategy \citep{athey2010structural, rafieian2021targeting}. We focus on the first week of our data which contains 38.6 million impressions and 446828 conversion events. We specifically focus on impressions generated by non-brand keywords.\footnote {There is no specific restriction that prevents us from using the brand keywords. However, those keywords are naturally thin markets where there is a dominant player. As we will discuss later, market thickness is an important consideration in our analysis. Focusing on non-brand keywords allows us to share insights into how market thickness affects the app store's decision on how much to explore. Non-brand keywords represent 52\% of all impressions generated and 27\% of the total revenue.} We focus on the top 1,800 non-brand keywords that account for 70\% of the total non-brand keyword revenue and generate 5.18 million impressions during the first week we choose as our analysis sample.

An important feature of our data is the extent of exploration in ad allocation through the Thompson Sampling algorithm that allows us to observe the conversion rates for a wide range of apps. This feature enables us to estimate the true conversion rates and use them in our counterfactual analysis. Although we focus on the sample of auctions run in the first week of our data, we use the entire six months of data to estimate conversion rates to increase the reliability of our estimates and shrink the confidence bounds around them. 

\subsection{Summary Statistics}
\label{ssec:summary}
We now present some summary statistics on our data. 

\subsubsection{Keyword-level Characteristics}
\label{sssec:keyword}
We begin by presenting the keyword-level characteristics. As outlined earlier, we focus on the top 1,800 non-brand keywords. For each keyword, we measure the following variables: (1) total number of impressions, (2) total number of conversions, (3) conversion rate, (4) total number of bidders, (5) number of bidders per impression, (6) total number of distinct ads shown, (7) total number of entrants, (8) total revenue, and (9) revenue per impression. Table \ref{tab:summary} provides a summary of how these keyword-level characteristics are distributed across the top 1,800 keywords. It is important to note that the reported revenue figures have been scaled by a constant factor for confidentiality.

\begin{table}[ht]
\centering
\footnotesize{
\begin{tabular}{lccccc} 
\hline\hline \\[-1.8ex]
\textbf{Variable} & \textbf{Mean} & \textbf{SD} & \textbf{Min} & \textbf{Median} & \textbf{Max}\\ \hline
\\[-1.8ex] 
\textbf{Total No. of Impressions} & 2880 & 5333 & 35 & 1434 & 99303 \\
\textbf{Total No. of Conversions} & 29.1 & 92.1 & 1 & 9 & 2592 \\
\textbf{Conversion Rate} & 0.022 & 0.048 & 0.0002 & 0.0058 & 0.49  \\
\textbf{Total No. of Bidders} & 38.2 & 7.7 & 18 & 37 & 74 \\
\textbf{Avg. No. of Bidders per Impression} & 11.6 & 7.7 & 1.1 & 10.2 &  40.6 \\
\textbf{Total No. of Distinct Ads Shown} & 31.4 & 8.2 & 2 & 32 & 69 \\
\textbf{Total No. of Entrant Bidders} & 21.6 & 11.7 & 3 & 20 &  64 \\
\textbf{Total Revenue} & 666 & 1468 & 8.3 & 233 & 29729  \\
\textbf{Avg. Revenue per Impression} & 0.93 & 2.65 & 0.0014 & 0.14 & 38.1\\[0.1cm]
\hline\hline
\end{tabular}
\caption{Summary Statistics of the Keyword-level Characteristics. \label{tab:summary}}}
\end{table}

As shown in Table \ref{tab:summary}, we observe substantial heterogeneity across keywords in their key characteristics. In particular, we note a few patterns in this table. First, conversion rates vary widely across keywords, ranging from as low as 0.0002 to as high as 0.49, consistent with prior literature that shows variability in popularity of keywords \citep{jerath2014consumer}. Second, we observe that the number of distinct ads shown also reveals considerable variability, with auctions dominated by as few as two distinct ads shown and others that allocate to as many as 69 bidders. This pattern suggests a substantial variation in overall market competitiveness. Finally, a comparison of total bidders and entrant bidders per keyword shows that entrants constitute a significant share of participation, with the median number of entrants exceeding half of the median number of bidders. This pattern emphasizes the prominence of the cold-start problem in our empirical setting.

\subsubsection{Exploration}
\label{sssec:exploration_summary}
We now turn to one of the key ingredients of our study: exploration. As shown in Table \ref{tab:summary}, many keywords have a large number of entrants who enter their market at some point. The app store's strategy to overcome this cold-start problem is to assign a relatively high prior mean to the quality score distribution of entrants, so these entrants could win auctions and the app store could gradually converge to their true quality score. We show the empirical rationale behind this exploration in Figure \ref{fig:cvrdiff} in Appendix \ref{appsec:rationale}. In this section, we show some evidence on the extent of exploration in our data and the motivation for exploration. 

\begin{figure}[h]
    \centering
    \includegraphics[width=0.5\linewidth]{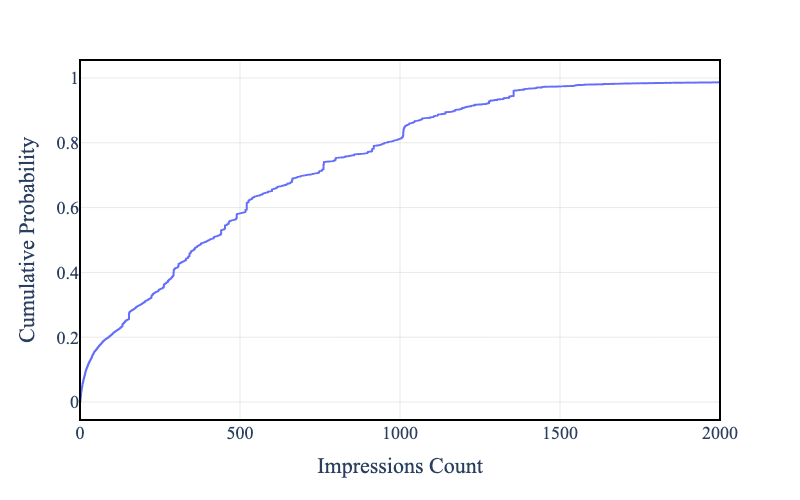}
    \caption{Empirical Cumulative Distribution Function of Impression Count for Entrants}
    \label{fig:entrant_cdf}
\end{figure}

Figure~\ref{fig:entrant_cdf} displays the empirical cumulative distribution function for the number of impressions received by entrants during the six-month period following their first bid. To capture the extent of exploration, we focus on all entrants who began bidding for the first time in the first week of data. They were then tracked over six months to record the total impressions allocated. As illustrated, 80\% of these entrants amassed at least 100 impressions, and 20\% exceeded 1,000 impressions, indicating a substantial level of exploration directed toward entrants. Later in our counterfactual analysis, we want to examine what would happen under different exploration policies. The extent of exploration in ad allocation allows us to use reliable estimates for entrants' conversion rates. 

\subsubsection{Market Thickness}
\label{sssec:thickness}
If a market is more competitive, entrants will face greater difficulty in securing impressions. As such, the choice of exploration policy becomes more sensitive to market thickness. In auction settings, market thickness can be quantified by the ratio of the second-highest bid to the highest bid, which reflects the fraction of the top bid that can be extracted as revenue in standard auction formats, such as second-price auctions. In our study, we define market thickness as the average ratio of the second-highest bidding score to the highest bidding score for a given keyword. By construction, thickness values lie between 0 and 1, with higher values indicating a more competitive auction environment.
\begin{figure}[h]
    \centering
    \includegraphics[width=0.5\linewidth]{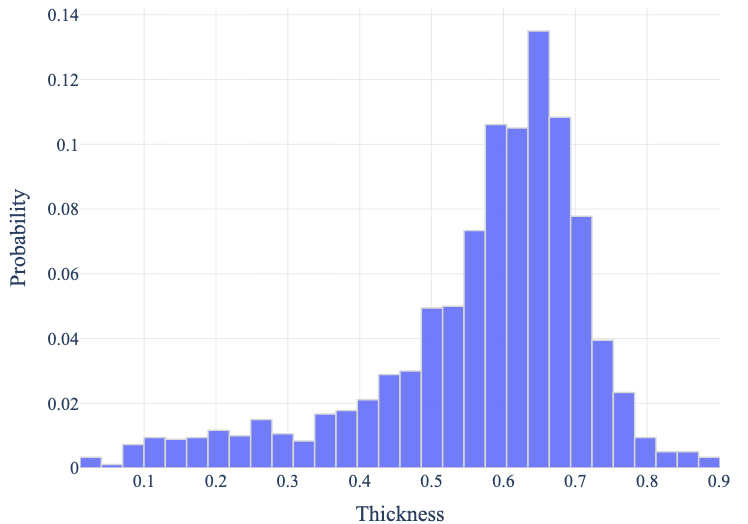}
    \caption{Histogram of the Market Thickness for Keywords}
    \label{fig:thickness}
\end{figure}

Figure \ref{fig:thickness} presents the histogram of market thickness in our sample data, revealing substantial heterogeneity in its distribution. The largest concentration in the distribution lies between 0.6 and 0.7, indicating that, for a significant portion of keywords, the top bidder extracts an informational rent of approximately 30-40\% of their valuation for an impression. We also observe instances of very thin markets, where the ratio of the second-highest bid to the highest bid is less than 0.1. Conversely, there are reasonably competitive markets with an average ratio close to 0.9. Using this measure of market thickness, we classify markets into three categories: the top 25\% of values are labeled as “thick” markets, the bottom 25\% as “thin” markets, and the interquartile range as “medium” markets—representing markets that are neither too competitive nor too thin.

\subsubsection{Role of Entrants in Market Outcomes}
\label{sssec:entrants_outcome}
We now examine the role of entrants in market revenue. An entrant can affect revenue in two ways. First, by taking the top place in the auction, the entrant wins the impression. Second, by finishing second, the entrant determines the price the winning bidder must pay. Conversely, if the entrant ends up in third place or lower, it does not influence the allocation or pricing of the auction. We present the fraction of auctions in which entrants occupy each of the top five positions in Appendix~\ref{appsec:entrant_positions}.

\begin{figure}[h]
    \centering
    \includegraphics[width=0.6\linewidth]{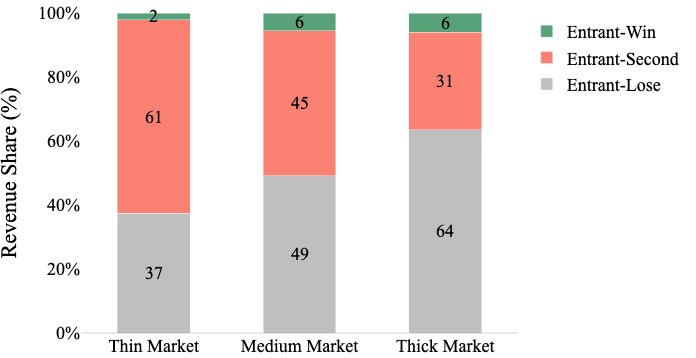}
    \caption{Revenue Share Based on Market Thickness and the Role of Entrants}
    \label{fig:revshare}
\end{figure}

To illustrate how entrants influence keyword revenue, we categorize revenue into three groups: (1) revenue from impressions won by entrants (Entrant-Win), (2) revenue from impressions where an entrant finishes second and influences pricing (Entrant-Second), and (3) revenue from impressions where incumbents take the top two places, leaving entrants no role in allocation or pricing (Entrant-None). Figure \ref{fig:revshare} presents these results. Although the revenue share from entrants winning impressions is relatively small, entrants significantly affect pricing by finishing second. Notably, the revenue share from entrants in second place is higher in thinner markets, where the gap between the highest and second-highest bidding scores is larger. The cross-keyword heterogeneity shown in Figure \ref{fig:revshare} underscores that the optimal exploration policy strongly depends on market thickness—a key concept we explore in our main analysis.

\subsection{Reduced-Form Evidence}
\label{ssec:reduced}
As discussed in $\S$\ref{sec:model}, auction outcomes are governed by two structural primitives: (1) bidders' private valuations for the event being sold, and (2) the true probability that the event occurs when the item is allocated to each bidder. In our setting, the former corresponds to advertisers' valuations for conversions and determines their bidding behavior, while the latter is the conversion rate. In this section, we first examine whether these structural primitives are fixed or time-varying in $\S$\ref{sssec:primitive_dynamics}. We then investigate whether there is any dynamic bidding by advertisers as a result of the platform's bandit learning of advertisers' quality scores in $\S$\ref{sssec:bidding_dynamics}.

\subsubsection{Model Primitives: Fixed or Time-Varying}
\label{sssec:primitive_dynamics}
\paragraph{Stability of conversion rates.}
We first examine whether conversion rates vary systematically over time. For each ad--keyword pair, we test whether the time trend in conversion outcomes is statistically different from zero. We present the regression specification and the distribution of the resulting $p$-values in Appendix~\ref{appsec:primitive_stability}. We observe a nearly uniform distribution of $p$-values. In particular, 5.23\% of the $p$-values are below 0.05. Thus, we do not find evidence that conversion rates change systematically over time.

\paragraph{Stability of bids.}
We next conduct a similar exercise for bids using our one-week bid sample. For each ad--keyword pair with within-pair bid variation, we test the null hypothesis that bids have no time trend against the alternative that the time trend is nonzero. Approximately 38\% of the pairs submit constant bids throughout the observation period, so the time-trend test is not defined for these pairs. We present the regression specification and the distribution of the resulting $p$-values in Appendix~\ref{appsec:primitive_stability}.

Unlike the results for conversion rates, we observe a deviation from the uniform distribution of $p$-values. Among the pairs for which the test is defined, 12\% have a $p$-value below 0.05, indicating the potential for time-varying valuations. Statistical significance, however, does not reveal whether the magnitude of these changes is economically meaningful.

To quantify the magnitude of bid variation, we therefore calculate the coefficient of variation separately for every ad--keyword pair:
\begin{equation}
    \operatorname{CV}_{ak}
    =
    \frac{\operatorname{SD}(b_{ak})}
    {\operatorname{Mean}(b_{ak})},
\end{equation}
where $\operatorname{SD}(b_{ak})$ and $\operatorname{Mean}(b_{ak})$ denote the standard deviation and mean of the bids submitted by advertiser $a$ for keyword $k$, respectively. Unlike the time-trend test, this measure captures the magnitude of bid variation relative to its average level.

\begin{figure}[!htp]
    \centering
    \includegraphics[width=0.5\textwidth]{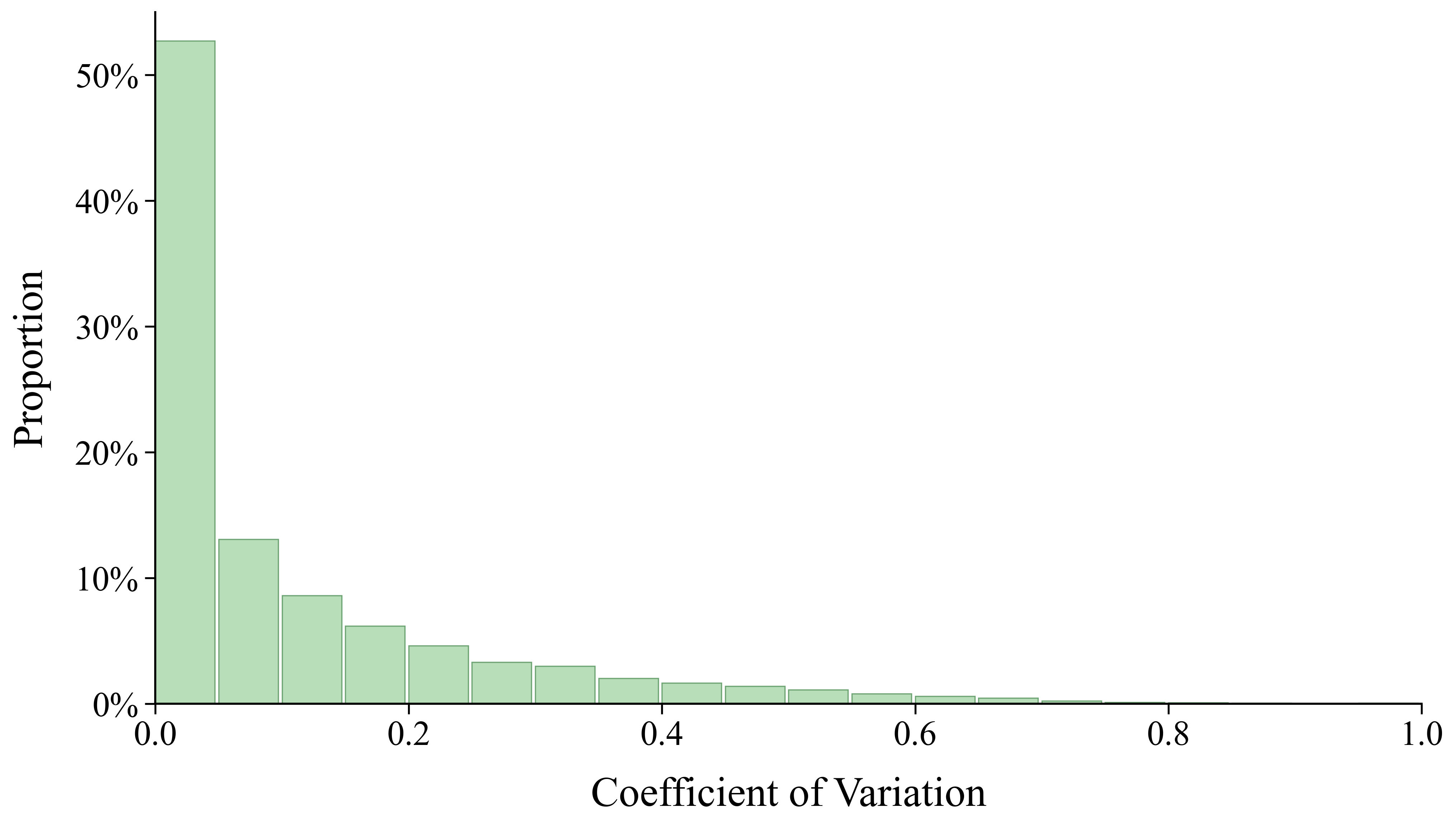}
    \caption{Distribution of the coefficient of variation in bids at the ad--keyword level}
    \label{fig:bid_cv}
\end{figure}

Figure~\ref{fig:bid_cv} presents the distribution of the coefficient of variation. The distribution is highly concentrated near zero: the median coefficient of variation is 0.039, and its 75th percentile is 0.154. Thus, although bids vary over time for some ad--keyword pairs, the magnitude of this variation is small for most pairs.

Taken together, these results suggest that treating advertisers' valuations as fixed over our one-week counterfactual period is a reasonable approximation. Nevertheless, in Section~\ref{ssec:time_varying_valuations}, we relax this assumption by allowing an advertiser's valuation in each auction to equal its contemporaneous submitted bid. That analysis shows that allowing valuations to vary over time does not materially change our results, either qualitatively or quantitatively.

\subsubsection{Bidding Dynamics}
\label{sssec:bidding_dynamics}
Even if advertisers' valuations per conversion remain fixed over time, they may still employ dynamic bidding strategies that maximize long-term utility. Such behavior can arise because of the platform's bandit learning of advertisers' quality scores: an advertiser's bid today may influence the evolution of these scores, thereby affecting future allocation and payment outcomes. To empirically test which strategy better aligns with the data, we consider different scenarios. If an advertiser’s bid does not affect quality score distributions, long-term utility maximization simplifies to per-period utility maximization. Thus, in settings where quality scores remain fixed over time (e.g., markets without new entrants), the distinction between these two utility-maximization assumptions becomes inconsequential.

However, the analysis becomes more nuanced when entrants enter the market and quality scores evolve dynamically. In these settings, current bidding behavior can influence the rate at which quality scores adjust, potentially affecting future auction outcomes. As such, long-term utility-maximization can lead to \textit{incumbents' bid-shading incentive} and \textit{entrants' overbidding incentives}. In the main text, we focus on the former and discuss the latter in Appendix \ref{appsec:overbidding}. 

Overestimating entrants' quality scores can result in situations where the winning incumbent ends up paying more than anticipated. In response, incumbents may reduce their bids temporarily, allowing entrants to win additional impressions. This, in turn, provides the auctioneer with more data to refine the entrants' quality scores. Over time, this improved accuracy in quality assessment is expected to lead to lower future prices for incumbents. Fortunately, we can empirically test these theoretical claims and determine whether bidders maximize their utility statically or dynamically. We begin with the incumbents' bid-shading incentive. According to this hypothesis, winning incumbents should reduce their bids when their costs suddenly increase, as a result of an entrant entering the market with a higher prior mean. To test this, we use data from such events and estimate the following model:
\begin{equation}
    \bar{b}_{a,k,h} = \beta_0 + \beta_1 \bar{b}_{a,k,h-1} + \beta_2 \text{CPC}_{a,k,h-1}+ \epsilon_{a,k,h},
\end{equation}
where $\bar{b}_{a,k,h}$ is incumbent $a$'s average bid in keyword $k$ at hour $h$, and $\text{CPC}_{a,k,h-1}$ is its average cost-per-conversion (CPC) in keyword $k$ in the previous time period $h-1$. We estimate this model and present the results in Column (1) of Table \ref{tab:dyntest}. As shown in this column, the CPC from the previous round does not appear to influence the incumbent's bid in the current period. Since the CPC in the prior period may be endogenous, we further investigate this relationship using an Instrumental Variable (IV) approach. Specifically, we use the entrant's quality score mean as the instrument to isolate the effect of the change in CPC due to the entrant’s prior quality score. The results of the 2SLS model are presented in Column (2) of Table \ref{tab:dyntest}, and we find the same conclusion: the increase in CPC resulting from the entrant’s higher-quality score does not lead to bid-shading behavior by the winning incumbent. This finding provides empirical evidence that supports our assumption of per-period utility maximization.

\begin{table}[t]
\caption{Regression Specifications to Test Static vs. Dynamic Bidding\label{tab:dyntest}}
\begin{minipage}{\columnwidth}
\begin{center}
\footnotesize{
\begin{tabular}[t]{lcc} 
\\[-1.8ex]\hline 
\hline \\[-1.8ex] 
 & \multicolumn{2}{c}{\textit{Dependent variable: Bid ($\bar{b}_{a,k,h}$)}} \\ 
\cline{2-3} 
\\[-1.8ex] & (1) OLS & (2) 2SLS  \\ 
\hline \\[-1.8ex] 
Lagged Bid ($\bar{b}_{a,k,h-1}$) & $0.9952^{***}$ & $0.9950^{***}$   \\
 & $(0.000)$ & $(0.000)$   \\[0.1cm]
Lagged Cost-per-Conversion ($\text{CPC}_{a,k,h-1}$) & $0.0012$ & $0.0019$   \\
 & $(0.001)$ & $(0.013)$  \\[0.1cm]
\hline \\[-1.8ex] 
$R^2$ & $0.991$ & $0.991$  \\
No. of Obs. & $92,750$ & $92,750$ \\

\hline \\[-1.8ex] 
\textit{Note:}  & \multicolumn{2}{r}{$^{*}$p$<$0.05; $^{**}$p$<$0.01; $^{***}$p$<$0.001} \\ 
\hline\hline
\end{tabular}
}
\end{center}
\end{minipage}
\end{table}

The regression analysis in Table \ref{tab:dyntest} together with the analysis in Appendix 
\ref{appsec:overbidding} suggest that advertisers exhibit little dynamic bidding behavior in our setting. An economic way to rationalize this is through static bidding, which means that the discount factor for advertisers is close to zero, $\delta \approx 0$. 

\section{Empirical Estimation and Results}
\label{sec:empirical_revised}
 
\subsection{Estimation Objectives}
\label{ssec:estimation_objectives}
 
The goal of our empirical analysis is to evaluate market outcomes, revenue and efficiency, under counterfactual TS-SP auction mechanisms that differ from the one observed in the data only in the prior distribution of quality scores for entrants. We start by formally characterizing these two outcomes. For any mechanism $M(p, e; \alpha^0, \beta^0)$, we denote the expected revenue by $\mathbb{E}[\texttt{Rev}(M)]$ and characterize it as follows:
\begin{equation}\label{eq:exp_rev}
    \mathbb{E}[\texttt{Rev}(M)] = \mathbb{E}\left[\sum_{k \in \mathcal{K}} \sum_{t=1}^{T_k} Y_{k,t}^M \left(\frac{\max\left\{r_k,\; \max_{a \in \mathcal{A}_{k,t} \setminus\{A_{k,t}^M\}} b_{a,k,t}^M \, q_{a,k,t}^M\right\}}{q_{A_{k,t}^M, k, t}^M}\right) \;\middle|\; M(p,e;\alpha^0,\beta^0) \right]
\end{equation}
where the expectation is taken over the randomness induced by updating the quality score distribution and the conversion outcome $Y_{k,t}^{M}$ is drawn from a Bernoulli distribution with parameter $\mu_{A_{k,t}^M,k,t}$. The superscript $M$ emphasizes that bids, quality scores, allocations, and conversions may all differ under a counterfactual mechanism.
 
We denote the expected efficiency or total surplus in the market for auction $M$ by $\mathbb{E}[\texttt{Eff}(M)]$ and define it as follows:
\begin{equation}\label{eq:exp_eff}
    \mathbb{E}[\texttt{Eff}(M)] = \mathbb{E}\left[\sum_{k \in \mathcal{K}} \sum_{t=1}^{T_k} Y_{k,t}^M \, v_{A_{k,t}^M, k, t} \;\middle|\; M(p,e;\alpha^0,\beta^0) \right]
\end{equation}
where $v_{A_{k,t}^M,k,t}$ is the private valuation for a conversion for the ad winning impression $t$ of keyword $k$ under mechanism $M$.
 
To carry out this evaluation, we need to simulate, for any counterfactual mechanism $M$, the full trajectory of bids $b_{a,k,t}^M$, quality score draws $q_{a,k,t}^M$, ad allocations $A_{k,t}^M$, and conversion outcomes $Y_{k,t}^M$. The structural primitives of our model are twofold:
\begin{enumerate}
    \item \textit{Advertisers' conversion rates} ($\mu_{a,k}$). The conversion rate for ad $a$ on keyword $k$ governs (i) the conversion outcome $Y_{k,t}^M \sim \text{Bern}(\mu_{a,k})$ whenever ad $a$ wins an impression, and (ii) the state transitions of the quality score distributions, since conversions and non-conversions drive the Bayesian updating of the Beta distribution parameters (Equation \eqref{eq:updating}). Because conversion rates are inherent to the ad--keyword pair and do not depend on which auction mechanism is in place, they are mechanism-invariant.
    \item \textit{Advertisers' valuations for conversion} ($v_{a,k}$). An advertiser's private valuation for a conversion on keyword $k$ determines their bidding behavior in every auction, so knowing $v_{a,k}$ pins down the bid advertiser $a$ submits under any counterfactual auction mechanism. Like conversion rates, valuations are a property of the advertiser--keyword pair and are invariant to the mechanism.
\end{enumerate}
Importantly, because both primitives are mechanism-invariant, they remain valid inputs for evaluating any TS-SP auction within the class we consider. The only element that changes across counterfactual mechanisms is the prior distribution $(\alpha^0, \beta^0)$ from which entrants' quality scores are initially drawn, which affects the trajectory of quality score distributions and hence the sequence of allocations, conversions, and payments. All other inputs to the simulation, the set of competing ads $\mathcal{A}_{k,t}$, the number of impressions per keyword $T_k$, advertiser valuations $v_{a,k}$, and conversion rates $\mu_{a,k}$, are held fixed at the values observed or estimated from the data.
 
\subsection{Identification Strategy and Assumptions}
\label{ssec:identification}
 
Our data offer a unique opportunity to separately identify both mechanism-invariant primitives of our model. On the one hand, they provide extensive exogenous variation in ad allocation, which allows for consistent estimation of conversion rates. On the other hand, the allocation is still through a second-price auction, which preserves enough strategic incentive for advertisers to allow for estimating their structural valuations for conversions. Had our data instead come from a fully randomized experiment, this exogenous variation would have come at the expense of strategic behavior, rendering advertisers' bids uninformative.

Our identification strategy relies on three empirically-supported assumptions. Our first assumption concerns conversion rates. Without it, learning the underlying conversion rate from the empirical distribution of outcomes would not be feasible, as the target itself would be shifting over time.
\begin{assumption}\label{as:constant_cvr}
Advertisers' conversion rates remain fixed over time, that is, $\mu_{a,k,t} = \mu_{a,k}$.
\end{assumption}
This assumption is widely adopted in the multi-armed bandit literature and provides the foundation for incorporating bandit frameworks into auctions to address the cold-start problem \citep{feng2023improved}. Furthermore, the empirical evidence presented in $\S$\ref{sssec:primitive_dynamics} corroborates this assumption. 
 
We impose two additional assumptions that together let us invert observed bids into valuations. The first is that valuations do not vary over time.
\begin{assumption}\label{as:constant_v}
Advertisers' private valuations for conversion do not change over time, that is, $v_{a,k,t} = v_{a,k}$.
\end{assumption}
Evidence presented in $\S$\ref{sssec:primitive_dynamics} empirically corroborates this assumption. 
Further, we focus on a span of one week in our data, so it is not realistic to assume that the valuation per conversion changes for an app within a week. 
 
Finally, we assume that bidders choose their bids to maximize per-period, rather than long-run, expected utility.
\begin{assumption}\label{as:static_revised}
Bidders choose their bids to maximize their per-period expected utility.
\end{assumption}
The alternative, that advertisers maximize long-run utility, would allow current bids to strategically influence the evolution of quality scores and hence future auction outcomes, breaking the link between bids and valuations. Evidence presented in $\S$\ref{sssec:bidding_dynamics} corroborates this assumption: we find that incumbents do not shade their bids in response to cost increases caused by entrants' inflated prior quality scores, and that entrants with high conversion rates do not overbid to accelerate learning of their true quality. In Appendix~\ref{appsec:dynamic}, we further relax this assumption and estimate a discount factor from our data, which comes back statistically indistinguishable from zero.

\subsection{Estimation of Mechanism-Invariant Primitives}
\label{ssec:estimation_primitives}
 
\subsubsection{Conversion Rates}
\label{sssec:cvr_estimation}
 
We estimate conversion rates using the sample analogue estimator:
\begin{equation}
    \widehat{\mu}_{a,k} = \frac{\sum_{t=1}^{T_k} \mathbbm{1}(a = {A}_{k,t}) Y_{k,t}}{\sum_{t=1}^{T_k} \mathbbm{1}(a = {A}_{k,t})}
\end{equation}
That is, for each ad--keyword pair, the estimated conversion rate is simply the fraction of impressions won by the ad that resulted in a conversion. This estimator is unbiased if the impressions allocated to an ad are randomly drawn from the overall impression distribution, an assumption that is reasonable given the posterior sampling-induced randomization in our Thompson Sampling setting.
 
Although we focus our counterfactual analysis on the first week of data, we use the entire six months of impression and conversion data to estimate conversion rates. This increases the reliability of our estimates and shrinks the confidence bounds around them. As shown in $\S$\ref{sssec:exploration_summary}, the extent of exploration in our data is substantial: 80\% of entrants accumulate at least 100 impressions and 20\% exceed 1,000 impressions over the six-month period, providing sufficient data for reliable estimation.
 
As an alternative approach, one could relax the random-impressions assumption and employ an Inverse Propensity Score (IPS) estimator that is unbiased under adaptively collected data \citep{horvitz1952generalization, hadad2021confidence, zhan2021off}. We present the details of the IPS approach in Appendix~\ref{appsec:ips} and confirm that both approaches yield highly similar estimates, reinforcing the validity of the sample analogue estimator.
 
\subsubsection{Advertisers' Valuations for Conversion}
\label{sssec:valuation_estimation}
 
The objective of this section is to recover advertisers' private valuations $v_{a,k}$ from observed bids $b_{a,k,t}$. Specifically, we ask whether the classical result that bidders bid truthfully ($b = v$) in second-price auctions carries over to our setting.
 
\paragraph{What is known.} \cite{vickrey1961counterspeculation} established that in a single-item second-price auction without score multipliers, truthful bidding is a weakly dominant strategy under any information structure: bidders need not know anything about other participants' values or strategies. The key insight is that a bidder's payment, conditional on winning, is determined entirely by others' bids, so one's own bid affects only whether one wins, not what one pays. Bidding one's true value is therefore optimal regardless of what others do or what one knows about them.
 
This result does not always extend to richer settings. In the Generalized Second-Price (GSP) auction with multiple slots, truthful bidding is no longer a dominant strategy \citep{edelman2007internet, varian2007position}. Similarly, as we discuss in Appendix~\ref{appsec:overbidding}, truthfulness breaks down in our setting if bidders are forward-looking, since a bid today can influence quality score evolution and hence future auction outcomes.
 
\paragraph{Truthfulness in our setting.} Our setting differs from the canonical Vickrey auction in two respects: the allocation is based on quality-adjusted bids $b_a q_a$ rather than bids alone, and each bidder's quality score draw is not known to other bidders, or even to the bidder itself, at the time of bidding. Section \ref{sssec:bidding} shows that neither difference is consequential for truthfulness. Because the platform draws quality scores in the Quality Score Sampling step before bids are used for allocation, each bidder's own quality score, and therefore its payment conditional on winning, is realized independently of its own bid, which is exactly the condition needed for the classical \citet{vickrey1961counterspeculation} argument to go through in this richer environment. Proposition~\ref{prop:truthful} formalizes this: under Assumption~\ref{as:static_revised}, truthful bidding is a weakly dominant strategy for every advertiser in every auction, and truthful bidding by all advertisers is an equilibrium of the TS-SP mechanism.
 
\paragraph{Identification.} Proposition~\ref{prop:truthful} alone links a bid to the valuation of the impression at which it was submitted, $b_{a,k,t}=v_{a,k,t}$, but does not by itself say that a single ad--keyword pair's bids can be pooled across auctions. Assumption~\ref{as:constant_v} supplies that missing step, and together the two give the valuation $v_{a,k}$ point identification from observed bids.
 
\begin{corollary}\label{cor:valuation_id}
Suppose Assumptions~\ref{as:static_revised} and~\ref{as:constant_v} hold and, as in Proposition~\ref{prop:truthful}, quality scores are determined independently of contemporaneous bids. Then for every advertiser $a$ and keyword $k$, $b_{a,k,t} = v_{a,k}$ at every impression $t$ with $a \in \mathcal{A}_{k,t}$; in particular, $v_{a,k}$ is point-identified from a single observed bid, and the sample-average-bid estimator
\begin{equation}
    \widehat{v}_{a,k} = \frac{\sum_{t=1}^{T_k} \mathbbm{1}(a \in \mathcal{A}_{k,t}) b_{a,k,t}}{\sum_{t=1}^{T_k} \mathbbm{1}(a \in \mathcal{A}_{k,t})}
\end{equation}
recovers $v_{a,k}$ from observed bids.
\end{corollary}

\paragraph{Estimation.} Corollary~\ref{cor:valuation_id} justifies estimating each advertiser's valuation as the average bid they submitted for that keyword. In the data, individual bids can deviate from this exact equality for reasons outside the model, such as bidding-interface rounding or brief adjustment lags after a valuation shock; averaging across a bidder's auctions on a keyword, as in $\widehat{v}_{a,k}$ above, smooths out this idiosyncratic noise. This identification strategy follows the approach used in prior literature that leverages data from second-price auctions with truthfulness as an equilibrium strategy \citep{athey2007nonparametric, celis2014buy}. 

\subsection{Results}
\label{ssec:results_revised}
\begin{figure}[!htbp]
    \centering
    \begin{subfigure}[t]{0.48\textwidth}
        \centering
        \includegraphics[width=\textwidth]{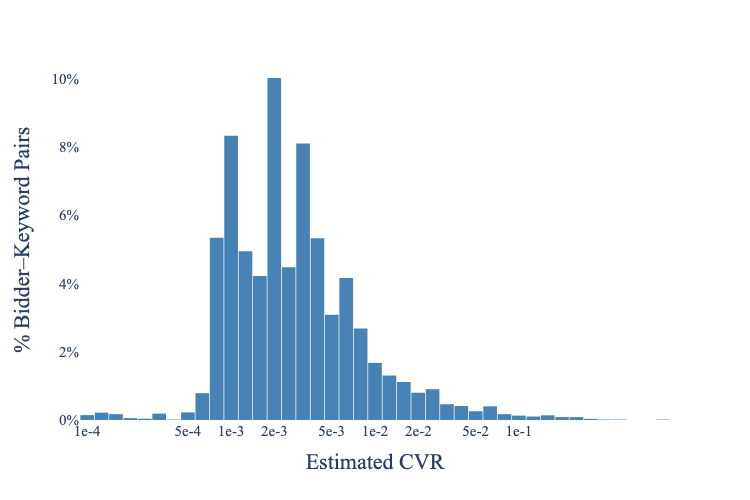}
        \caption{Estimated conversion rates}
        \label{fig:cvr_bidder_keyword_dist}
    \end{subfigure}
    \hfill
    \begin{subfigure}[t]{0.48\textwidth}
        \centering
        \includegraphics[width=\textwidth]{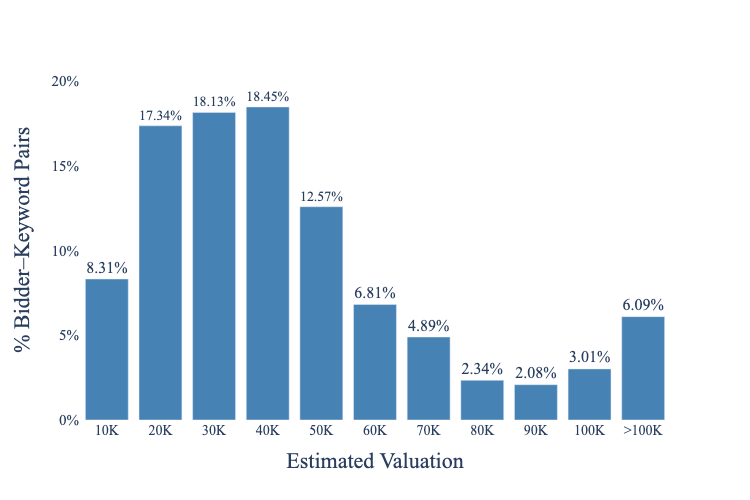}
        \caption{Estimated valuations}
        \label{fig:valuation_bidder_keyword_dist}
    \end{subfigure}
    \caption{Distributions of Estimated Primitives Across Bidder--Keyword Pairs}
    \label{fig:bidder_keyword_primitive_distributions}
    \vspace{0.25em}
    \begin{minipage}{0.95\textwidth}
    \footnotesize
    \textit{Notes:} Panel~(a) excludes pairs with zero estimated conversion rates and uses a logarithmic x-axis. In Panel~(b), labels give the upper endpoint of each 10K-wide bin, and the final bin contains all valuations above 100K.
    \end{minipage}
\end{figure}

\paragraph{Conversion rates.}\label{sssec:cvr_results}
We first examine the distribution of estimated conversion rates across bidder--keyword pairs. 29.43\% of unique bidder--keyword pairs have an estimated conversion rate of zero during the six-month estimation window. Panel~(a) of Figure~\ref{fig:bidder_keyword_primitive_distributions} plots the distribution of $\widehat{\mu}_{a,k}$ for pairs with strictly positive estimated conversion rates. The x-axis is shown on a log scale to account for the strong right skew in the positive part of the distribution.

\paragraph{Valuations.}\label{sssec:valuation_results}
Panel~(b) of Figure~\ref{fig:bidder_keyword_primitive_distributions} presents the distribution of estimated valuations across bidder--keyword pairs. To make the distribution easier to read, we group valuations into 10K-wide bins up to 100K and combine all valuations above 100K into a single top bin. The x-axis labels denote the upper endpoint of each bin; for example, the bar labeled 60K contains all valuations in $(50\text{K},60\text{K}]$.

For both primitives, Appendix~\ref{appsec:keyword_level_primitives} reports the distributions of the mean and median estimates computed within each keyword.

\section{Counterfactual Analysis}
\label{sec:results}

The objective of this section is to examine, qualitatively and quantitatively, how the optimal exploration policy is impacted by the fact that the problem is not a standard bandit problem, but a combined auction-bandit problem. 

Our strategy to formally capture the role of auctions involves comparing two metrics: efficiency and revenue. Our measure of efficiency in a given auction is defined as the total value generated in the auction that is shared between the platform and the advertisers. One can show that maximizing efficiency is equivalent to solving a weighted bandit problem. This is because the efficiency metric is a function of bidders' valuation only and invariant to the price of the item. As such, under a direct revelation mechanism where bidders truthfully report their valuations, efficiency maximization reduces to a weighted bandit problem: the auctioneer can treat the reported valuations as weights for arms with unknown rewards (CVR in our setting). On the other hand, maximizing the revenue depends not only on which bidder the item is allocated to, but also how much the bidder eventually pays. This ``full'' version of the problem, hence, involves both the auction and the bandit aspects. As a result, comparing the revenue-optimal exploration policy (i.e., optimal prior on entrants) against the efficient policy would allow us to isolate the role of the auctions in our auction-bandit problem.

We carry out the comparative analysis in two stages. First, we assume the firm has a uniform prior across each group of thin, thick, and mid-sized auctions. We will then turn to studying how the comparison between revenue- and efficiency-maximizing policies is impacted by allowing the firm to customize the prior in a more granular fashion.

\subsection{Market Outcomes Under Uniform Priors}
\label{sec:rev_eff}

\subsubsection{Revenue-Efficiency Tradeoff}
\label{sssec:frontier}

\begin{figure}[h]
    \centering
    \begin{subfigure}[t]{0.48\textwidth}
        \centering
        \includegraphics[width=\linewidth]{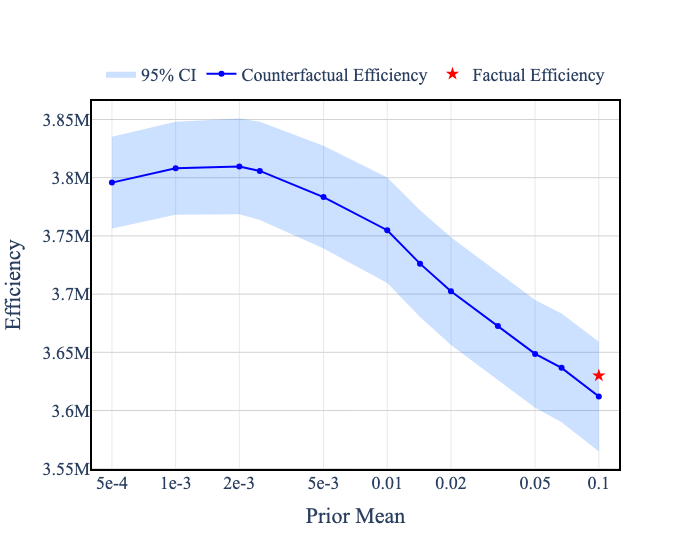}
        \caption{Efficiency}
        \label{fig:efficiency agsinst uniform prior}
    \end{subfigure}
    \hfill
    \begin{subfigure}[t]{0.48\textwidth}
        \centering
        \includegraphics[width=\linewidth]{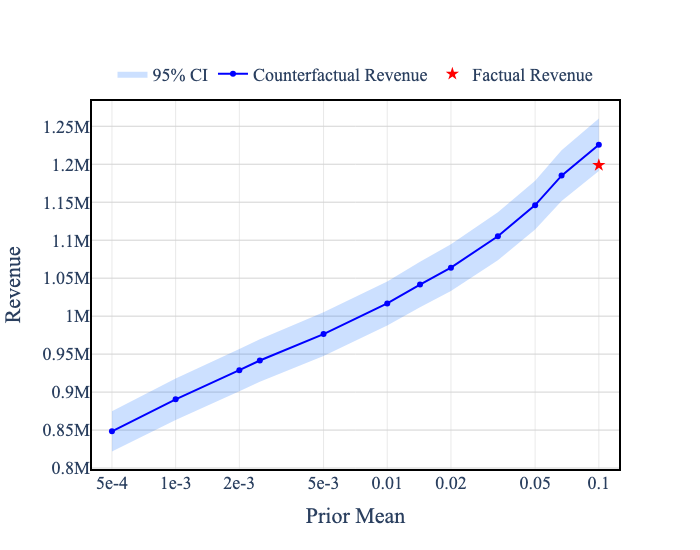}
        \caption{Revenue}
        \label{fig:rev agsinst uniform prior}
    \end{subfigure}
    \caption{Efficiency and Revenue Across Prior Means}
    \vspace{0.25em}
    \begin{minipage}{0.95\textwidth}
    \footnotesize
    \textit{Notes:} The horizontal axes show prior means on a logarithmic scale. Solid lines show mean counterfactual outcomes, and shaded regions show the mean plus or minus 1.96 standard deviations across simulation runs. Red stars indicate factual outcomes: observed revenue and efficiency computed from observed allocations and estimated per-conversion valuations.
    \end{minipage}
\end{figure}

Figure \ref{fig:efficiency agsinst uniform prior} shows how efficiency varies as we vary the prior mean ${1}/{(1+\beta)}$ on the CVR, by varying $\beta$. As shown in this figure, the efficiency-maximizing prior mean is 0.002. To see whether this policy involves a degree of exploration: the prior mean exceeds the median CVR among entrants in our dataset (0.0013), though it lies below the mean (0.0035), so the platform starts the typical new ad with an optimistic belief about its CVR. Efficiency is relatively flat for prior means up to 0.002, so modest exploration is cheap. Beyond this level, however, over-exploration of entrants sharply reduces efficiency by misallocating the position: efficiency falls by about 5\% as the prior mean rises to 0.1.

Figure \ref{fig:rev agsinst uniform prior} shows how revenue varies as we vary the prior mean. As can be seen from the figure, the revenue-maximizing prior mean is 0.1, at the end of the range examined.\footnote{\label{fn:prior_cap}As the figure suggests, revenue would likely increase further with prior means above 0.1. We exclude such values from our main analysis to remain conservative with our counterfactual analysis, because sufficiently high prior means may alter incumbents' incentives. In particular, incumbents facing high entrant scores may optimally bid lower and lose some rounds, allowing entrants to win and have their scores revised downward, before returning to truthful bidding once the second-highest bid score has fallen. We find no evidence of such forward-looking behavior at the observed prior mean of 0.1, so Figure \ref{fig:rev agsinst uniform prior} is valid because it focuses on counterfactual cases with weaker dynamic incentives. Evaluating higher prior means, however, would require a richer model of bidder behavior. In $\S$\ref{ssec:robustness_checks}, we estimate a dynamic structural bidding model with forward-looking bidders and use it to extend the range of prior means considered. Our qualitative results remain unchanged.} This stands in contrast with the efficiency-maximizing prior.

The contrast between the prior means suggests that the auction component of the auction-bandit problem plays an empirically non-trivial role. To the extent that the result of our analysis may generalize to other online advertising settings, this result is suggestive that the auction-bandit problem is of general empirical relevance, and that further theoretical and empirical studies of this problem would be of applied value. This contrast between the revenue- and efficiency-maximizing priors is indicative that there is a trade-off between maximizing the two objects. Quantifying this trade-off is a task that we turn to next.

\begin{figure}[h]
    \centering
    \includegraphics[width=0.5\linewidth]{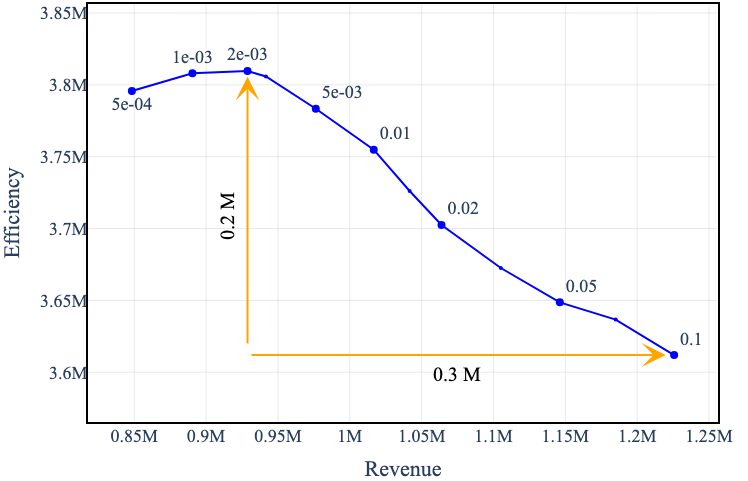}
    \caption{Revenue--Efficiency Tradeoff Under Uniform Priors}
    \label{fig: rev efficiency Pareto curve uniform}
    \vspace{0.25em}
    \begin{minipage}{0.95\textwidth}
    \footnotesize
    \textit{Notes:} Each point represents the outcomes under a different prior.
    \end{minipage}
\end{figure}

Figure \ref{fig: rev efficiency Pareto curve uniform} examines possible trade-offs between revenue and efficiency. This figure plots efficiency against revenue as we vary the prior. As shown by the figure, and as expected from the previous analysis, there is no trade-off for lower priors. If a prior is below the efficiency-maximizing level, increasing it improves both the efficiency and the revenue. The trade-off appears once we reach an efficiency-maximizing level. Past this threshold, an increase in the prior improves the revenue at the expense of efficiency.

It is worth examining the empirical magnitude of the revenue-efficiency tradeoff. As Figure \ref{fig: rev efficiency Pareto curve uniform}  shows, the range of the revenue as we trace a range of priors is wider than that of the efficiency. Efficiency ranges between 3.6M and 3.8M whereas revenue ranges between 0.8M and 1.2M. The latter range is wider than the former both in absolute terms (0.4M vs. 0.2M) and in relative terms (50\% vs. 6\%). This, once again, is suggestive that the auction component of the auction-bandit problem is of substantial empirical relevance. This observation motivates a deeper analysis of the mechanisms behind the tradeoff, a task which we turn to next.

\subsubsection{Discussion of the Underlying Mechanisms}\label{subsubsec: mechanisms, uniform}

Why does a higher prior mean on entrants' CVR achieve a higher revenue? And why does it come at the expense of efficiency?

\begin{figure}[h]
    \centering

    \begin{subfigure}[t]{0.48\linewidth}
        \centering
        \includegraphics[width=\linewidth]{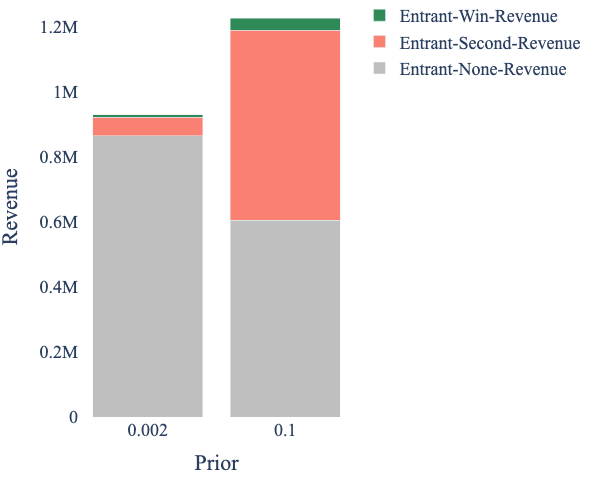}
        \caption{Revenue composition}
        \label{fig:total rev decomposition}
    \end{subfigure}
    \hfill
    \begin{subfigure}[t]{0.48\linewidth}
        \centering
        \includegraphics[width=\linewidth]{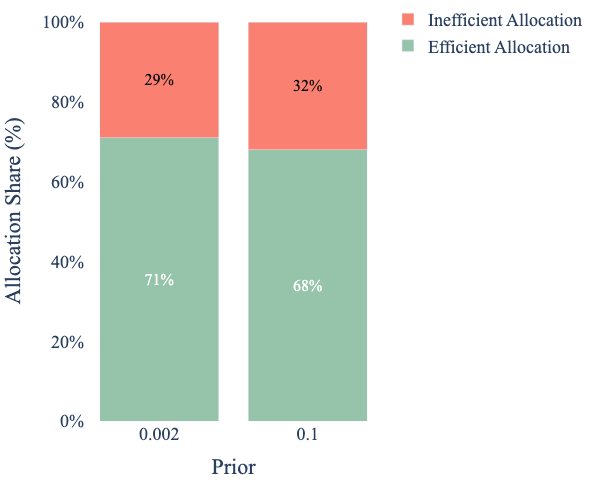}
        \caption{Allocation efficiency}
        \label{fig:percent efficient}
    \end{subfigure}

    \caption{Revenue Composition and Allocation Efficiency Under Revenue- and Efficiency-Maximizing Priors}
    \label{fig:rev-eff-subplots}
    \vspace{0.25em}
    \begin{minipage}{0.95\textwidth}
    \footnotesize
    \textit{Notes:} Panel (a) decomposes revenue by entrants' auction positions. Panel (b) reports the share of auctions with an ex-post efficient allocation.
    \end{minipage}
\end{figure}

Figure \ref{fig:total rev decomposition} illustrates part of the answer. The figure presents stacked bar plots of total revenue under revenue- and efficiency-maximizing priors. Each total-revenue amount is decomposed into three components: (i) revenue from auctions in which the entrant was allocated the object, (ii) revenue from auctions in which the entrant was not allocated the object but finished with the second highest score and, hence, influenced the revenue extracted from the winner, and (iii) auctions in which the entrant ranked 3rd or lower, thereby impacting neither the allocation of the ad nor the revenue extracted.

A comparison between the efficiency- and revenue-maximizing policies shows that the latter generates significant revenue by placing the entrant in the second place, thereby impacting the revenue without changing the allocation. This is consistent with the theoretical observation by \cite{akbarpour2020credible} in a different context, that the auctioneer has the incentive to induce payment rules akin to a first-price auction in a setting where the auction rule is second-price. The difference relative to the analysis by \cite{akbarpour2020credible} is that they study lack of truthfulness by the auctioneer, whereas in our case the seller truthfully runs a second-price auction, tuning the entrant prior optimally.

Relatedly, for the auctioneer, it is a consequence to truthfully amplify entrant bids as a means to extract more revenue from the winner. The consequence entails the possibility that the entrant may indeed win, leading to an increased likelihood for inefficient allocation. This can be seen in Figure \ref{fig:percent efficient}. Of course such a compromise in efficiency indirectly hurts the revenue. That said, as long as the revenue effect of the efficiency compromise does not outweigh the direct revenue boost from the cases where the entrant is placed second, a policy of increased prior for the entrants may indeed be revenue-optimal, as we find in our empirical application.

\begin{figure}[h]
    \centering
    \includegraphics[width=0.75\linewidth]{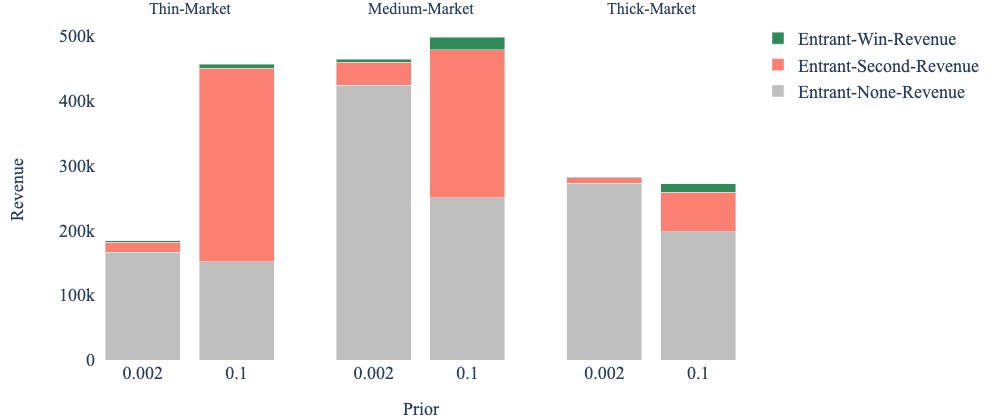}
    \caption{Revenue Composition by Market Thickness}
    \label{fig:total rev decomposition_by thickness}
    \vspace{0.25em}
    \begin{minipage}{0.95\textwidth}
    \footnotesize
    \textit{Notes:} Bars decompose revenue by entrant role under revenue-maximizing and efficiency-maximizing TS--SP policies.
    \end{minipage}
\end{figure}

\noindent \textbf{Implications for market thickness.} We finish this discussion by noting that the above mechanism points to a possible role to be played by market thickness. In thinner markets, there is a higher likelihood that the score for the winner and the rest of the bidders are far apart. This, in turn, makes it more likely that a boost to the prior on entrants can place them second and extract a higher revenue from the winner, without posing as pronounced a risk of damaging the efficiency of the allocation. Figure \ref{fig:total rev decomposition_by thickness} confirms this expectation. We partition the dataset into thin, mid-size, and thick markets; and we replicate the revenue-components analysis in Figure \ref{fig:total rev decomposition} for each of these partitions. As the figure shows, the revenue boost from changing the prior is indeed the highest in thinner markets, confirming our expectation.

\subsection{Customized Exploration}
\label{ssec:customized}
Our analysis of market thickness in the previous section suggests that the distinction between the auction-bandit problem and the standard bandit problem may be more pronounced depending on the features of the auction (such as thickness). Thus, it would be worth studying how further customization (beyond thickness) may impact the performance of the solution on the fronts of efficiency and revenue as well as the tradeoff between them.

To this end, we build on our analysis of Pareto frontiers in the revenue-efficiency space. More specifically, for each of our three market-thickness categories, we plot the frontier emerging from uniform policies alongside the frontier arising from full customization. The former frontier can be traced by varying $\beta$ in the prior and obtaining a revenue-efficiency pair for each level of $\beta$ examined. The latter frontier is more nuanced. For each keyword, we form a convex combination $V(\tau,\beta^0_{k})$ of revenue and efficiency: 

$$V(\tau,\beta^0_{k}):=\tau\times \mathbb{E}[\widehat{\texttt{Rev}}_{k}\big(M(p,e;\alpha^0,\beta^0_k)\big)]+(1-\tau)\times \mathbb{E}[\widehat{\texttt{Eff}}_{k}\big(M(p,e;\alpha^0,\beta^0_k)\big)]$$
where $\tau$ is the weight on revenue in the convex combination and $\beta^0_{k}$ is the keyword specific prior policy that we vary to optimize $V(\tau,\beta^0_{k})$.\footnote{Similar to the main analysis, we maintain all $\alpha^0_{a,k}$ at $\alpha^0=1$. Also note that we are varying $\beta^0_{a,k}$ only by 
 keyword $k$ and keeping it fixed within each keyword across ads $a$.}

For each value of $\tau$, we choose $\beta^0_{k}$ such that the convex combination $V(\tau,\beta^0_{k})$ is maximized. We record the resultant $\widehat{\texttt{Rev}}_{k}$ and $\widehat{\texttt{Eff}}_{k}$. We do this by tracing values of $\tau$ from 0 to 1, which yields a Pareto frontier curve over attainable revenue and efficiency levels.

The three panels of Figure \ref{fig:Pareto frontiers, personalized policies} present the results. There are three lessons worth describing from these graphs. First, customization can deliver Pareto improvement in the efficiency-revenue at any thickness level and upon any uniform prior. This is perhaps not surprising; but it is interesting to observe that non-trivial gains can be made even for thick markets. As the right panel of the figure shows, customization can deliver, roughly, a 5\% revenue increase upon the revenue-maximizing uniform prior. Note that this improvement in revenue cannot be entirely attributed to the advantage of the customized solution in matching keyword-specific priors on the CVR (i.e., the advantage of the customized solving the bandit problem). To see why, note that the most efficient point on the customized Pareto curve delivers an about 5\% \textit{lower} revenue relative to the revenue-maximizing  uniform prior. As a result, the distinction between the auction-bandit problem and the pure bandit problem is relevant even in the case of thick markets.

\begin{figure}[h]
    \centering
    \includegraphics[width=1\linewidth]{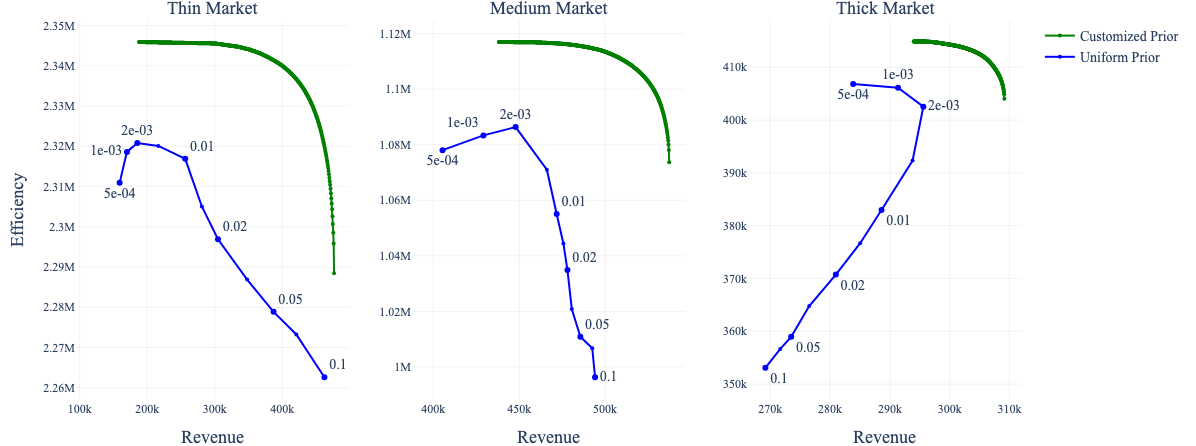}
    \caption{Revenue--Efficiency Tradeoffs Under Customized and Uniform Priors}
    \label{fig:Pareto frontiers, personalized policies}
    \vspace{0.25em}
    \begin{minipage}{0.95\textwidth}
    \footnotesize
    \textit{Notes:} Uniform policies use a common prior within each market-thickness group; customized policies allow keyword-specific priors.
    \end{minipage}
\end{figure}

The second lesson is that the customized Pareto curves in all three thickness groups are visibly more concave relative to the uniform Pareto curves. We interpret this to mean that customization ``eases'' the tension between the auction component and the bandit component of the auction-bandit problem. In other words, customization allows the platform to increase the prior only for auctions where such increase helps with revenue, and to keep the prior selection focused on efficiency in other cases. To visually see this ease of tension and its relation to concavity, observe that on the customized Pareto curve, it is easier to pick a point (i.e., a ``prior policy'') that achieves a revenue approximating that curve's maximum revenue and, at the same time, an efficiency level approximating the curve's maximum efficiency.

Relatedly, our third lesson is that, in the case of all thickness groups, the customized Pareto curve seems to have an almost horizontal part. That is, there is a non-trivial degree of revenue improvement that can be achieved at a negligible cost to efficiency. We expect that this happens because there are auctions with strong incumbents with substantially larger ``true'' scores than all other participants, including entrants. In such auctions, revenue can be improved with little efficiency cost. We suspect that as one goes from left to right along the almost-horizontal segment of each Pareto curve, the optimal policy is changing course only for these specific types of auctions. This hypothesis is consistent with the observation that the Pareto frontier for thin markets (which by definition have strong incumbents) seems to be more pronounced than that in the other two frontiers: for those auctions, revenue can be almost tripled at an efficiency cost of less than 1\%.

In sum, our analysis recommends customization. It does so not only because customization (by construction) helps improve on revenue and efficiency individually, but also because it helps ease the trade-off between the two. Customization especially helps mitigate the trade-off under thinner markets.

\section{Extensions and Robustness Checks}
\label{sec:extension}

\subsection{Alternative Bandit Algorithm: Upper Confidence Bound}
\label{ssec:ucb}
So far, our analysis has focused on Thompson Sampling (TS) as the primary bandit algorithm for learning quality scores in the auction-bandit problem. In particular, we have shown that TS’s exploration policy can have a non-trivial impact on market outcomes. However, TS is only one approach to balancing the exploration-exploitation trade-off. In this section, we extend our analysis to another prominent bandit algorithm with strong regret guarantees: the Upper Confidence Bound (UCB) algorithm. Our goal is to compare market outcomes under a UCB-based auction with our main findings for TS-SP. We begin by formally defining the UCB exploration policy and describing its integration into an auction. We then compare the performance of auctions using UCB and TS and examine the mechanisms underlying their differences.
Similar to TS, we first define the exploration policy under UCB:
\begin{defn}
The \textbf{UCB Exploration Policy} with exploration parameter $\rho\ge0$ is defined by the quality score estimate for each ad $a\in\mathcal{A}$ on keyword $k\in\mathcal{K}$ at time $t$:
\[
q_{a,k,t}
\;=\;
\underbrace{\frac{C_{a,k,t}}{I_{a,k,t}+1}}_{\text{empirical conversion rate}}
\;+\;
\underbrace{\rho \,\sqrt{\frac{\log (t+1)}{\,I_{a,k,t}+1\,}}}_{\text{exploration bonus}}
\]
where $I_{a,k,t}$ and $C_{a,k,t}$ are the total number of impressions and conversions the ad $a$ has received on keyword $k$ up to time $t-1$, respectively, and $\rho$ controls the degree of exploration (with $\rho=0$ reducing to pure exploitation).
\end{defn}
In the generic bandit formulation, the Upper Confidence Bound (UCB) algorithm follows the principle of \textit{optimism in the face of uncertainty}, selecting the action with the highest upper bound on its estimated reward. This upper bound combines the mean estimate with an exploration bonus, encouraging the algorithm to try actions with high potential or little data. As data accumulate, the bounds narrow and UCB stops exploring an arm once its upper bound is lower than the lower bound of another arm. We now describe how we can integrate it in a second-price auction as follows:

\begin{defn}
Under the \textbf{UCB Second‐Price (UCB‐SP)} auction mechanism, quality score estimates are according to the UCB Exploration Policy and the allocation and payment rules are the same as defined under TS-SP in Definition \ref{def:tssp}.
\end{defn}

While TS is probabilistic, UCB is deterministic and explores in an optimistic manner. As such, we expect TS to tend to explore for a longer period of time, which can have important consequences in terms of market outcomes.
Under Thompson Sampling, the sampled quality scores asymptotically converge to the conversion rates for every ad as posterior distributions concentrate; formally $\mathbb{E}[q^{\mathrm{TS}}_{a,k,t}]\to \mu_{a,k}$. Under UCB, the winner receives almost all impressions asymptotically, so its bonus vanishes and the quality score $q^{\mathrm{UCB}}$ converges to its $\mu$; however, non-winners are pulled only until their upper confidence bounds fall just below the winner’s mean with high probability. Their bonuses therefore remain strictly positive on average over finite horizons, so their quality scores tend to be upward biased relative to $\mu$. In a second-price mechanism, this compresses the gap between the top two scores and raises prices; hence, we expect higher revenue under UCB.

\begin{figure}[h]
    \centering
    \includegraphics[width=0.6\linewidth]{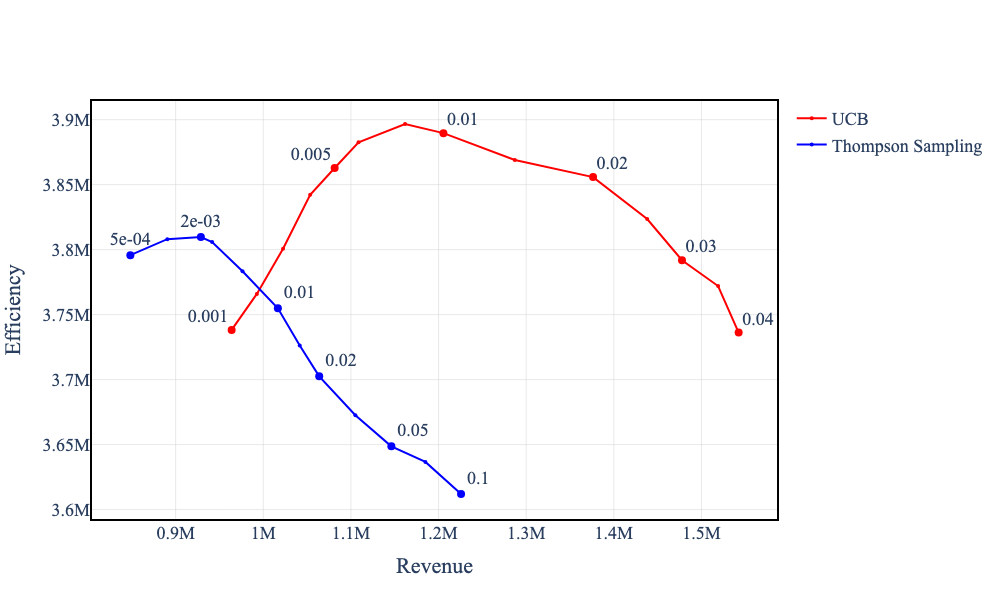}
    \caption{Revenue--Efficiency Tradeoffs Under UCB (red) and TS (blue)}
    \label{fig: ucb_ts_exp_level}
    \vspace{0.25em}
    \begin{minipage}{0.95\textwidth}
    \footnotesize
    \textit{Notes:} Point labels indicate the exploration parameter $\rho$ under UCB and the prior mean under TS.
    \end{minipage}
\end{figure}

Figure~\ref{fig: ucb_ts_exp_level} plots the revenue-efficiency frontier for UCB (varying $\rho$) against TS (varying the prior). To maintain the validity of our counterfactual estimates, we only display the UCB policies that explore the entrants no more than what we observed in the data (i.e., 18\% entrant impression share; see Appendix Figure~\ref{fig:entrant_place}). Across the range we consider, UCB Pareto-dominates TS. While efficiency is slightly improved under UCB, the revenue gain is large in both percentage and magnitude. UCB with parameter $\rho = 0.04$ achieves about 25\% more revenue compared to TS with prior $0.1$, while both allocate 18\% of impressions to entrants.
To understand this dominance, we examine how the quality scores assigned under the two mechanisms deviate from the true conversion rates. We find that UCB's revenue advantage is driven by a positive bias in nonwinners' quality scores that persists throughout the counterfactual horizon. This optimism elevates the second-highest auction scores and, consequently, prices. Appendix~\ref{appsec:ucb_quality_bias} presents the detailed analysis.

\subsection{Robustness checks}
\label{ssec:robustness_checks}

\paragraph{Budget constraints.}\label{ssec:budget_constraint}
Our baseline counterfactuals do not impose explicit budget limits. Because changing the entrant prior can also change advertisers' spending, we examine whether our results are robust to daily spending caps. Since advertisers' true budgets are unobserved, we construct a conservative, data-driven cap for each ad--keyword pair using its maximum observed daily expenditure. We then rerun the TS--SP counterfactuals while removing an advertiser from the remaining auctions of a day once it reaches this cap. Appendix~\ref{appsec:budget_constraint} provides the exact definition and implementation details.

The budget-constrained counterfactuals generate lower revenue levels than the unconstrained baseline, while preserving the strong positive relationship between the prior mean and revenue. The relationship between the prior and efficiency is also very similar across the two specifications; see Appendix Figure~\ref{fig:revenue_efficiency_robust}. Thus, our main qualitative conclusion is unchanged: higher priors increase revenue at the cost of efficiency.

\paragraph{Time-varying valuations.}\label{ssec:time_varying_valuations}
Our main counterfactuals treat each advertiser's valuation for a keyword as fixed and estimate it using the advertiser's average submitted bid. We relax this assumption by allowing valuations to vary across auctions and using the exact contemporaneous submitted bid as the valuation per conversion in each auction: $v_{a,k,t}=b_{a,k,t}$.
We then rerun the TS--SP counterfactuals while holding all other components of the simulation unchanged. Appendix~\ref{appsec:time_varying_valuations} presents the results and corresponding figure. The alternative specification modestly shifts the estimated levels of efficiency and revenue but preserves their qualitative relationships with the entrant prior and leaves our main conclusions unchanged.

\paragraph{Strategic incentives of forward-looking bidders.}\label{ssec:forward_looking}
Our counterfactual analysis recommends that the platform assign high priors to entrants, since the resulting inflation of entrants' scores raises the prices paid by winning incumbents. A natural concern is that incumbents may respond strategically to such a policy. Under the static bidding maintained in Assumption~\ref{as:static_revised}, they have no reason to: the Vickrey logic of Proposition~\ref{prop:truthful} applies period by period, so truthful bidding remains optimal regardless of the prior. A forward-looking incumbent, however, has a strategic response available. By shading its bid and deliberately losing to a prior-inflated entrant, it lets the entrant's score adjust downward, and then returns to winning at a lower price. If incumbents behave this way, the revenue gains from high priors could be smaller than our static counterfactuals suggest.

To assess this concern, Appendix~\ref{appsec:dynamic} relaxes static bidding and estimates a dynamic structural model of incumbent bidding. The estimator nests the forward-simulation approach of \cite{bajari2007estimating} (BBL) inside an outer loop that searches over the discount factor $\delta$, recovering keyword-incumbent-specific $(\hat v_a, \hat\delta)$ from the revealed preference of observed bids over counterfactual bid deviations. Two findings emerge. First, bidders are indeed static bidders: $93.6\%$ of keyword-incumbent pairs are estimated at $\hat\delta=0$, and the pooled criterion across keywords is likewise minimized at $\delta=0$. This structural result is consistent with the reduced-form evidence in Section~\ref{sssec:bidding_dynamics}, where Table~\ref{tab:dyntest} showed that incumbents do not shade their bids in response to cost increases caused by high-prior entrants. Second, even if incumbents were fully patient, our recommendation would stand. We rerun the counterfactuals of Section~\ref{sec:results} with incumbents bidding according to the dynamic policy at $\delta=0.9999$. Revenue still rises with the prior throughout and ends only about $9\%$ below the static counterfactual at the highest prior we consider. The reason is advertisers' entry rate: a new strong bidder crosses into the top of the score distribution roughly once every $70$ auctions, so the price relief an incumbent buys by suppressing the current runner-up's score is short-lived and rarely worth the forgone margins. Moreover, a higher prior accelerates the arrival of entrants, so the exploratory policy partially protects itself against the very strategic response that might undermine it.

\section{Managerial and Policy Implications}

Our empirical analysis has key implications for managers and regulators. On the managerial front, the main lesson for platforms is that the assignment of a prior to an entrant's CVR in ad auctions should be done with an eye on the auction aspect of the problem. In more specific terms, as we show empirically, it might be profitable for a platform to set the CVR prior at a level at least an order of magnitude larger than the CVR range that the majority of bidders are empirically observed to have. The profitability of having a CVR prior so disparate from the true CVR range is difficult to explain with learning-based (i.e., bandit theory) mechanisms only. But as shown in our analysis, auction-based reasons for high CVR priors are theoretically expected and empirically relevant. In this light, our empirical findings may be interpreted to suggest that ``prior setting'' in the cold-start problem is an \textit{implicit pricing tool} at the disposal of the platforms. Such a view toward prior-setting is largely absent from the literature, which tends to focus on (explicit) reserve prices.

Our empirical findings also carry policy implications. From a regulatory perspective, a CVR prior set substantially above the true empirical CVR range could be viewed as an “abuse of the learning process.” As discussed earlier, the reason higher priors can be optimal is that they enable platforms to charge the winner an amount close to its bid, thereby effectively approximating a first-price auction within a second-price auction environment where bidders submit their true valuations without shading. A direct version of this adverse incentive by auctioneers (i.e., running a first-price auction in stealth while announcing to bidders that the auction is second-price) has been mentioned in the literature \citep{akbarpour2020credible}. Our empirical results suggest that a regulatory concern about such direct credibility issues should also extend to concerns about the role of priors in the generalized auction-bandit problem that the platform solves.

If ``excessively high'' levels of CVR priors are to be considered concerning, what regulatory actions constitute a solution? A formal answer to this question is beyond our scope. We speculate, however, that one possible direction is transparency. A possible regulation is to require platforms to disclose a minimal information content that would shed light on whether  it is setting priors to boost revenue. The determination of what information is minimal but useful to disclose is an avenue for future research. One should also note that informational regulations may be met with resistance. Platforms often avoid sharing detailed information about auctions citing concerns that bidders might use the information to ``game the system''. 

We leave a formal and comprehensive treatment of such regulatory issues to future research. We do emphasize, however, that our empirical results serve to motivate such future studies and a broader policy discussion around whether/how to regulate exploration strategies by platforms. We expect the policy debate around the pricing transparency of advertising auctions to only grow in importance with the rise of Generative AI (due to an expected increase both in the number and types of advertisers and in the types of ads displayed on platforms), which amplifies the need for effective cold-start strategies. This should further increase the importance of the regulatory oversight of online ads, which includes (but is not limited to) the cold start problem.

\section{Conclusion}

This paper empirically examined the interactions between the bandit problem and the auction problem. Applied to a context of online advertisements on a platform based in Asia, our framework provides the first empirical analysis of the combined auction-bandit problem. We find that the platform's revenue-optimal level of exploration in this problem is substantially larger than the exploration level if only the bandit aspect of the problem were considered. We further investigate the mechanism behind this additional exploration. We find that the additional exploration (i.e., assigning a higher prior to the ``conversion rate'' of an entrant bidder) boosts revenue by increasing the ``effective market thickness,'' through elevating the overall score of the entrant bidder and placing it below the winning incumbent by a small margin.

Our analysis suggests that the auction-bandit problem is empirically relevant, thereby providing further motivation for future research in the area, both theoretical and empirical. Nevertheless, our analysis has some limitations that point to fruitful directions for future research. First, we abstract away from budget pacing and autobidding mechanisms. In our empirical context, advertisers managed budgets manually, and we found no evidence of automated pacing behavior. However, pacing is now common in many platforms, and its interaction with exploration could meaningfully alter both efficiency and revenue outcomes. Understanding how auction-bandit mechanisms perform in the presence of pacing algorithms remains an important open question.

Second, our main analysis focuses primarily on exploration policies that are no more aggressive than those observed in the data to preserve the credibility of our counterfactuals but leaves the performance of more aggressive exploration policies largely underexplored. At sufficiently high levels of exploration, incumbent bidders may have an incentive to shade their bids in early rounds, allowing entrants to win and thereby receive downward revisions to their quality scores. Once entrants' scores have been sufficiently reduced, incumbents may then return to higher bids in later periods, facing weaker competitive pressure. We find no evidence of such behavior in our data, including when we extend our framework to a dynamic model that explicitly captures these forward-looking incentives. Nevertheless, large-scale randomized experiments evaluating alternative exploration regimes in the field represent a promising direction for future research.

\section*{Competing Interests Declaration}
Author(s) have no competing interests to declare.

\singlespacing
\bibliographystyle{chicago}
\bibliography{references}

\begin{thebibliography}{}

\bibitem[\protect\citeauthoryear{Bajari, Benkard, and Levin}{Bajari et~al.}{2007}]{bajari2007estimating}
Bajari, P., C.~L. Benkard, and J.~Levin (2007).
\newblock Estimating dynamic models of imperfect competition.
\newblock {\em Econometrica\/}~{\em 75\/}(5), 1331--1370.

\bibitem[\protect\citeauthoryear{Komarova, Sanches, Silva~Junior, and Srisuma}{Komarova et~al.}{2018}]{komarova2018joint}
Komarova, T., F.~Sanches, D.~Silva~Junior, and S.~Srisuma (2018).
\newblock Joint analysis of the discount factor and payoff parameters in dynamic discrete choice models.
\newblock {\em Quantitative Economics\/}~{\em 9\/}(3), 1153--1194.

\bibitem[\protect\citeauthoryear{Magnac and Thesmar}{Magnac and Thesmar}{2002}]{magnac2002identifying}
Magnac, T. and D.~Thesmar (2002).
\newblock Identifying dynamic discrete decision processes.
\newblock {\em Econometrica\/}~{\em 70\/}(2), 801--816.

\bibitem[\protect\citeauthoryear{Vickrey}{Vickrey}{1961}]{vickrey1961counterspeculation}
Vickrey, W. (1961).
\newblock Counterspeculation, auctions, and competitive sealed tenders.
\newblock {\em The Journal of finance\/}~{\em 16\/}(1), 8--37.

\end{thebibliography}


\begin{thebibliography}{}

\bibitem[\protect\citeauthoryear{Akbarpour and Li}{Akbarpour and Li}{2020}]{akbarpour2020credible}
Akbarpour, M. and S.~Li (2020).
\newblock Credible auctions: A trilemma.
\newblock {\em Econometrica\/}~{\em 88\/}(2), 425--467.

\bibitem[\protect\citeauthoryear{Athey and Haile}{Athey and Haile}{2007}]{athey2007nonparametric}
Athey, S. and P.~A. Haile (2007).
\newblock Nonparametric approaches to auctions.
\newblock {\em Handbook of econometrics\/}~{\em 6}, 3847--3965.

\bibitem[\protect\citeauthoryear{Athey and Nekipelov}{Athey and Nekipelov}{2010}]{athey2010structural}
Athey, S. and D.~Nekipelov (2010).
\newblock A structural model of sponsored search advertising auctions.
\newblock In {\em Sixth ad auctions workshop}, Volume~15, pp.\ ~5.

\bibitem[\protect\citeauthoryear{Babaioff, Kleinberg, and Slivkins}{Babaioff et~al.}{2015}]{babaioff2015truthful}
Babaioff, M., R.~D. Kleinberg, and A.~Slivkins (2015).
\newblock Truthful mechanisms with implicit payment computation.
\newblock {\em Journal of the ACM (JACM)\/}~{\em 62\/}(2), 1--37.

\bibitem[\protect\citeauthoryear{Babaioff, Sharma, and Slivkins}{Babaioff et~al.}{2009}]{babaioff2009characterizing}
Babaioff, M., Y.~Sharma, and A.~Slivkins (2009).
\newblock Characterizing truthful multi-armed bandit mechanisms.
\newblock In {\em Proceedings of the 10th ACM conference on Electronic commerce}, pp.\  79--88.

\bibitem[\protect\citeauthoryear{Bajari, Benkard, and Levin}{Bajari et~al.}{2007}]{bajari2007estimating}
Bajari, P., C.~L. Benkard, and J.~Levin (2007).
\newblock Estimating dynamic models of imperfect competition.
\newblock {\em Econometrica\/}~{\em 75\/}(5), 1331--1370.

\bibitem[\protect\citeauthoryear{Celis, Lewis, Mobius, and Nazerzadeh}{Celis et~al.}{2014}]{celis2014buy}
Celis, L.~E., G.~Lewis, M.~Mobius, and H.~Nazerzadeh (2014).
\newblock Buy-it-now or take-a-chance: Price discrimination through randomized auctions.
\newblock {\em Management Science\/}~{\em 60\/}(12), 2927--2948.

\bibitem[\protect\citeauthoryear{Chan and Park}{Chan and Park}{2015}]{chan2015consumer}
Chan, T.~Y. and Y.-H. Park (2015).
\newblock Consumer search activities and the value of ad positions in sponsored search advertising.
\newblock {\em Marketing Science\/}~{\em 34\/}(4), 606--623.

\bibitem[\protect\citeauthoryear{Chapelle and Li}{Chapelle and Li}{2011}]{chapelle2011empirical}
Chapelle, O. and L.~Li (2011).
\newblock An empirical evaluation of thompson sampling.
\newblock {\em Advances in neural information processing systems\/}~{\em 24}.

\bibitem[\protect\citeauthoryear{Choi and Mela}{Choi and Mela}{2019}]{choi2019monetizing}
Choi, H. and C.~F. Mela (2019).
\newblock Monetizing online marketplaces.
\newblock {\em Marketing Science\/}~{\em 38\/}(6), 948--972.

\bibitem[\protect\citeauthoryear{Devanur and Kakade}{Devanur and Kakade}{2009}]{devanur2009price}
Devanur, N.~R. and S.~M. Kakade (2009).
\newblock The price of truthfulness for pay-per-click auctions.
\newblock In {\em Proceedings of the 10th ACM conference on Electronic commerce}, pp.\  99--106.

\bibitem[\protect\citeauthoryear{Edelman, Ostrovsky, and Schwarz}{Edelman et~al.}{2007}]{edelman2007internet}
Edelman, B., M.~Ostrovsky, and M.~Schwarz (2007).
\newblock Internet advertising and the generalized second-price auction: Selling billions of dollars worth of keywords.
\newblock {\em American economic review\/}~{\em 97\/}(1), 242--259.

\bibitem[\protect\citeauthoryear{Feng, Liaw, and Zhou}{Feng et~al.}{2023}]{feng2023improved}
Feng, Z., C.~Liaw, and Z.~Zhou (2023).
\newblock Improved online learning algorithms for ctr prediction in ad auctions.
\newblock In {\em International Conference on Machine Learning}, pp.\  9921--9937. PMLR.

\bibitem[\protect\citeauthoryear{Hadad, Hirshberg, Zhan, Wager, and Athey}{Hadad et~al.}{2021}]{hadad2021confidence}
Hadad, V., D.~A. Hirshberg, R.~Zhan, S.~Wager, and S.~Athey (2021).
\newblock Confidence intervals for policy evaluation in adaptive experiments.
\newblock {\em Proceedings of the national academy of sciences\/}~{\em 118\/}(15), e2014602118.

\bibitem[\protect\citeauthoryear{Hauser, Urban, Liberali, and Braun}{Hauser et~al.}{2009}]{hauser2009website}
Hauser, J.~R., G.~L. Urban, G.~Liberali, and M.~Braun (2009).
\newblock Website morphing.
\newblock {\em Marketing Science\/}~{\em 28\/}(2), 202--223.

\bibitem[\protect\citeauthoryear{Horvitz and Thompson}{Horvitz and Thompson}{1952}]{horvitz1952generalization}
Horvitz, D.~G. and D.~J. Thompson (1952).
\newblock A generalization of sampling without replacement from a finite universe.
\newblock {\em Journal of the American statistical Association\/}~{\em 47\/}(260), 663--685.

\bibitem[\protect\citeauthoryear{Jain, Li, Loghmani, Mason, and Yoganarasimhan}{Jain et~al.}{2024}]{jain2024effective}
Jain, L., Z.~Li, E.~Loghmani, B.~Mason, and H.~Yoganarasimhan (2024).
\newblock Effective adaptive exploration of prices and promotions in choice-based demand models.
\newblock {\em Marketing Science\/}~{\em 43\/}(5), 1002--1030.

\bibitem[\protect\citeauthoryear{Jerath, Ma, and Park}{Jerath et~al.}{2014}]{jerath2014consumer}
Jerath, K., L.~Ma, and Y.-H. Park (2014).
\newblock Consumer click behavior at a search engine: The role of keyword popularity.
\newblock {\em Journal of Marketing Research\/}~{\em 51\/}(4), 480--486.

\bibitem[\protect\citeauthoryear{Kim and Pal}{Kim and Pal}{2025}]{kim2025nonparametric}
Kim, D. and P.~Pal (2025).
\newblock Nonparametric estimation of sponsored search auctions and impact of ad quality on search revenue.
\newblock {\em Management Science\/}.

\bibitem[\protect\citeauthoryear{Lahaie and Pennock}{Lahaie and Pennock}{2007}]{lahaie2007revenue}
Lahaie, S. and D.~M. Pennock (2007).
\newblock Revenue analysis of a family of ranking rules for keyword auctions.
\newblock In {\em Proceedings of the 8th ACM Conference on Electronic Commerce}, pp.\  50--56.

\bibitem[\protect\citeauthoryear{Lahaie, Pennock, Saberi, and Vohra}{Lahaie et~al.}{2007}]{lahaie2007sponsored}
Lahaie, S., D.~M. Pennock, A.~Saberi, and R.~V. Vohra (2007).
\newblock Sponsored search auctions.
\newblock {\em Algorithmic game theory\/}~{\em 1}, 699--716.

\bibitem[\protect\citeauthoryear{Lattimore and Szepesv{\'a}ri}{Lattimore and Szepesv{\'a}ri}{2020}]{lattimore2020bandit}
Lattimore, T. and C.~Szepesv{\'a}ri (2020).
\newblock {\em Bandit algorithms}.
\newblock Cambridge University Press.

\bibitem[\protect\citeauthoryear{Levin and Milgrom}{Levin and Milgrom}{2010}]{levin2010online}
Levin, J. and P.~Milgrom (2010).
\newblock Online advertising: Heterogeneity and conflation in market design.
\newblock {\em American Economic Review\/}~{\em 100\/}(2), 603--607.

\bibitem[\protect\citeauthoryear{Misra, Schwartz, and Abernethy}{Misra et~al.}{2019}]{misra2019dynamic}
Misra, K., E.~M. Schwartz, and J.~Abernethy (2019).
\newblock Dynamic online pricing with incomplete information using multiarmed bandit experiments.
\newblock {\em Marketing Science\/}~{\em 38\/}(2), 226--252.

\bibitem[\protect\citeauthoryear{Myerson}{Myerson}{1981}]{myerson1981optimal}
Myerson, R.~B. (1981).
\newblock Optimal auction design.
\newblock {\em Mathematics of operations research\/}~{\em 6\/}(1), 58--73.

\bibitem[\protect\citeauthoryear{Ostrovsky and Schwarz}{Ostrovsky and Schwarz}{2023}]{ostrovsky2023reserve}
Ostrovsky, M. and M.~Schwarz (2023).
\newblock Reserve prices in internet advertising auctions: A field experiment.
\newblock {\em Journal of Political Economy\/}~{\em 131\/}(12), 3352--3376.

\bibitem[\protect\citeauthoryear{Rafieian}{Rafieian}{2020}]{rafieian2020revenue}
Rafieian, O. (2020).
\newblock Revenue-optimal dynamic auctions for adaptive ad sequencing.
\newblock Technical report, Working paper: Cornell University.

\bibitem[\protect\citeauthoryear{Rafieian and Yoganarasimhan}{Rafieian and Yoganarasimhan}{2021}]{rafieian2021targeting}
Rafieian, O. and H.~Yoganarasimhan (2021).
\newblock Targeting and privacy in mobile advertising.
\newblock {\em Marketing Science\/}~{\em 40\/}(2), 193--218.

\bibitem[\protect\citeauthoryear{Rafieian and Zuo}{Rafieian and Zuo}{2026}]{rafieian2026rational}
Rafieian, O. and S.~Zuo (2026).
\newblock From rational reliance to impeded learning: The information advantage of personalized algorithms.
\newblock SSRN Working Paper.

\bibitem[\protect\citeauthoryear{Schwartz, Bradlow, and Fader}{Schwartz et~al.}{2017}]{schwartz2017customer}
Schwartz, E.~M., E.~T. Bradlow, and P.~S. Fader (2017).
\newblock Customer acquisition via display advertising using multi-armed bandit experiments.
\newblock {\em Marketing Science\/}~{\em 36\/}(4), 500--522.

\bibitem[\protect\citeauthoryear{Stourm and Bax}{Stourm and Bax}{2017}]{stourm2017incorporating}
Stourm, V. and E.~Bax (2017).
\newblock Incorporating hidden costs of annoying ads in display auctions.
\newblock {\em International Journal of Research in Marketing\/}~{\em 34\/}(3), 622--640.

\bibitem[\protect\citeauthoryear{Varian}{Varian}{2007}]{varian2007position}
Varian, H.~R. (2007).
\newblock Position auctions.
\newblock {\em international Journal of industrial Organization\/}~{\em 25\/}(6), 1163--1178.

\bibitem[\protect\citeauthoryear{Vickrey}{Vickrey}{1961}]{vickrey1961counterspeculation}
Vickrey, W. (1961).
\newblock Counterspeculation, auctions, and competitive sealed tenders.
\newblock {\em The Journal of finance\/}~{\em 16\/}(1), 8--37.

\bibitem[\protect\citeauthoryear{Wilbur, Xu, and Kempe}{Wilbur et~al.}{2013}]{wilbur2013correcting}
Wilbur, K.~C., L.~Xu, and D.~Kempe (2013).
\newblock Correcting audience externalities in television advertising.
\newblock {\em Marketing Science\/}~{\em 32\/}(6), 892--912.

\bibitem[\protect\citeauthoryear{Yao and Mela}{Yao and Mela}{2011}]{yao2011dynamic}
Yao, S. and C.~F. Mela (2011).
\newblock A dynamic model of sponsored search advertising.
\newblock {\em Marketing Science\/}~{\em 30\/}(3), 447--468.

\bibitem[\protect\citeauthoryear{Ye, Yoganarasimhan, and Zheng}{Ye et~al.}{2025}]{ye2025lola}
Ye, Z., H.~Yoganarasimhan, and Y.~Zheng (2025).
\newblock Lola: Llm-assisted online learning algorithm for content experiments.
\newblock {\em Marketing Science\/}~{\em 44\/}(5), 995--1016.

\bibitem[\protect\citeauthoryear{Ye, Zhang, Zhang, Zhang, Chen, and Xu}{Ye et~al.}{2022}]{ye2022cold}
Ye, Z., D.~J. Zhang, H.~Zhang, R.~Zhang, X.~Chen, and Z.~Xu (2022).
\newblock Cold start to improve market thickness on online advertising platforms: Data-driven algorithms and field experiments.
\newblock {\em Management Science\/}.

\bibitem[\protect\citeauthoryear{Zhan, Hadad, Hirshberg, and Athey}{Zhan et~al.}{2021}]{zhan2021off}
Zhan, R., V.~Hadad, D.~A. Hirshberg, and S.~Athey (2021).
\newblock Off-policy evaluation via adaptive weighting with data from contextual bandits.
\newblock In {\em Proceedings of the 27th ACM SIGKDD Conference on Knowledge Discovery \& Data Mining}, pp.\  2125--2135.

\end{thebibliography}
\newpage
\appendix

\begin{appendices}
\begin{bibunit}[chicago]

\section{Proof for Proposition \ref{prop:truthful}}
\label{appsec:proof}
\begin{proof}
Fix $a$, $k$, $t$, and any rivals' bidding policy profile $\sigma_{-a}$. Since $\delta = 0$, the Bellman equation \eqref{eq:bellman} reduces to $V^{\sigma_{-a}}_{a,k,t}(s_{k,t}) = \max_{b_{a,k,t}} \bar{\pi}^{\sigma_{-a}}_{a,k,t}(b_{a,k,t}, s_{k,t})$. Because $\bar{\pi}^{\sigma_{-a}}_{a,k,t}$ averages over the quality-score draws and rivals' bids, it suffices to show that $b_{a,k,t} = v_{a,k,t}$ maximizes $a$'s profit at every realization of these variables.

Fix any such realization and define the rival price and the reserve price, both expressed in bid units,
\[
e_{a,k,t}^{\text{rival}} = \frac{\max_{a' \in \mathcal{A}_{k,t} \setminus\{a\}} b_{a',k,t}\, q_{a',k,t}}{q_{a,k,t}}, \qquad e_{a,k,t}^{\text{reserve}} = \frac{r_k}{q_{a,k,t}},
\]
and let $e_{a,k,t} = \max\{e_{a,k,t}^{\text{rival}}, e_{a,k,t}^{\text{reserve}}\}$ denote the minimum bid advertiser $a$ would need to submit to win the impression; whenever $a$ wins, $e_{a,k,t}$ equals its per-conversion price under Definition~\ref{def:tssp}. By assumption, $q_{a,k,t}$ does not depend on $b_{a,k,t}$, and neither do $r_k$, rivals' quality scores, or rivals' bids, which $\sigma_{-a}$ generates from $s_{k,t}$. Hence $e_{a,k,t}$ is fixed from $a$'s perspective when choosing its own bid. Because quality scores are drawn from continuous Beta distributions, ties occur with probability zero, so $a$ wins if and only if $b_{a,k,t} > e_{a,k,t}$. Averaging over the conversion outcome alone, $a$'s profit at this realization is
\[
\mu_{a,k,t}\left(v_{a,k,t} - e_{a,k,t}\right)\mathbbm{1}\{b_{a,k,t} > e_{a,k,t}\},
\]
and averaging this expression over realizations yields $\bar{\pi}^{\sigma_{-a}}_{a,k,t}(b_{a,k,t}, s_{k,t})$ in \eqref{eq:stage_utility}, because $\mathbb{E}[(v_{a,k,t} - e_{a,k,t})\mathbbm{1}\{A_{k,t} = a\}] = (v_{a,k,t} - \mathbb{E}[e_{a,k,t} \mid A_{k,t} = a])\Pr(A_{k,t} = a)$, with all terms conditional on $b_{a,k,t}$, $s_{k,t}$, and $\sigma_{-a}$.

The profit at a fixed realization is exactly the payoff of a bidder in a standard single-item second-price auction with reserve price $e_{a,k,t}$, scaled by the constant $\mu_{a,k,t}$. This is the setting of \citet{vickrey1961counterspeculation}'s classical result that truthful bidding is a weakly dominant strategy in a second-price auction, extended in the standard way to accommodate a reserve price: bidding $b_{a,k,t} = v_{a,k,t}$ wins exactly when winning is profitable ($v_{a,k,t} > e_{a,k,t}$) and loses exactly when winning would be unprofitable ($v_{a,k,t} < e_{a,k,t}$). No alternative bid improves on either outcome: overbidding past $e_{a,k,t}$ when $v_{a,k,t} < e_{a,k,t}$ only buys a strictly negative payoff, and underbidding below $e_{a,k,t}$ when $v_{a,k,t} > e_{a,k,t}$ only forgoes a strictly positive one. Because this holds at every realization, truthful bidding also maximizes $\bar{\pi}^{\sigma_{-a}}_{a,k,t}(b_{a,k,t}, s_{k,t})$, for every $\sigma_{-a}$. Truthful bidding is therefore weakly dominant, and because the argument holds for every advertiser regardless of rivals' strategies, truthful bidding by all advertisers is an equilibrium.
\end{proof}

\clearpage
\section{Rationale Behind Exploration}
\label{appsec:rationale}
\begin{figure}[h]
    \centering
    \includegraphics[width=0.5\linewidth]{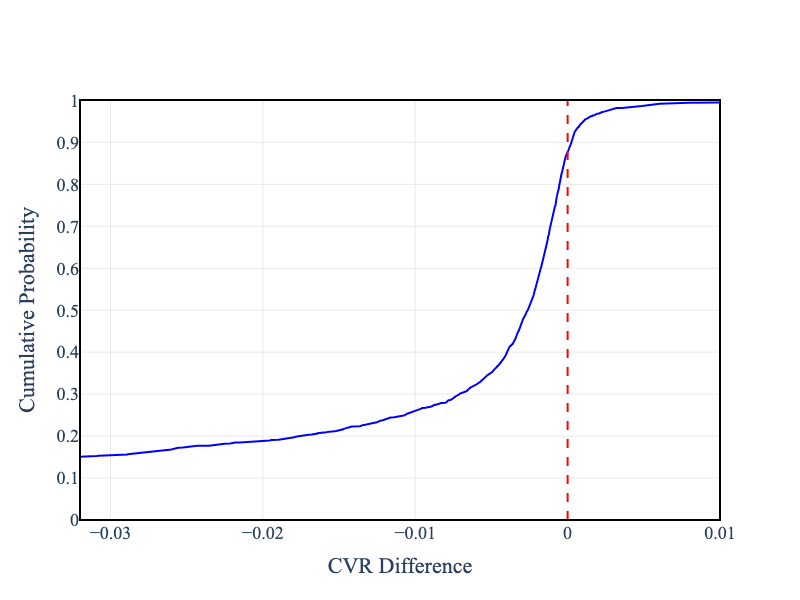}
    \caption{Empirical CDF of the Difference Between Average Conversion Rates of Entrants and Incumbents}
    \label{fig:cvrdiff}
\end{figure}

In this section, we turn to a more fundamental question: why does an app store engage in exploration? As highlighted in our theoretical framework, the primary motivation for exploration is the potential to discover high-quality ads that create value for the app store. To empirically assess this argument, we compare the aggregated conversion rates of entrants and incumbents for each keyword over the six month period. A higher conversion rate for entrants relative to incumbents indicates greater value derived from exploring new entrants. Figure \ref{fig:cvrdiff} presents the empirical cumulative distribution function (CDF) of the difference in conversion rates between entrants and incumbents for each keyword. The differences in average conversion rates highlight the importance of exploration in the auction mechanism. Specifically, for 12\% of keywords, entrants achieve a higher average conversion rate than incumbents. This finding suggests that while incumbents generally dominate due to their past performance and revenue generation, entrants can occasionally outperform them in certain contexts. These results underscore the necessity of continued exploration to identify high-performing entrants and prevent the auction mechanism from prematurely settling on suboptimal allocations driven solely by historical performance.

\clearpage
\section{Entrants' Positions in Auctions}
\label{appsec:entrant_positions}

\begin{figure}[H]
    \centering
    \includegraphics[width=0.6\linewidth]{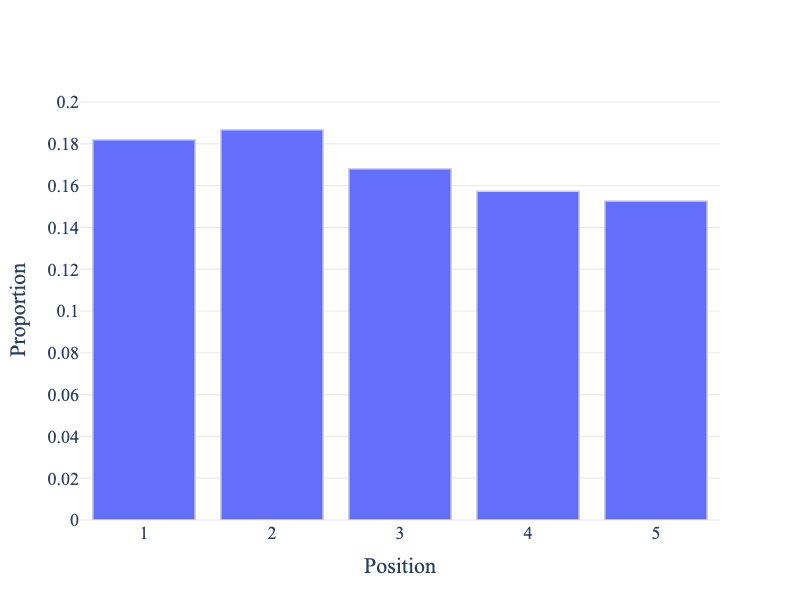}
    \caption{Entrant Shares by Auction Position}
    \label{fig:entrant_place}
    \vspace{0.25em}
    \begin{minipage}{0.95\textwidth}
    \footnotesize
    \textit{Notes:} Bars report the fraction of auctions in which an entrant occupies each position.
    \end{minipage}
\end{figure}

This appendix supplements the discussion in Section~\ref{sssec:entrants_outcome} by reporting how frequently entrants occupy each of the top five positions in the sample auctions. Although only the first position results in an impression, the lower positions indicate how often entrants can influence pricing. Figure~\ref{fig:entrant_place} shows the fraction of auctions in which an entrant occupies each position. Entrants occupy approximately 18\% of first positions, indicating that they receive a substantial share of impressions, and approximately 19\% of second positions, in which case their scores directly determine the winner's price. Entrants therefore play a meaningful role in both allocation and pricing outcomes.

\clearpage
\section{Time-Trend Diagnostics for Conversion Rates and Bids}
\label{appsec:primitive_stability}

This appendix presents the time-trend specifications and the distributions of the resulting $p$-values discussed in Section~\ref{sssec:primitive_dynamics}.

For conversion rates, let $Y_{a,k,t}$ equal one if an impression allocated to advertiser $a$ on keyword $k$ results in a conversion, and zero otherwise. For each ad--keyword pair, we separately estimate
\begin{equation}
    \operatorname{logit}\!\left(\Pr(Y_{a,k,t}=1)\right)
    =
    \alpha_{a,k}
    +
    \beta_{a,k}d_t
    \label{eq:cvr_time_trend}
\end{equation}
where $d_t$ denotes calendar time and $\beta_{a,k}$ captures the time trend in the probability of conversion. For each pair, we test the null hypothesis $H_0:\beta_{a,k}=0$.

For bids, we use the one-week bid sample and separately estimate the following regression for each ad--keyword pair:
\begin{equation}
    \log(b_{a,k,t})
    =
    \alpha_{a,k}
    +
    \gamma_{a,k}d_t
    +
    \eta_{a,k,t},
    \label{eq:bid_time_trend}
\end{equation}
where $\gamma_{a,k}$ captures the time trend in bids. We test the null hypothesis $H_0:\gamma_{a,k}=0$. For ad--keyword pairs with constant bids, the associated $p$-value are not defined.

\begin{figure}[htbp]
    \centering
    \begin{subfigure}[t]{0.48\textwidth}
        \centering
        \includegraphics[width=\linewidth]{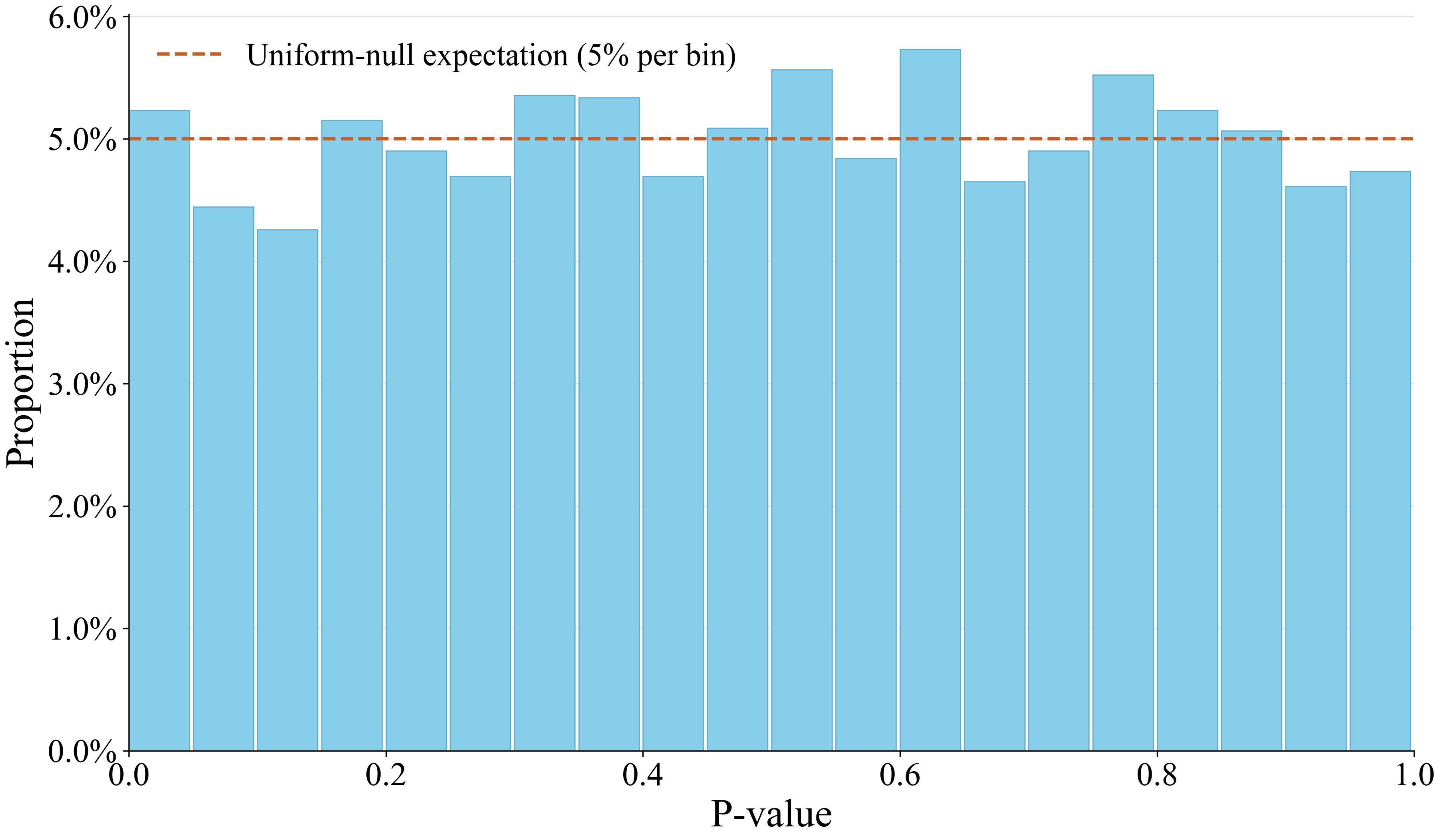}
        \caption{Conversion rates}
        \label{fig:cvr_time_pvalues}
    \end{subfigure}
    \hfill
    \begin{subfigure}[t]{0.48\textwidth}
        \centering
        \includegraphics[width=\linewidth]{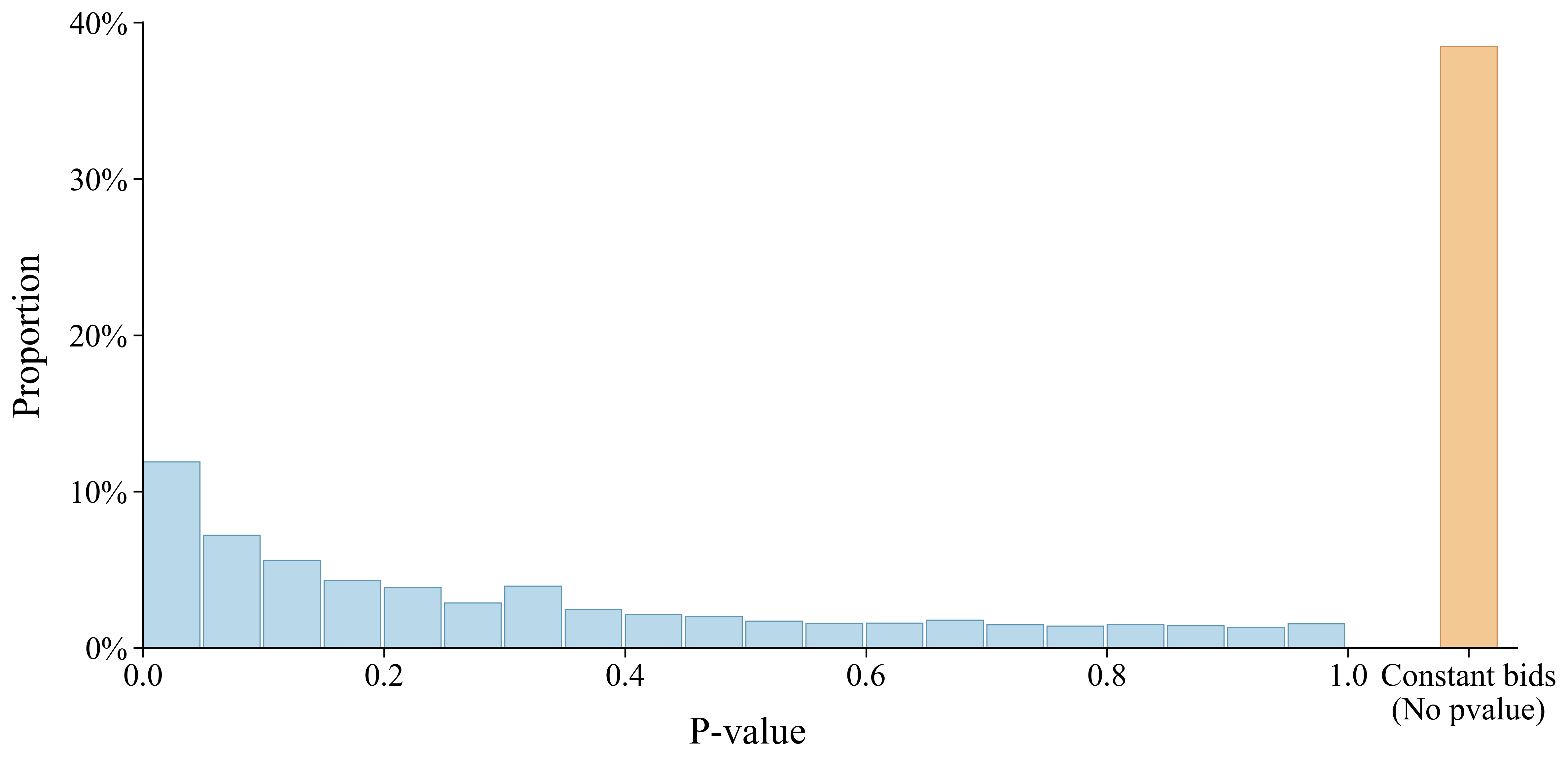}
        \caption{Bids}
        \label{fig:bid_time_pvalues}
    \end{subfigure}
    \caption{Time-Trend Tests for Conversion Rates and Bids}
    \label{fig:primitive_stability_pvalues}
    \vspace{0.25em}
    \begin{minipage}{0.95\textwidth}
    \footnotesize
    \textit{Notes:} Each panel shows the distribution of $p$-values from separate ad--keyword-level time-trend regressions. The light-orange bar in panel (b) identifies pairs with constant bids.
    \end{minipage}
\end{figure}

Figure~\ref{fig:primitive_stability_pvalues} presents the resulting distributions. Panel~(a) reports the tests for conversion outcomes. Panel~(b) reports the tests for bids and separately identifies the ad--keyword pairs with constant bids.

\clearpage
\section{Examining Entrants' Overbidding Incentives}
\label{appsec:overbidding}

In this section, we discuss \textit{entrants' overbidding incentives}, which can be justified under long-term utility maximization by bidders. High-quality entrants may bid above their true valuation in the early stages, especially if their prior quality score underestimates their actual conversion rate, in order to accelerate the revelation of their true quality. This strategy helps them win impressions, which leads to the auctioneer learning their true quality scores. As a result, they can lower their bid in later periods and extract higher total utility compared to a scenario in which they only maximize per-period utility.

We empirically examine the overbidding incentives of entrants, which suggest that an entrant with a quality score higher than the prior mean may have an incentive to overbid in order to allow the auctioneer to learn its true quality score. To test this empirically, we first identify the entrants in our data whose quality scores exceed the prior mean. We then estimate the following regressions:
\begin{gather}
    \bar{b}_{a,k,h} = \alpha  + \gamma_1 \cdot \text{Hour}_{h} + \tau_{a,k} +  \epsilon'_{a,k,h} \\
    \bar{b}_{a,k,h} = \alpha  + \gamma_2 \cdot \bar{q}_{a,k,h} + \tau_{a,k} +  \epsilon''_{a,k,h}
\end{gather}
where $\text{Hour}_{h}$ is the numerical variable for the hour that measures the time trend, and $\bar{q}_{a,k,h}$ is the average quality score draw for ad $a$ in keyword $k$ at hour $h$. If high-quality entrants' overbidding is indeed a deliberate strategy, we would expect these entrants to place higher bids in the early periods and subsequently lower their bids over time (indicated by a negative $\gamma_1$) or when assigned lower prior quality scores (indicated by a negative $\gamma_2$). We present the estimates for these regressions in Columns (1) and (2) of Table \ref{tab:dyntest1}. As shown in these columns, the coefficients are not statistically significant, which provides support for the per-period utility maximization assumption over the long-term utility maximization hypothesis.

\begin{table}[h]
\caption{Regression Specifications to Test Static vs. Dynamic Bidding Based on Entrants' Overbidding Incentives\label{tab:dyntest1}}
\begin{minipage}{\columnwidth}
\begin{center}
\footnotesize{
\begin{tabular}[t]{lcc} 
\\[-1.8ex]\hline 
\hline \\[-1.8ex] 
 & \multicolumn{2}{c}{\textit{Dependent variable: Bid ($\bar{b}_{a,k,h}$)}} \\ 
\cline{2-3} 
\\[-1.8ex] & (1) OLS & (2) OLS \\ 
\hline \\[-1.8ex] 
Hour & $-2.2934$ &  \\
  & $(2.340)$ &  \\[0.1cm]
Mean Quality Score  &  & $-670.7890$ \\
   &  & $(1921.237)$ \\[0.1cm]
\hline \\[-1.8ex] 
Keyword-Ad FE    &  \checkmark &  \checkmark \\
$R^2$ &  $0.905$ & $0.905$ \\
No. of Obs. &  $12,641$ & $12,641$ \\

\hline \\[-1.8ex] 
\textit{Note:}  & \multicolumn{2}{r}{$^{*}$p$<$0.05; $^{**}$p$<$0.01; $^{***}$p$<$0.001} \\ 
\hline\hline
\end{tabular}
}
\end{center}
\end{minipage}
\end{table}

\newpage 

\section{Inverse Propensity Score Estimator for Conversion Rates}
\label{appsec:ips}
In $\S$\ref{sssec:cvr_estimation}, we presented a sample analogue estimator for conversion rates and discussed that the estimator would be unbiased if impressions allocated to an ad are randomly drawn from the overall impression distribution. In this section, we relax this assumption and employ an Inverse Propensity Score (IPS) estimator:
\begin{equation}
    \widehat{\mu}_{a,k}^{\text{IPS}} = \frac{1}{T_k} \sum_{t=1}^{T_k} \frac{\mathbbm{1}(a = A_{k,t}) Y_{k,t}}{\widehat{\pi}(a \mid B_{k,t}, I_{k,t}, C_{k,t}; \alpha^0,\beta^0)},
\end{equation}
where $\widehat{\pi}(a \mid B_{k,t}, I_{k,t}, C_{k,t}; \alpha^0,\beta^0)$ represents the estimated propensity score for ad $a$ to be shown in impression $t$ of keyword $k$. We confirm that both approaches yield highly similar estimates, reinforcing the validity of our assumption that impressions allocated to each ad approximate a random selection of impressions.

\newpage

\section{Within-Keyword Distributions of Estimated Primitives}
\label{appsec:keyword_level_primitives}

This appendix supplements the results in Section~\ref{ssec:results_revised} by describing how the estimated primitives vary across keywords. For each keyword, we compute the mean and median estimate across the bidders participating in that keyword.

\subsection{Conversion Rates}

\begin{figure}[!htbp]
    \centering
    \begin{subfigure}[t]{0.48\textwidth}
        \centering
        \includegraphics[width=\textwidth]{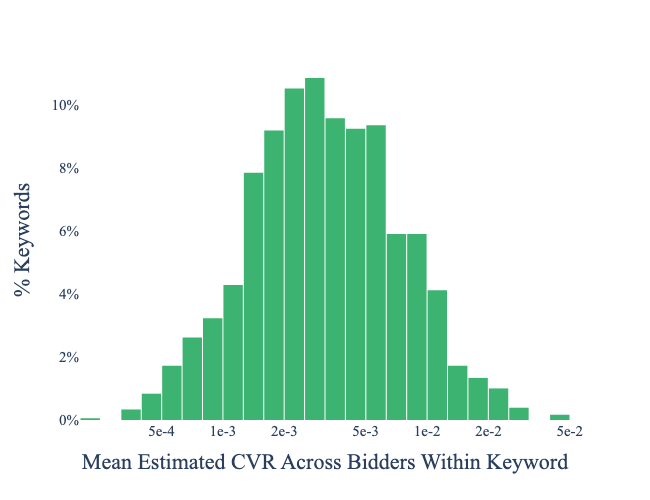}
        \caption{Mean conversion rate}
        \label{fig:cvr_keyword_mean}
    \end{subfigure}
    \hfill
    \begin{subfigure}[t]{0.48\textwidth}
        \centering
        \includegraphics[width=\textwidth]{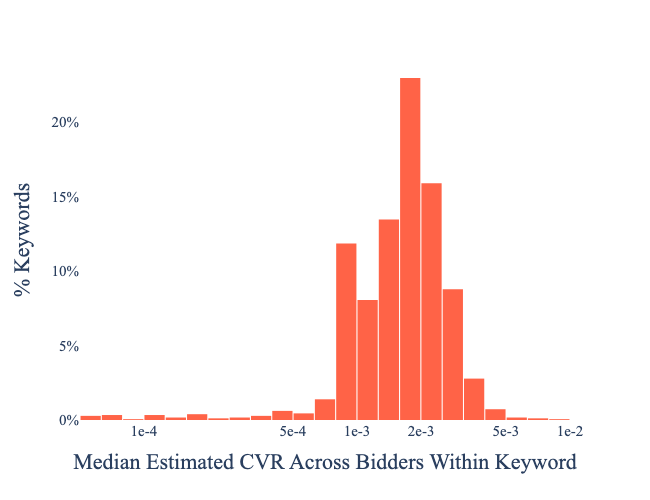}
        \caption{Median conversion rate}
        \label{fig:cvr_keyword_median}
    \end{subfigure}
    \caption{Keyword-Level Conversion-Rate Distributions}
    \label{fig:cvr_keyword_mean_median}
    \vspace{0.25em}
    \begin{minipage}{0.95\textwidth}
    \footnotesize
    \textit{Notes:} Means and medians are computed across bidders within each keyword. The histograms show their distributions across keywords. Both horizontal axes use a logarithmic scale.
    \end{minipage}
\end{figure}

Figure~\ref{fig:cvr_keyword_mean_median} presents the distributions of the within-keyword mean and median estimated conversion rates. Both distributions are concentrated at low conversion-rate levels, consistent with the bidder--keyword-level evidence in Figure~\ref{fig:cvr_bidder_keyword_dist}. The distribution of medians is shifted further toward low values, reflecting the large number of bidder--keyword pairs with zero or near-zero estimated conversion rates.

\subsection{Valuations}

\begin{figure}[!htbp]
    \centering
    \begin{subfigure}[t]{0.48\textwidth}
        \centering
        \includegraphics[width=\textwidth]{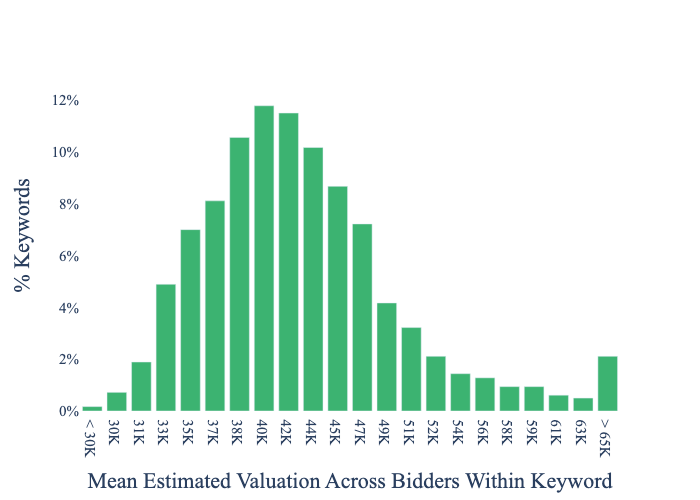}
        \caption{Mean valuation}
        \label{fig:valuation_keyword_mean}
    \end{subfigure}
    \hfill
    \begin{subfigure}[t]{0.48\textwidth}
        \centering
        \includegraphics[width=\textwidth]{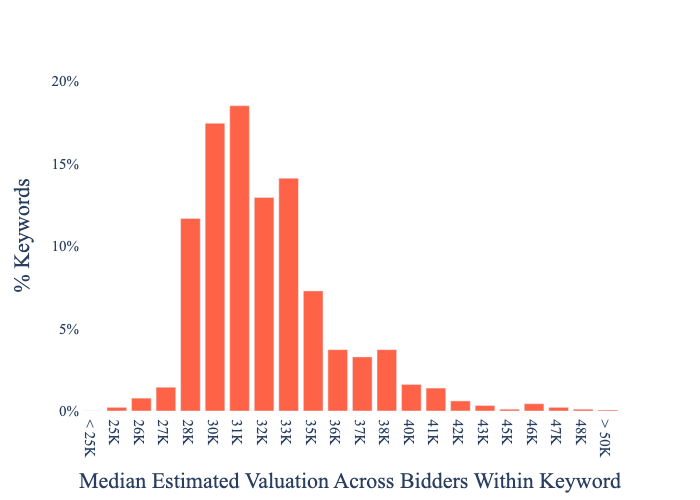}
        \caption{Median valuation}
        \label{fig:valuation_keyword_median}
    \end{subfigure}
    \caption{Keyword-Level Valuation Distributions}
    \label{fig:valuation_keyword_mean_median}
    \vspace{0.25em}
    \begin{minipage}{0.95\textwidth}
    \footnotesize
    \textit{Notes:} Means and medians are computed across bidders within each keyword. The histograms show their distributions across keywords.
    \end{minipage}
\end{figure}

Figure~\ref{fig:valuation_keyword_mean_median} presents the distributions of the within-keyword mean and median estimated valuations. Although bidder-level valuations exhibit a long right tail, the keyword-level means and medians are more tightly distributed.

\clearpage

\section{Supplementary Analysis of UCB Counterfactuals}
\label{appsec:ucb_quality_bias}

In Section~\ref{ssec:ucb}, we show that UCB outperforms TS in our counterfactuals. Here, we explore the mechanism behind this result by examining quality-score bias, defined as the difference between an advertiser's assigned quality score and its true conversion rate.

\begin{figure}[htbp]
    \centering
    \begin{subfigure}[b]{0.49\textwidth}
        \includegraphics[width=\linewidth]{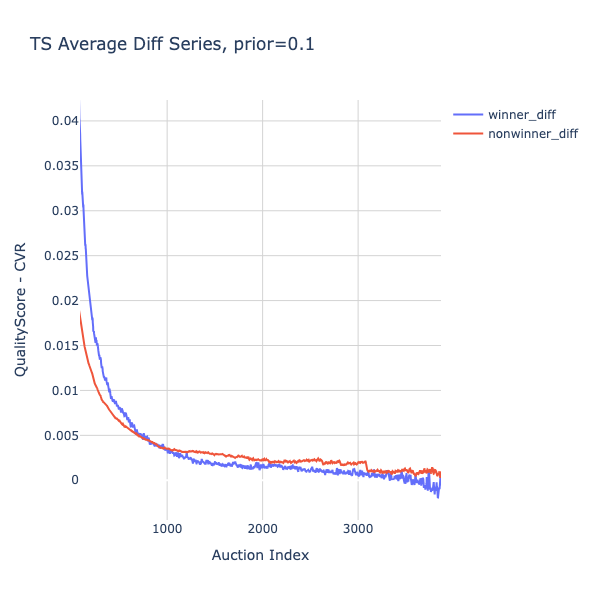}
        \caption{Thompson Sampling (prior mean $=0.1$)}
        \label{fig:sub1}
    \end{subfigure}
    \hfill
    \begin{subfigure}[b]{0.49\textwidth}
        \includegraphics[width=\linewidth]{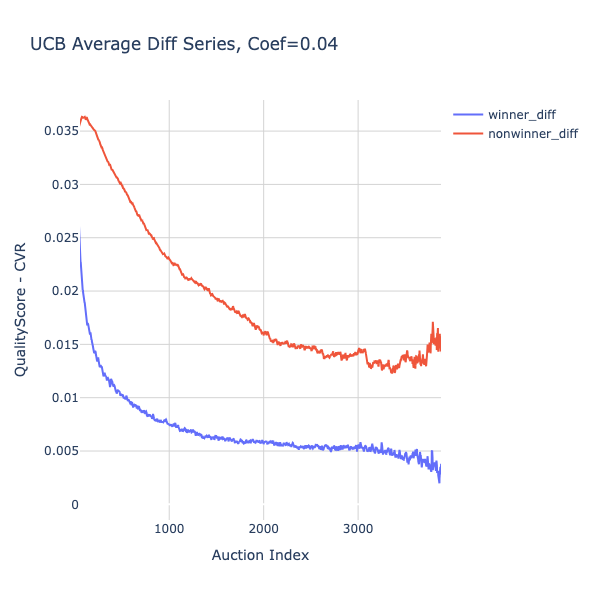}
        \caption{UCB ($\rho=0.04$)}
        \label{fig:sub2}
    \end{subfigure}
    \caption{Winner and Nonwinner Quality-Score Bias}
    \label{fig:diff_series}
\end{figure}

We compare TS with the mean prior $0.1$ (same prior as in the data) and UCB with $\rho = 0.04$. Let $\mu_{a,k}$ be the true CVR, $q^{M}_{a,k,t}$ the quality score under mechanism $M$, and $A^M_{k,t}$ the winner on keyword $k$ at time $t$. We define the winner bias and the nonwinner bias as follows:
\[
\text{w-bias}^{M}_{t}
=\frac{1}{|\mathcal{K}|}\sum_{k}\big(q^{M}_{A^M_{k,t},k,t}-\mu_{A^M_{k,t},k}\big),\quad
\text{nw-bias}^{M}_{t}
=\frac{1}{|\mathcal{K}|}\sum_{k}\frac{1}{|\mathcal{A}_{k,t}|-1}\sum_{a\neq A^M_{k,t}}\big(q^{M}_{a,k,t}-\mu_{a,k}\big)
\]
Figure~\ref{fig:diff_series} plots these time series: Thompson Sampling in the left panel (prior mean \(0.1\)) and UCB in the right panel (\(\rho=0.04\)).
The $y$-axis reports the average bias, the difference between quality scores and CVR averaged across keywords. Under TS, both biases decay toward zero, consistent with the posterior concentration. Under UCB, the winner bias also decays, but the \emph{nonwinner bias remains positive} throughout the horizon—precisely the optimism that elevates second-highest scores and prices. This result explains why UCB achieves higher revenue than TS in our setting.

\clearpage

\section{Budget-Constrained Counterfactuals}
\label{appsec:budget_constraint}

This appendix presents the construction and implementation of the daily spending caps used in the robustness check summarized in Section~\ref{ssec:budget_constraint}. Throughout this exercise, we maintain the per-period truthful-bidding assumption. Because we do not observe advertisers' true budget capacities, we impose a conservative and nonparametric constraint based on their observed spending. Let $S_d(a,k)$ denote the total payment by advertiser $a$ on keyword $k$ and day $d$. We set the daily cap for each ad--keyword pair $(a,k)$ equal to its maximum observed daily expenditure over the sample window:
\begin{equation}
    \bar{B}_{a,k}=\max_d S_d(a,k).
    \label{eq:daily_budget_cap}
\end{equation}
This construction assumes that any daily budget is weakly above the largest daily expenditure observed for that pair and remains constant over the seven-day counterfactual window.

We rerun the counterfactuals from Section~\ref{sec:results}, varying the entrant CVR prior while tracking cumulative spending for every ad--keyword pair within each day. Once advertiser $a$ reaches $\bar{B}_{a,k}$ on keyword $k$, we remove it from subsequent auctions for that keyword for the remainder of the day. The advertiser re-enters at the beginning of the next day with its posterior carried forward. All other components of the TS--SP mechanism remain unchanged.

\begin{figure}[htbp]
    \centering
    \includegraphics[width=0.8\linewidth]{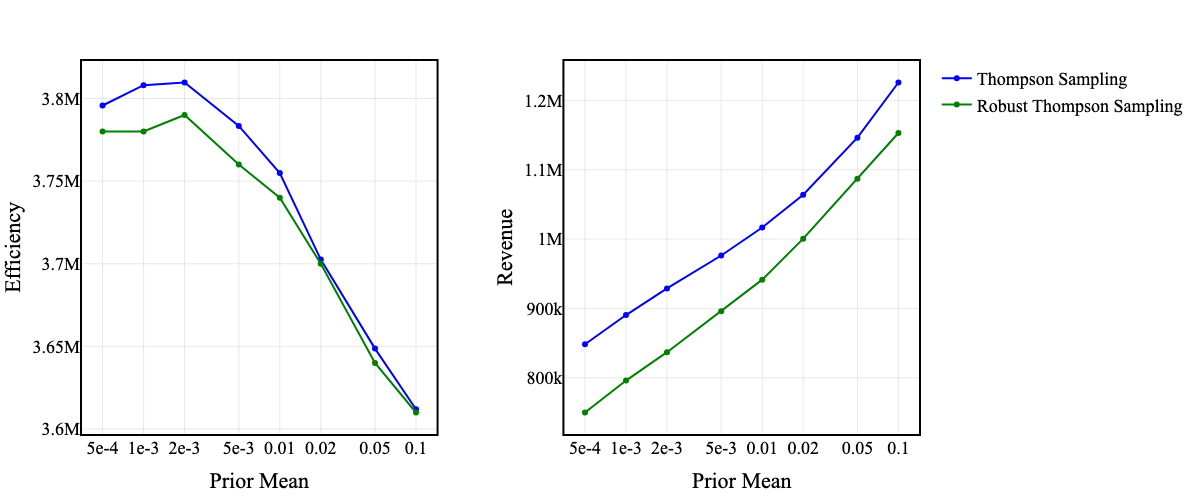}
    \caption{Counterfactual Outcomes with Budget Constraints}
    \label{fig:revenue_efficiency_robust}
    \vspace{0.25em}
    \begin{minipage}{0.95\textwidth}
    \footnotesize
    \textit{Notes:} The left panel shows efficiency and the right panel revenue across entrant prior means. Blue denotes unconstrained Thompson Sampling; green denotes Thompson Sampling with daily ad--keyword spending caps equal to each pair's maximum observed daily expenditure.
    \end{minipage}
\end{figure}

Figure~\ref{fig:revenue_efficiency_robust} compares the baseline unconstrained projections with the budget-constrained counterfactuals. The spending caps reduce revenue levels but preserve the positive relationship between the entrant prior and revenue. They also leave the relationship between the prior and efficiency largely unchanged.

\newpage

\section{Counterfactuals with Time-Varying Valuations}
\label{appsec:time_varying_valuations}

In the main analysis, we use the average bid submitted by advertiser $a$ on keyword $k$ as its structural valuation, $\widehat{v}_{a,k}$, as described in Section~\ref{sssec:valuation_estimation}. The baseline simulation therefore assigns the same valuation to advertiser $a$ whenever it participates in an auction for keyword $k$. To relax this restriction, we instead use the exact contemporaneous submitted bid as the bidder's valuation per conversion:
\[
v_{a,k,t}=b_{a,k,t}.
\]
We rerun the full set of TS--SP counterfactuals while holding all other components of the simulation unchanged.

\begin{figure}[htbp]
    \centering
    \includegraphics[width=0.8\linewidth]{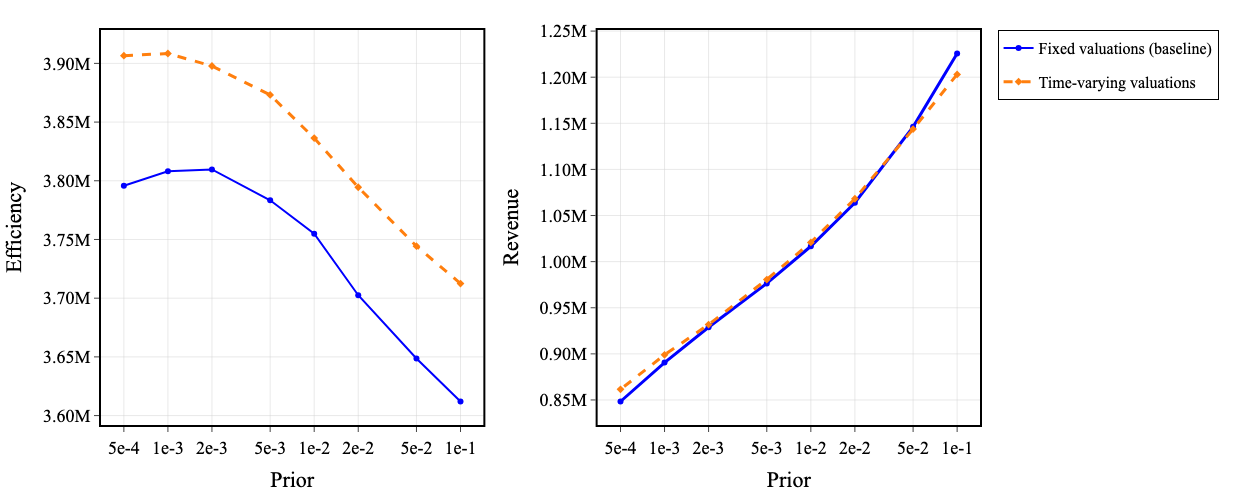}
    \caption{Counterfactual Outcomes with Time-Varying Valuations}
    \label{fig:time_varying_valuations}
    \vspace{0.25em}
    \begin{minipage}{0.95\textwidth}
    \footnotesize
    \textit{Notes:} The left panel shows efficiency and the right panel revenue. Solid blue lines use fixed valuations estimated from average submitted bids. Dashed orange lines use contemporaneous submitted bids as time-varying valuations. All other simulation inputs remain unchanged.
    \end{minipage}
\end{figure}

Figure~\ref{fig:time_varying_valuations} compares the fixed- and time-varying-valuation specifications. The left panel shows that the time-varying specification produces an approximately 2--3\% higher level of measured efficiency while preserving its qualitative relationship with the prior. This level difference reflects positive selection: bidders are more likely to win in periods when their contemporaneous valuations are relatively high, and the time-varying specification assigns those higher valuations to the resulting allocations.

The right panel shows that the revenue curves nearly overlap. The difference between the two specifications is less than 2\% at every prior mean considered, and both produce the same revenue-maximizing prior within the displayed grid. Taken together, the alternative specification modestly changes the quantitative results but preserves the qualitative relationships between the entrant prior, revenue, and efficiency that underpin our main conclusions.

\newpage

\section{Dynamic Structural Model and Estimation}
\label{appsec:dynamic}

In the  main specification of our model, we proved that truthful bidding is a weakly dominant strategy under the static-bidding assumption. This argument was justified by a la \cite{vickrey1961counterspeculation}. Key among the features of our environment that enables a Vickrey-style argument is another assumption that we have made: we have assumed that each bidder, in each period $t$, cares only about maximizing its payoff in that period and is, hence, indifferent towards what will transpire in the future. We justified this static bidding assumption by referring to evidence analyzed in section~\ref{sssec:bidding_dynamics}.

In this appendix, we relax the static bidding assumption and estimate a dynamic structural model of bidder behavior. If bidders are assumed patient, they might have an incentive to strategically manipulate their bids and influence the platform's learning process. 

The most important manifestation of such dynamic scenarios for our application is the following scenario: an incumbent expects to win the auction (perhaps due to past recent wins) but there is a recent entrant with a high score who is expected to finish a close second. If the entrant's high score is due to the high prior assigned to it by the platform (i.e., the very revenue extraction mechanism at the center of this paper's argument) and if the incumbent has fine-grained information on this score, then the incumbent has the means to break this scenario: by placing lower bids and allowing the entrant to win for a period of time, the incumbent will create the opportunity for the entrant's score to get adjusted down. Once the entrant's score shrinks adequately, the incumbent can get back to bidding truthfully and winning the auction. Note that with $\delta=0$, which was the assumption in the main text of the paper, the incumbent did not have the incentive to exhibit this behavior. But with $\delta>0$, such an incentive exists.

\subsection{Environment and Primitives}
\label{app:dynamic-environment}

Fix a keyword and index auctions by $t=1,\dots,T$ in chronological order. Unlike in the main text, we drop the keyword subscript \(k\) for brevity. Let $\mathcal{A}$ denote the set of potential bidders that appear for this keyword, and let $\mathcal{A}_t\subseteq\mathcal{A}$ be the subset of bidders eligible to participate in auction $t$. We treat the participation process $\{\mathcal{A}_t\}_{t=1}^T$ and the platform's reserve prices $\{r_t\}_{t=1}^T$ as exogenous and observed. In our setting, the reserve price is a fixed number. If $a\notin\mathcal{A}_t$, we restrict $b_{a,t}=0$, meaning the bidder cannot win and its learning state does not evolve in that period.

\subsubsection{State Evolution and Platform Learning}
For each bidder $a\in\mathcal{A}$, the learning state is summarized by $x_{a,t}\equiv (C_{a,t}, I_{a,t})$, where $I_{a,t}$ and $C_{a,t}$ are cumulative impressions and conversions, respectively, up to the start of auction $t$. Given common Beta prior hyperparameters $(\alpha^0,\beta^0)$, the platform draws the quality score $q_{a,t}$ from its posterior distribution of bidder $a$'s conversion rate at time $t$: $q_{a,t} \mid x_{a,t} \sim \mathrm{Beta}(\alpha_{a,t},\beta_{a,t})$, where $\alpha_{a,t}=\alpha^0+C_{a,t}$ and $\beta_{a,t}=\beta^0+(I_{a,t}-C_{a,t})$. The posterior mean is:
\[
\bar q_{a,t}\equiv \mathbb{E}[q_{a,t}\mid x_{a,t}] = \frac{\alpha^0+C_{a,t}}{\alpha^0+\beta^0+I_{a,t}}.
\]
The economy-wide state is $s_t \equiv \{x_{a,t}\}_{a\in\mathcal{A}}$. The combination of bids $\mathbf{b}_t$, the auction mechanism, and the realized conversion outcomes induces a Markov transition kernel $$P(s_{t+1}\mid s_t,\mathbf{b}_t).$$ Because the state updates conditionally on winning, an exogenous deviation in bidding alters the allocation, shifting the learning trajectory and future competitive dynamics.

\subsubsection{Payoffs and primitives for the incumbent}
Fix an incumbent bidder\footnote{Throughout this appendix, an \emph{entrant} is a bidder with no impressions on the keyword before its first win, and an \emph{incumbent} is any other bidder. A typical keyword has about 17 incumbents. We estimate the dynamic model for one incumbent per keyword, which we call the \emph{main incumbent}: the bidder with the highest true score on that keyword. This is the bidder for whom a forward-looking loss is most plausibly an investment, since it is the one most often paying a price set by a prior-inflated runner-up.} $a\in\mathcal{A}$. Bidder $a$ receives flow utility only when it wins and a conversion occurs. Let $v_a>0$ denote bidder $a$'s value per conversion, and let $\delta\in[0,1)$ denote its discount factor. Denote the auction winner by $A_t$ and the conversion outcome by $Y_t\in\{0,1\}$. Let $e_t$ be the per-conversion price by the winner in auction $t$. Bidder $a$'s one-period payoff is
\[
\pi_a(s_t,\mathbf{b}_t,Y_t;v_a)
=\mathbf{1}\{A_t=a\}\cdot Y_t \cdot (v_a-e_t).
\]

The discount factor $\delta$ governs the importance of future payoffs relative to current ones. This is economically important because a bid at time $t$ can change who wins today, which changes whose state updates, and therefore changes future auction outcomes and future payoffs. Given a bidding policy profile $\sigma$, bidder $a$'s total expected discounted payoff from state $s_t$ is
\[
V^\sigma(s_t;\delta,v_a)
\equiv
\mathbb{E}\!\left[\sum_{\tau=0}^{\infty}\delta^\tau\,
\pi_a(s_{t+\tau},\mathbf{b}_{t+\tau},Y_{t+\tau};v_a)
\ \Big|\ s_t,\sigma\right],
\]
where $\delta$ and $v_a$ are the structural parameters. When $\delta$ is close to zero, bidder $a$ behaves approximately myopically; when $\delta$ is large, it internalizes the effect of current bidding on future learning and future competition.

\subsection{Estimation}

A key simplification makes the estimation feasible: the state transitions are known without estimation. The platform's Beta prior hyperparameters $(\alpha^0,\beta^0)$, provided by the company, fully specify how quality scores update after each auction, and the true conversion rates, recovered from the data in Section~\ref{sssec:cvr_estimation}, supply the Bernoulli outcomes that trigger those updates. Together, these pin down how the state evolves under any bid profile, so we can simulate the market forward without having to estimate transitions from data.

The estimation itself has a nested structure. The \emph{inner loop} takes the discount factor $\delta$ as given and finds the value per conversion $v_a$ that best rationalizes the incumbent's observed bidding: specifically, the $v_a$ that minimizes the total extent to which one-shot deviations would have improved the incumbent's payoff. To run this comparison, the inner loop needs to simulate forward from any state, which requires knowing how all bidders bid as a function of the state; we estimate these bid policies in a preliminary first stage directly from the data. The \emph{outer loop} then searches over $\delta$: for each candidate discount factor, the inner loop finds its best-fit $v_a$, and the outer loop selects the $\delta$ that --- paired with its best-fit $v_a$ --- produces the smallest total violation of optimality.

\subsubsection{Pre-Estimation: State Transitions and Bid Policies}

The inner loop simulates the market forward, which requires tracking each bidder's state and its transitions. 

\subsubsection*{State Transitions}

The state transitions $P(s_{t+1}\mid s_t,\mathbf{b}_t)$ require no estimation:
the kernel is fully determined by objects already in hand, and we sample from
it by simulation rather than tabulating it. Given the state $s_t$ and the bid
profile $\mathbf{b}_t$, each bidder's quality score is drawn from the Beta
posterior implied by its state $(C_{a,t}, I_{a,t})$ and the known prior
$(\alpha^0,\beta^0)$, and the winner is the bidder with the highest
quality-adjusted bid clearing the reserve. The winner's conversion outcome is
then a Bernoulli draw with its conversion rate estimated in
$\S$\ref{sssec:cvr_estimation}, and the platform's updating rule increments the
winner's state by $(Y_t, 1)$, leaving all other bidders' states unchanged.
Each step of the forward simulation therefore draws quality scores and a
conversion outcome and writes down the next state directly.
\subsubsection*{State-Contingent Bid Policies}

Our model is a dynamic game. As a result, it requires us not only to learn the state transitions, but also to learn policies of different players as a function of the state. We will later use these learned bid policies and impose an equilibrium assumption a la BBL to estimate time discounting.

Our state space is very large. Hence, some states are not visited often enough for us to be able to learn the policies of different bidders at them. As a result, we only look at the space of policies that are defined on a coarsening of the original state space. That is, we approximate the information relevant to bidder $a$  by the triple

$$z_{a,t} \equiv (C_{a,t},\, I_{a,t},\, \kappa_{a,t})$$ 

The first two components capture bidder $a$'s cumulative conversions and impressions. The last component $\kappa_{a,t}$ is a predetermined competitiveness proxy based on the lagged \emph{maximum score of other bidders}. 

Transitions of this state depend on three inputs: the true conversion rates $\{\mu_a\}$, the platform's Beta prior $(\alpha^0, \beta^0)$, and the bid policies of all bidders. The first two are already available: conversion rates are estimated in Section~\ref{sssec:cvr_estimation} and the prior is provided by the company. We therefore estimate the bid policies here, before entering the nested loop. Following \citet{bajari2007estimating} (hereafter BBL), we treat observed bidding behavior as an equilibrium policy profile and estimate a flexible reduced-form bid function from the data.

\paragraph{Policy profile.}
For each bidder $a\in\mathcal{A}$, let $\sigma_a(z)$ denote a stationary bid policy mapping a low-dimensional state summary $z$ into bids. In the empirical implementation we use a binned (nonparametric) approximation, so that $\hat\sigma_a$ is constant within bins of $z$.

Define the (ex post) maximum competing score faced by bidder $a$ in auction $t$ as
\[
m_{a,t}\equiv
\max_{a'\in\mathcal{A}_t\setminus\{a\}}
\bigl(b_{a',t}\,q_{a',t}\bigr),
\]
with the convention that the maximum is zero if there are no competing eligible bidders.\footnote{If reserve eligibility is imposed before scoring, then the maximum is taken over reserve-eligible competitors only.}

Because $m_{a,t}$ depends mechanically on within-auction outcomes, we use a predetermined proxy in the state summary, namely the last observed value carried forward:
\[
\kappa_{a,t}\equiv m_{a,\tau(a,t)},\qquad
\tau(a,t)=\max\{t'<t:\ a\in\mathcal{A}_{t'}\},
\]
with an initialization rule when bidder $a$ has not appeared previously.

\paragraph{Binned policy estimator.}
For each bidder $a$, restrict to auctions where $a$ is present and submits a positive bid. Discretize $z_{a,t}$ into bins and estimate the policy as a bin-level average bid:
\[
\hat\sigma_a(z)
=\frac{1}{N_a(z)}\sum_{t:\ a\in\mathcal{A}_t,\ z_{a,t}\in \mathrm{bin}(z)} b_{a,t},
\]
where $N_a(z)$ is the number of observations for bidder $a$ in the bin. This delivers an estimated stationary policy profile $\hat\sigma=\{\hat\sigma_a\}_{a\in\mathcal{A}}$.
In the implementation, the bin cutoffs are chosen from the empirical quantiles of the bidder-keyword sample. When the policy is evaluated at a state outside the observed support, we apply boundary clipping before applying the bin rule.\footnote{We apply boundary clipping for numerical stability and to avoid extrapolating trends beyond the support of the observed data.} %

With the estimated policy profile $\hat\sigma$ in hand, all three inputs to the state transitions are determined. That is, we can now compute all of the transition probability distributions. We are now ready to proceed to the nested loop.

\subsubsection{Inner Loop: Forward Simulation and BBL Criterion}
The inner loop implements the BBL approach to recovering structural parameters. The key advantage of BBL is that it avoids solving the full dynamic programming problem: rather than computing the value function globally by solving a Bellman equation, it evaluates continuation values only at observed states by simulating forward under the estimated policy profile $\hat\sigma$. This is computationally tractable precisely because $\hat\sigma$ and the state transitions are already in hand from the pre-estimation step.

With the policy profile $\hat\sigma$ in hand, the inner loop proceeds as follows. Given a candidate parameter pair $(\delta,v_a)$ and the estimated policy profile $\hat\sigma$, define bidder $a$'s continuation value under $\hat\sigma$ as
\[
V^{\hat\sigma}(s;\delta,v_a)
\equiv
\mathbb{E}\!\left[\sum_{\tau=0}^{\infty}\delta^\tau\,
\pi_a(s_{t+\tau},\mathbf{b}_{t+\tau},Y_{t+\tau};v_a)
\ \Big|\ s_t=s,\ \hat\sigma\right].
\]
In practice, we approximate $V^{\hat\sigma}$ by forward simulation, truncating the horizon at $H$. We denote this approximation by $\widehat V^{obs}$:
\[
\widehat V^{obs}(s_t;\delta,v_a)
=
\sum_{\tau=0}^{H-1}\delta^\tau\,
\widehat \pi_{a,t+\tau}(v_a),
\]
where $\widehat \pi_{a,t+\tau}(v_a)$ denotes bidder $a$'s profit at simulated auction $t+\tau$, averaged across simulation runs. The simulated path applies: (i) bids given by $\hat\sigma_a(z_{a,t+\tau})$ for each $a\in\mathcal{A}_{t+\tau}$, (ii) allocation and payment under the auction rule, and (iii) state updating under the platform's learning rule. We use common random numbers across baseline and counterfactual simulations.

The inner loop's objective is the BBL criterion: it identifies structural parameters by enforcing that observed behavior is (approximately) optimal relative to one-shot deviations, given the estimated policy profile and the known state transitions.

\paragraph{A single deviation per decision point.}
Fix a decision point $t$ where bidder $a$ is present. Let the baseline path use bidder $a$'s observed action at time $t$, while all other bidders use $\hat\sigma$ at time $t$ and all bidders use $\hat\sigma$ from time $t+1$ onward. We consider exactly one economically meaningful one-shot deviation at $t$:
\[
b^{dev}_{a,t}\equiv
\begin{cases}
0, & \text{if } A_t=a \text{ under the baseline action (``force-lose'')},\\[6pt]
\dfrac{\max\{r_t,\ m_{a,t}+\varepsilon\}}{q_{a,t}},
& \text{if } A_t\neq a \text{ under the baseline action (``force-win'')},
\end{cases}
\]
where $\varepsilon>0$ is a small number, $m_{a,t}$ is the realized maximum score of the other bidders in auction $t$, and $q_{a,t}$ is bidder $a$'s realized quality draw in auction $t$.

Define the one-shot deviation policy profile $\hat\sigma^{(t)}$ that changes only bidder $a$'s action in period $t$:
\[
b_{a,t}=b^{dev}_{a,t},
\qquad
b_{a,t+\tau}=\hat\sigma_a(z_{a,t+\tau})\ \ \forall \tau\ge 1,
\qquad
b_{a',t+\tau}=\hat\sigma_{a'}(z_{a',t+\tau})\ \ \forall a' \neq a,\ \forall \tau\ge 0.
\]
Although the deviation is one-shot in actions, it can change the winner and therefore the next-period state, which then changes future bids endogenously through the state dependence of $\hat\sigma$.

\paragraph{Inequality and violations.}
Equilibrium optimality implies
\[
V^{\hat\sigma}(s_t;\delta,v_a)\ \ge\ V^{\hat\sigma^{(t)}}(s_t;\delta,v_a).
\]
We estimate both sides by forward simulation, producing $\widehat V^{obs}(s_t;\delta,v_a)$ and $\widehat V^{dev}(s_t;\delta,v_a)$. Define the deviation gain and violation:
\[
g_t(\delta,v_a)\equiv \widehat V^{dev}(s_t;\delta,v_a)-\widehat V^{obs}(s_t;\delta,v_a),
\qquad
viol_t(\delta,v_a)\equiv \max\{0,\ g_t(\delta,v_a)\}.
\]
Aggregating over a set of decision points $\mathcal{T}$ yields the BBL criterion:
\[
Q(\delta,v_a)\equiv \frac{1}{|\mathcal{T}|}\sum_{t\in\mathcal{T}} viol_t(\delta,v_a)^2.
\]
Given a discount factor $\delta$, the inner loop minimizes $Q(\delta,v_a)$ over $v_a$, returning the value per conversion that best rationalizes the incumbent's observed bids at that discount factor.

\subsubsection{Outer Loop: Estimating $\delta$}
The outer loop searches over $\delta$ on a grid.\footnote{\label{fn:delta_id}A natural concern is that discount factors are not identified in dynamic discrete choice models \citep{magnac2002identifying}. That result relies on a per-period payoff that is an unrestricted function of the state, which can absorb the continuation value for any $\delta$. Our payoff is instead linear in a single unknown parameter, $\pi_a=\mathbf{1}\{A_t=a\}\,Y_t\,(v_a-e_t)$, with the price observed and entering with a known coefficient; the belief states affect payoffs only through future prices and allocations. \citet{komarova2018joint} show that $\delta$ and the payoff parameters are generically point identified in this linear-in-parameters class, and Section~\ref{app:transitions-results} documents the variation that provides the identifying power in our data: at the median keyword, losing to an entrant lowers the price the incumbent expects to face in the next auction by about $2.7\%$, whereas losing to a non-entrant lowers it by only $0.4\%$, so the continuation gain from losing is sizable at decision points where the runner-up is an entrant and close to zero at the others; since $v_a$ enters the per-period profit identically at both types of decision points, a difference in the incumbent's willingness to lose across them can be rationalized only through $\delta$.} For each candidate $\delta$, the inner loop returns the best-fit value per conversion together with the resulting minimized criterion:
\[
v^*(\delta)\in\arg\min_{v_a>0} Q(\delta,v_a),
\qquad
Q_{\mathrm{prof}}(\delta)\equiv Q\bigl(\delta,v^*(\delta)\bigr).
\]
The outer loop then selects the discount factor with the smallest profiled criterion and reports the valuation associated with it:
\[
\hat\delta\in\arg\min_{\delta\in[0,1)} Q_{\mathrm{prof}}(\delta),
\qquad
\hat v_a\equiv v^*(\hat\delta).
\]
The pair $(\hat\delta,\hat v_a)$ is therefore the joint minimizer of $Q(\delta,v_a)$ over both parameters; the nested structure is how we compute it, not a restriction on it. The profiled criterion $Q_{\mathrm{prof}}(\delta)$ is also the object we plot in Section~\ref{app:delta-results} and average across keywords in the common-$\delta$ estimator below.

In addition, we estimate a \emph{common} discount factor across keywords by allowing each keyword to have its own profiled valuation while restricting the discount factor to be the same across all keyword-incumbent pairs. Let \(\mathcal K\) denote the set of keyword-incumbent pairs in the estimation sample. The common discount factor is defined as
\[
\hat\delta^{\,\mathrm{common}}
\in
\arg\min_{\delta\in[0,1)}
\frac{1}{|\mathcal K|}
\sum_{k\in\mathcal K}
Q_{k,\mathrm{prof}}(\delta).
\]
Thus, \(\hat\delta^{\,\mathrm{common}}\) is the discount factor that minimizes the average profiled criterion across keywords.

\subsection{Results and Interpretation}
We implement the BBL estimation procedure across all queries and corresponding dominant incumbents. By evaluating the optimality of observed bids against counterfactual deviations, we recover point estimates for the discount factor ($\hat{\delta}$) and valuation ($\hat{v}_a$) for each keyword-incumbent pair.

Per the estimation routine described in the previous section, the objects to report for each query should entail the state transitions, bidding policies, as well as the parameters  ($\hat{\delta}$) and  ($\hat{v}_a$) of the dynamic value functions. Given the large number of objects, however, we find that reporting them all in detail would be cumbersome. We instead focus on two objects in reporting, which we believe are of the most economic significance. We start by providing statistics that summarize the state transitions (as opposed to reporting a large matrix). We then move to reporting the estimated discount factors.\footnote{That is, we skip the reporting of estimated bid policies by all bidders, as well as the reporting of estimated values $\hat{v}_a$.}

\subsubsection{Estimated state transitions}
\label{app:transitions-results}

Given that the size of the state space is large, reporting a full transition matrix would not be informative. We, hence, focus instead on summary statistics that shed light on the economics of whether dynamic considerations alter the key recommendation we made using a static model in the main text.

Specifically, we examine two types of statistics. First, we look at how an incumbent's loss in an auction changes, in expectation, the price the incumbent would face in the subsequent auction. Since the incumbent's price is the second-highest score scaled by the incumbent's own quality draw, this is the price-denominated analogue of asking how a loss moves the future second-highest score. Such an effect arises if the second-highest score is artificially high due to the platform's prior: by losing to that bidder, the incumbent provides an opportunity for the score to get adjusted down towards the bidder's true CVR and away from the platform's originally high prior. If the incumbent is forward looking, and if the impact of a loss on its future price is non-trivial, this will provide an incentive for the incumbent to underbid and lose on purpose.

The second type of statistic that we examine is focused on understanding the rate of entry of new bidders. It matters because it relates to the dynamic argument built in the previous paragraph. Suppose an adequately forward looking incumbent has the dynamic incentive to lose to the current second-highest-score bidder in order to lower that score in the future, thereby enjoy a larger continuation payoff when it gets back to bidding high. The intensity of this dynamic incentive depends on how long the incumbent believes the second-highest-score bidder will remain second-highest before another high-score bidder enters and renders the incumbent foresighted losing streak payoff-irrelevant. This expected time period is, in turn, a function of the entry rates in the market.

We turn to reporting these statistics next.

\subsubsection*{Impact of a loss on the incumbent's future price}

To construct this statistic, we focus on auctions in which the incumbent $a$ is present but loses to bidder $a'=A_t\neq a$. The dynamic consequence of the loss is that bidder $a'$'s posterior parameters update after its win. We isolate this channel by comparing the observed post-win state, $(\alpha_{a',t+1},\beta_{a',t+1})$, to a counterfactual in which bidder $a'$ enters the next auction with its pre-win parameters, $(\alpha_{a',t},\beta_{a',t})$. All other objects --- the remaining bidders' states, bids, the reserve price, and the next auction's participant set --- are held fixed.

For a given draw of qualities in auction $t+1$, the incumbent's potential price-per-conversion is $\max\{r_{t+1},\, m_{a,t+1}\}/q_{a,t+1}$, where $m_{a,t+1}$ is the maximum competing score defined earlier and $q_{a,t+1}$ is the incumbent's own quality draw. We compute the expectation of this price under the observed and counterfactual states for bidder $a'$, and report the post-loss price change as the percentage change in the expected price; a negative value means the loss lowers the price the incumbent expects to face in the next auction. We compute this measure for every incumbent-loss event, average within keyword, and condition on the identity of the winning bidder: an entrant (a bidder with zero impressions before winning) versus a non-entrant.

\begin{figure}[htp!]
    \centering
    \includegraphics[width=0.6\linewidth]{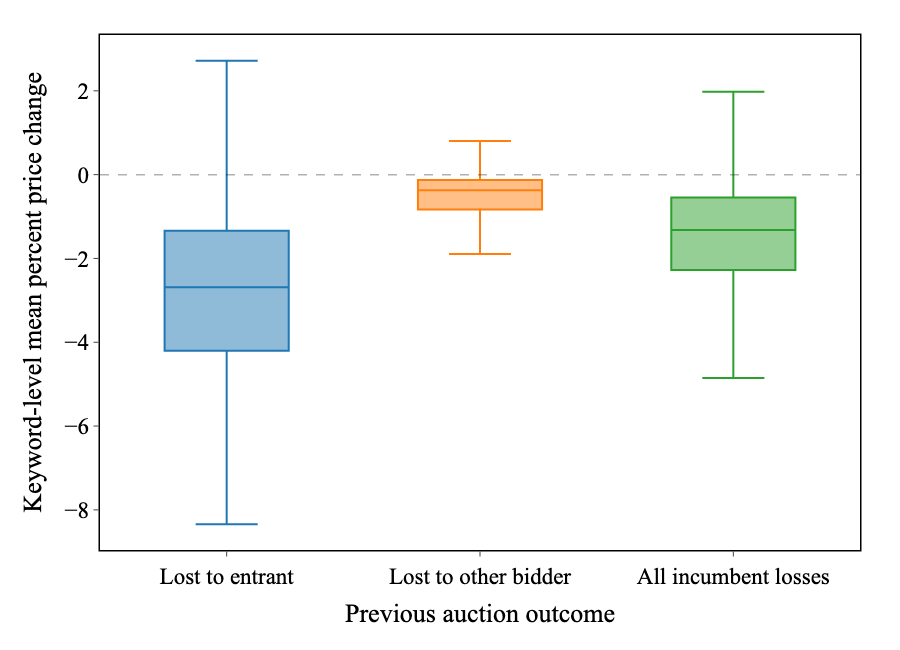}
    \caption{Post-Loss Changes in the Incumbent's Expected Price}
    \label{fig:loss_lift_distribution}
    \vspace{0.25em}
    \begin{minipage}{0.95\textwidth}
    \footnotesize
    \textit{Notes:} The distributions show keyword-level mean percentage changes in the incumbent's expected next-auction price following losses to entrants, losses to non-entrants, and all losses pooled. Boxes show the interquartile range ($IQR=Q_3-Q_1$), with the median marked inside; whiskers extend to the most extreme observations within $[Q_1-1.5\,IQR,\;Q_3+1.5\,IQR]$. Individual observations outside the whiskers are not displayed for readability.
    \end{minipage}
\end{figure}

Figure~\ref{fig:loss_lift_distribution} plots the distribution of keyword-level average post-loss price changes. Losing generally lowers the incumbent's expected future price, but the magnitude depends sharply on who the incumbent loses to. The median keyword-level price change after losing to an entrant is approximately $-2.69\%$ (interquartile range $-4.20\%$ to $-1.34\%$), compared to only $-0.37\%$ (interquartile range $-0.83\%$ to $-0.13\%$) after losing to a non-entrant; pooling all losses, the median is $-1.32\%$.

\paragraph{Interpretation.} This pattern is consistent with the mechanism motivating this appendix: the price relief from losing is concentrated precisely where the platform's prior inflates the competing score, namely entrants. As a result, a sufficiently patient incumbent would have some incentive to lose the current auction in the hope of paying a lower price in future auctions it will win. This variation is what identifies the discount factor $\delta$ in our estimation (see Footnote~\ref{fn:delta_id}).

\subsubsection*{Analysis of entry rates}

The impact of the current bid on future prices, documented above, is what helps estimate the discount factor $\delta$. This matters for policy: if $\delta$ is large, the platform may be discouraged from following the exploratory policy (i.e., the high prior) that we recommend, since patient incumbents would fight it by losing many times on purpose and thereby adjusting competitors' scores down.

However, in addition to $\delta$, there is another factor that determines whether --- and to what extent --- the platform should be concerned about incumbents' foresighted reaction to a high prior: the churn rate. The dynamic incentive discussed above is most pronounced when the incumbent expects its current closest competitors (as measured by score) to remain its closest competitors for a long time. The incentive is mitigated if the incumbent regularly expects a new strong entrant --- perhaps strong precisely because of the head start given to it by the platform's high prior --- in the near future. In that case, losing to the current second-ranked bidder is an investment that does not yield a durable return, and is hence discouraged.

Two statistics reported earlier in the paper combine to measure this churn rate; Table~\ref{tab:churn} collects the relevant numbers. First, entry is frequent: the median keyword features 20 entrants over roughly 1,400 auctions, i.e., a new entrant arrives about once every 70 auctions. Second, arriving entrants almost surely make their way to the top: virtually all entrants eventually receive impressions --- that is, finish first in at least one auction --- and 80\% of them do so at least 100 times.

\begin{table}[htp!]
\centering
\footnotesize{
\begin{tabular}{lcc}
\hline\hline \\[-1.8ex]
\textbf{Statistic} & \textbf{Value} & \textbf{Source}\\ \hline
\\[-1.8ex]
Entrants per keyword (median) & 20 & Table~\ref{tab:summary} \\
Auctions per keyword (median) & 1,434 & Table~\ref{tab:summary} \\
\quad \textit{Implied entry rate} & \textit{one per $\sim$70 auctions} & \\[0.1cm]
Entrants eventually winning at least one auction & $\approx$ 100\% & Figure~\ref{fig:entrant_cdf} \\
Entrants winning at least 100 auctions & 80\% & Figure~\ref{fig:entrant_cdf} \\[0.1cm]
\quad \textit{Implied rate of a new bidder reaching the top rank} & \textit{one per $\sim$70 auctions} & \textit{steady state}\\
\hline\hline
\end{tabular}
\caption{Ingredients of the churn-rate calculation.\label{tab:churn}}}
\end{table}

Combining these two statistics through a steady-state argument yields the churn rate of interest. Since essentially every entrant eventually crosses into the winning rank, the rate at which \emph{some} new bidder crosses into the top of the score distribution must, in steady state, equal the entry rate itself --- about once every 70 auctions --- regardless of the delay between an entrant's arrival and its first win.\footnote{The steady-state argument requires that entry into these keywords has been ongoing at similar rates before our sample window, so that crossings by earlier entrants arrive during the sample at the same rate as new entries. Given that the platform and these keywords long predate our sample, we view this as innocuous. The calculation is also conservative in that it counts only crossings into the \emph{first} rank; adding entrants who reach the second-highest score would make churn look even faster.}

\paragraph{Interpretation.} A back-of-envelope calculation combines the two statistics reported in this section and illustrates what is at stake. Deliberately losing one auction costs the incumbent one period's expected margin, $\mu_a(v_a-e)$; in return, it enjoys the price relief documented above --- roughly $1.3\%$ of the price, or $2.7\%$ when the runner-up is an entrant --- for at most the ${\sim}70$ auctions that churn allows before a fresh entrant renders the investment obsolete. So the deviation pays if and only if the relief, cumulated over up to $70$ discounted periods, exceeds the markup $(v_a-e)/e$. At the markup implied by typical observed market thickness (on the order of $50\%$), this requires a per-auction discount factor above roughly $0.98$ ($0.95$ when the runner-up is an entrant). Two lessons follow. First, because auctions on a keyword are only minutes apart, any calendar-time rate of time preference maps into a per-auction $\delta$ indistinguishable from one; whether incumbents engage in foresighted losing is therefore a question about behavioral patience --- precisely the $\delta$ we estimate next, and find to be near zero. Second, churn bounds the temptation rather than eliminating it: even a fully patient incumbent would earn at most a few times one period's margin from the maneuver, and this cap tightens mechanically as the platform raises its prior, since a higher prior accelerates the arrival of strong entrants.

\subsubsection{Estimated discount factors}
\label{app:delta-results}

\begin{figure}[htp!]
    \centering
    \includegraphics[width=0.5\linewidth]{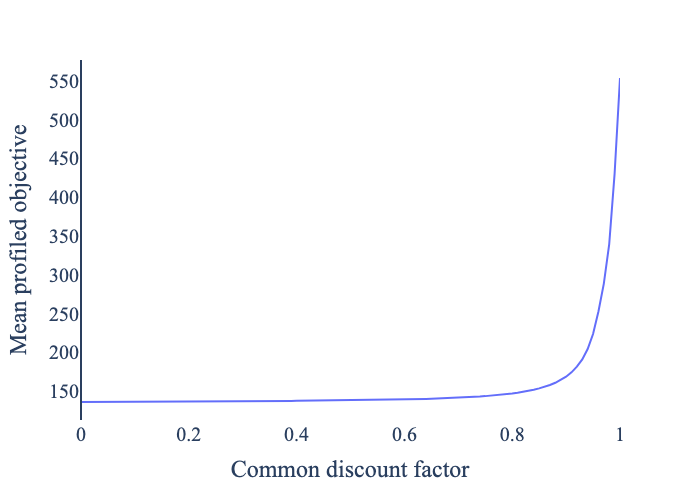}
    \caption{Profiled Objective for the Common Discount Factor}
    \label{fig:common_delta_profile}
    \vspace{0.25em}
    \begin{minipage}{0.95\textwidth}
    \footnotesize
    \textit{Notes:} For each discount factor, the criterion is minimized over each incumbent's valuation and averaged across keyword--incumbent pairs. The objective is evaluated on a grid of 101 points uniformly spaced from $0$ to $0.9999$.
    \end{minipage}
\end{figure}

We begin by estimating a common discount factor across all keywords. Figure \ref{fig:common_delta_profile} plots the average profiled objective,
\[
\frac{1}{|\mathcal K|}\sum_{k\in\mathcal K} Q_{k,\mathrm{prof}}(\delta),
\]
against \(\delta\). The average criterion is minimized at \(\delta=0\), implying that the pooled objective across all keywords favors a static benchmark. This pooled result, however, can mask heterogeneity across keyword-incumbent pairs.

\begin{figure}[htp!]
    \centering
    \includegraphics[width=0.4\linewidth]{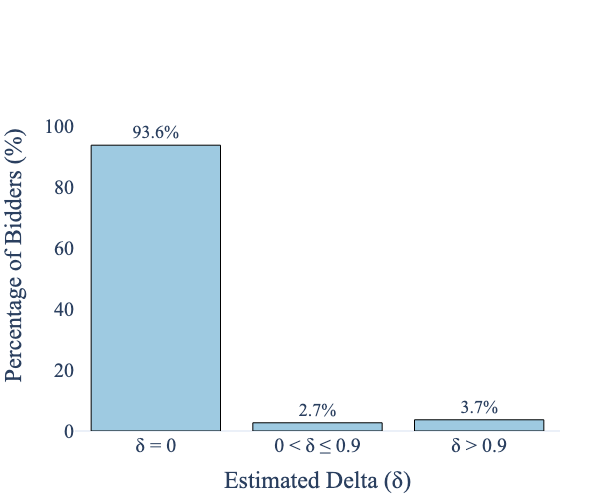}
    \caption{Distribution of Estimated Discount Factors}
    \label{fig:delta_dist}

    \vspace{0.25em}
    \begin{minipage}{0.95\textwidth}
    \footnotesize
    \textit{Notes:} Each estimate corresponds to one keyword--incumbent pair. The discount factor is selected from a grid of 11 points ranging from $0$ to $0.9999$.
    \end{minipage}
\end{figure}

We therefore next turn to the distribution of pair-specific discount factor estimates.
Figure \ref{fig:delta_dist} displays the distribution of the estimated discount factors across the sample. The results allow us to categorize bidding behavior, distinguishing between static agents ($\hat{\delta} \approx 0$) who optimize only for immediate payoffs, and forward-looking agents who internalize the dynamic value of learning and platform state updates.

As the figure shows, the estimates are concentrated at zero: $93.6\%$ of keyword-incumbent pairs are estimated to be fully static, while $3.7\%$ are assigned a high discount factor ($\hat{\delta}>0.9$). The latter segment accounts for about $6.3\%$ of revenue and about $3.3\%$ of conversions in the set of queries we evaluate. With these estimates in hand, we next turn to counterfactuals.

\subsection{Counterfactuals}
\label{app:dynamic-counterfactuals}

This section uses the estimates of the previous section to extend the counterfactual analysis of Section~\ref{sec:results} in two directions. First, we relax the static-bidding assumption on which the main counterfactuals rest: incumbents now choose bids dynamically, either at their estimated discount factors or at a discount factor near one. Second, we extend the range of prior means beyond $0.1$, the upper end of the range examined in the main text. Recall that the main text stops at $0.1$ precisely because higher priors are where an incumbent's incentive to bid strategically becomes a concern (footnote~\ref{fn:prior_cap}); with a dynamic bidding model in hand, we can evaluate those priors directly. The remainder of this section describes how we compute the incumbent's dynamic bidding policy, lays out the three specifications we compare, and reports the results.

The environment and primitives are those of Section~\ref{app:dynamic-environment}, evaluated for each keyword at the incumbent's estimated parameters $(\hat v_a,\hat\delta)$. We write $\pi_a(s_t,b_{a,t})$ for the incumbent's expected per-period profit: $\mu_a(\hat v_a-e_t)$ if the incumbent wins the auction at per-conversion price $e_t$, and zero otherwise, where $\mu_a$ is the incumbent's true conversion rate estimated in Section~\ref{sssec:cvr_estimation}. In principle, because the horizon is finite, the incumbent's dynamic problem could be solved by backward induction. However, the state space is large because the state includes the histories of all bidders in the keyword. We therefore compute the incumbent's policy by an approximate policy-iteration procedure based on \(n\)-step temporal-difference (TD) updates.

We initialize the incumbent's bidding policy at the static value-bidding rule, $b^{(0)}_{a,t}=\hat v_a$. Given a current policy \(b^{(m)}\), we evaluate two local deviations at each reached state: winning the current auction and losing it.\footnote{Reducing the continuous bid choice to this binary comparison is without loss of optimality. Because the auction is second-price, the incumbent's bid affects neither the price it pays upon winning, $e_t=\max\{r_t,\,m_{a,t}\}/q_{a,t}$, where $m_{a,t}$ is the highest competing score and $q_{a,t}$ the incumbent's own quality draw, nor the platform's state update, which depends only on the winner's identity; and in this counterfactual the other bidders' bids are constant, so they cannot respond to the bid level either. Consequently, all losing bids yield the same payoff of zero and the same continuation path, and all winning bids yield the same per-period payoff $\mu_a(\hat v_a-e_t)$ and the same continuation path. The bid level matters only through the win/lose outcome, and the willingness-to-pay rule derived below therefore fully characterizes optimal bidding.} After forcing action \(\omega \in \{win,lose\}\) at auction \(t\), we approximate the continuation value by the \(n\)-step TD estimate

\[
\widehat V^{\omega,m}_{t+1}(s_{t+1}^{\omega})
=
\sum_{\ell=1}^{n-1}
\hat\delta^{\ell-1}
\widehat\pi_a(s_{t+\ell}^{\omega}, b^{(m)}_{a,t+\ell})
+
\hat\delta^{n-1}
\widehat V^{m}_{t+n}(s_{t+n}^{\omega}),
\]
where \(\widehat\pi_a\) denotes the incumbent's per-period profit averaged across simulation runs, and \(s_{t+\ell}^{\omega}\) is the simulated state reached after taking action \(\omega\) at \(t\) and then following the current policy \(b^{(m)}\). The final term \(\widehat V^{m}_{t+n}(s_{t+n}^{\omega})\) is the bootstrap value under the current policy.

The value of winning the current auction at price \(e_t\) is then \(\mu_a(\hat v_a-e_t)+\hat\delta\,\widehat V^{win,m}_{t+1}(s_{t+1}^{win})\), while the value of losing is \(\hat\delta\,\widehat V^{lose,m}_{t+1}(s_{t+1}^{lose})\). The price at which the incumbent is indifferent between the two defines its dynamic willingness to pay, which we take as the updated policy:

\[
b^{(m+1)}_{a,t}
=
\hat v_a
+
\frac{\hat\delta}{\mu_a}
\left[
\widehat V^{win,m}_{t+1}(s_{t+1}^{win})
-
\widehat V^{lose,m}_{t+1}(s_{t+1}^{lose})
\right].
\]

When losing today improves the incumbent's future prospects --- for instance, by allowing an entrant's prior-inflated score to adjust downward --- the bracketed term is negative and the incumbent shades its bid below $\hat v_a$. We update the incumbent's bid using this expression and repeat the policy-improvement step until the policy converges.

We compare three counterfactual specifications. They differ in the discount factor assigned to incumbents and, crucially, in whether that discount factor is estimated or imposed:
\begin{enumerate}
    \item \emph{Main counterfactual} ($\delta=0$, imposed). This is the specification of the main text. All advertisers, incumbents included, bid their valuations in every auction, with valuations inferred from average observed bids under the static-bidding assumption. The discount factor is not estimated; it is fixed at zero by assumption.
    \item \emph{Dynamic model} ($\delta=\hat\delta$, estimated). Each keyword's main incumbent uses its estimated pair $(\hat v_a,\hat\delta)$ from Section~\ref{app:delta-results} and bids according to the policy-iteration rule above. This is the only specification in which the discount factor is taken from the data.
    \item \emph{Forward-looking model} ($\delta=0.9999$, imposed). This is the dynamic model with the discount factor set to $0.9999$ for every incumbent, so that all incumbents are treated as fully patient. It retains the estimated valuations $\hat v_a$, so the comparison with the dynamic model isolates the role of patience alone.
\end{enumerate}
In all three specifications, non-incumbent bidders use the constant bids of the main counterfactuals. Comparing the first two specifications asks whether the estimated dynamics change the conclusions of the main text; comparing the last two asks how much they could change them if incumbents were as patient as possible.

\begin{figure}[!ht]
    \centering

    \begin{subfigure}[t]{0.48\textwidth}
        \centering
        \includegraphics[width=\linewidth]{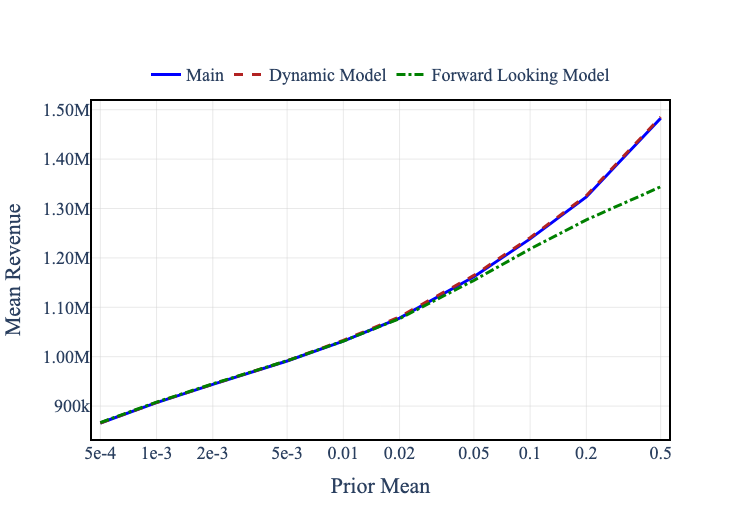}
        \caption{Revenue}
        \label{fig:forward-looking-counterfactuals-revenue}
    \end{subfigure}
    \hfill
    \begin{subfigure}[t]{0.48\textwidth}
        \centering
        \includegraphics[width=\linewidth]{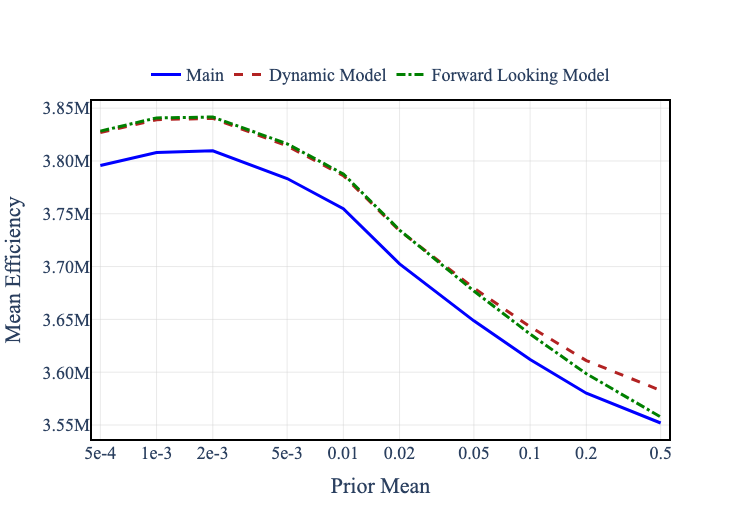}
        \caption{Efficiency}
        \label{fig:forward-looking-counterfactuals-efficiency}
    \end{subfigure}

    \caption{Counterfactual Outcomes Under Alternative Bidding Models}
    \label{fig:forward-looking-counterfactuals}
    \vspace{0.25em}
    \begin{minipage}{0.95\textwidth}
    \footnotesize
    \textit{Notes:} The main counterfactual imposes $\delta=0$ and uses valuations estimated from average bids. The dynamic model uses the main incumbent's estimated valuation and discount factor for each keyword. The forward-looking model retains these estimated valuations but imposes $\delta=0.9999$. Each specification is evaluated across entrant prior means.
    \end{minipage}
\end{figure}

Figure \ref{fig:forward-looking-counterfactuals} reports revenue and efficiency across prior means. The dynamic model is close to the main counterfactual, while the forward-looking model shows the effect of stronger dynamic incentives.

\subsubsection*{Interpretation}

Panel (a) carries two messages. First, the dynamic model is nearly indistinguishable from the main counterfactual. This is a direct implication of the estimates in Section~\ref{app:delta-results}: with $93.6\%$ of keyword-incumbent pairs estimated to be fully static, feeding the estimated $(\hat v_a,\hat\delta)$ into the dynamic bidding rule reproduces essentially static behavior. The value of this exercise is internal consistency --- the dynamic machinery, evaluated at the estimated primitives, confirms that dynamics do not alter the revenue analysis of the main text.

Second, and more importantly, even the forward-looking model --- in which every incumbent is fully patient, with a discount factor above the break-even threshold computed in Section~\ref{app:transitions-results} --- does not overturn the revenue gains from higher priors. Revenue under this model separates from the main counterfactual only at high prior means, precisely where the platform's prior inflates entrants' scores and gives incumbents something to fight; yet it continues to rise with the prior throughout, ending roughly $9\%$ below the main counterfactual at the highest prior we consider. The reason revenue is dampened rather than curbed is churn. As documented in Section~\ref{app:transitions-results}, deliberately losing in order to dampen a competitor's score is an investment with a short payback window: a new strong bidder crosses into the top of the score distribution roughly once every $70$ auctions, at which point the dampened score of the previous rival becomes payoff-irrelevant. Moreover, a higher prior itself exacerbates this force: the higher the platform's prior, the stronger arriving entrants start out, and the sooner an incumbent's investment in suppressing the current runner-up is rendered obsolete. In this sense, the exploratory policy partially protects itself --- raising the prior strengthens the very churn that disciplines incumbents' dynamic incentives to resist it.

Panel~(b) shows that all three specifications produce the same qualitative pattern: efficiency is maximized at a relatively modest prior and declines as the prior becomes more optimistic. The level difference between the main and dynamic curves largely reflects their different valuation primitives. The main counterfactual infers valuations from average observed bids under the static truth-telling assumption, whereas the dynamic specifications use the structural estimates from Appendix~\ref{appsec:dynamic}. By contrast, the dynamic and fully forward-looking specifications use the same valuations and differ only in their discount factors. These curves are nearly identical at low priors, where entrants receive little advantage and incumbents have weak incentives to shade their bids. They separate at higher priors, where fully patient incumbents increasingly shade their bids and relinquish some auctions to prior-inflated entrants, modestly reducing allocative efficiency.

\clearpage
\putbib[references]
\end{bibunit}
\end{appendices}

\end{document}